\documentclass[11pt]{article}
\usepackage{amsmath}
\usepackage{amssymb}
\usepackage{relsize}
\usepackage[bottom]{footmisc}
\usepackage[numbers,sort]{natbib}
\usepackage[textwidth=460pt,textheight=670pt]{geometry}
\usepackage{graphicx}
\usepackage{psfrag}
\usepackage{xspace}
\usepackage[bf,small]{caption2}
\setcaptionwidth{.9\textwidth}
\makeatletter
\@addtoreset{equation}{section}

\makeatother

\newcommand\bulletskip{1.2ex}
\newcommand\bulletsep{\vskip 1.7ex}
\newcommand\eqn[1]{(\ref{#1})}

\DeclareMathOperator{\Tr}{Tr}

\DeclareMathOperator{\sgn}{sgn}
\def\x{{\bf x}}
\def\y{{\bf y}}

\def\Hk{{\cal H}_{\rm kin}}
\def\Hd{{\cal H}_{\rm diff}}
\def\S{{\cal S}}

\def\D{{\cal D}}
\def\V{{\cal V}}
\def\N{{\iota}}
\def\hH{\hat{H}}
\def\hO{\hat{O}}
\def\d{{\rm d}}
\providecommand{\href}[2]{#2}

\newcommand{\wsl}{weak~$\!^*\!$~limit\xspace}
\newcommand{\wst}{weak~$\!^*\!$~topology\xspace}

\begin{document}
\pagestyle{empty}
\begin{flushright}
AEI-2004-129\\
hep-th/0501114\\
January 14th, 2005
\end{flushright}
\vskip 7ex

\begin{center}
\begin{minipage}{.9\textwidth}
{\huge\bf Loop quantum gravity: an outside view}\\[6ex]
{\large\bf Hermann Nicolai, Kasper Peeters and Marija Zamaklar}\\[5ex]
Max-Planck-Institut f\"ur Gravitationsphysik\\
Albert-Einstein-Institut\\
Am M\"uhlenberg 1\\
14476 Golm, GERMANY\\[3ex]
{\tt hermann.nicolai}, 
{\tt kasper.peeters},
{\tt marija.zamaklar@aei.mpg.de}
\vskip 7ex

{\bf Abstract:}\\[1ex] 
We review aspects of loop quantum gravity in a pedagogical manner, with the 
aim of enabling a precise but critical assessment of its achievements so
far. We emphasise that the off-shell (`strong') closure of the constraint 
algebra is a crucial test of quantum space-time covariance, and thereby of 
the consistency, of the theory. Special attention is paid to the appearance 
of a large number of ambiguities, in particular in the formulation of 
the Hamiltonian constraint. Developing suitable approximation methods 
to establish a connection with classical gravity on the one hand, 
and with the physics of elementary particles on the other, remains 
a major challenge.
\end{minipage}
\vskip 5ex

\end{center}
\newpage
\pagestyle{plain}
\hrule
\tableofcontents
\bigskip\medskip
\hrule
\bigskip\medskip

\section{Key questions}

When four-dimensional Einstein gravity is quantised canonically, in a
perturbation series in the Newton constant around flat space-time,
divergences arise at two-loop order. An impressive calculation by
Goroff and Sagnotti~\cite{Goroff:1985sz,Goroff:1985th} has
demonstrated that in order to obtain a finite \mbox{S-matrix}, the
action should contain a counterterm
\begin{equation}
\label{e:2loopdiv}
\Gamma^{(2)}_{\text{div}} = \frac{1}{\epsilon}\frac{209}{2880}
\frac{1}{(16\pi^2)^2}
\int\!{\rm d}^4 x\, \sqrt{g}\, C_{\mu\nu\rho\sigma} 
                C^{\rho\sigma\lambda\tau} C_{\lambda\tau}{}^{\mu\nu}\,,
\end{equation}
a result which was later confirmed in an independent background-field
calculation by van de Ven~\cite{vandeVen:1991gw}. The usual conclusion
drawn from this result, and from the fact that the coupling constant
is dimensionful, is that an infinite number of counterterms is needed. 
It is generally agreed that this non-renormalisability renders perturbatively 
quantised Einstein gravity meaningless as a fundamental theory because
an infinite number of parameters would be required to make any physical 
prediction.

However, this still leaves open the possibility that Einstein gravity
\emph{can} be quantised consistently, but that it is simply the
perturbation series in Newton's constant which is ill-defined.  This
possibility has been raised and advocated not only in the context of
canonical quantisation, but also, and independently, in the context of
suggestions that there may exist a non-trivial fixed point of the
renormalisation group in Einstein's
theory~\cite{wein4,wein3,Lauscher:2001rz}. Implicitly, it also
underlies the path integral approach to Euclidean quantum
gravity~\cite{Gibbons:1976ue,hawk3}, which provides a possible 
framework for the discussion of semi-classical
states~\cite{Hartle:1983ai}. Indeed, to date there is no proof that
such a quantisation which does \emph{not} make use of a series expansion
around a fixed background is guaranteed to fail. However, quantising
without relying on perturbation theory around a free theory is hard.

Loop quantum gravity, or LQG for short, is an attempt to quantise Einstein 
gravity non-perturbatively. In contrast to string theory, which posits 
that the Einstein-Hilbert action is only an effective low energy approximation 
to some other, more fundamental, underlying theory, LQG takes Einstein's 
theory in \emph{four} spacetime dimensions as the basic starting point.
It is a Hamiltonian approach, which is background independent in the
sense that its basic quantities and concepts do not presuppose the
existence of a given spatial background metric. In comparison to the 
older geometrodynamics approach (which is also formally background
independent) it makes use of many new conceptual and technical ingredients. 
A key role is played by the Ashtekar variables, which allow the reformulation 
of gravity in terms of connections and holonomies, and which (at least in 
the form originally proposed) greatly simplify the constraints. A related
key feature is the use of spin networks, which in turn requires other 
mathematical ingredients, such as non-separable (`polymer') Hilbert spaces 
and representations of operators, which are not weakly continuous.
Undoubtedly, novel concepts and ingredients such as these will be 
necessary in order to circumvent the problems of perturbatively quantised 
gravity, but, as for any other approach to quantum gravity, it is important 
not to loose track of the physical questions that one is trying to answer.

The goal of the present paper is to review some essential properties of
loop quantum gravity in an easily accessible way from a non-specialist's 
perspective, and with a non-LQG audience in mind. It is \emph{not meant 
to be a comprehensive review of the subject}; readers who want to know
more about the very latest developments in this field are instead 
referred to a number of excellent recent reviews describing the story 
from the specialist's point of view~%
\cite{Gambini:1996ik,Thiemann:2001yy,Ashtekar:2004eh,b_rove1,Perez:2004hj}.
Rather, we would like to provide an \emph{entr\'ee} for `outsiders', 
and to focus on the outstanding problems as we perceive them, and thereby 
initiate and enable an informed debate between the proponents of the 
different approaches. Accordingly, we will take the liberty to omit 
mathematical details that in our opinion are not truly essential 
to understand the physical consequences of the formalism (at least 
not on a first perusal of the literature), and to describe some results 
`our own way'. At the same time, as we move along, we will try to 
make precise and clearly state the questions that are often raised 
about the LQG programme (for earlier reviews which mention some of these 
concerns, see~\cite{Horowitz:1991ty,Kuchar:1993ne,Carlip:2001wq}).

\medskip 

In order to focus the discussion, and for the reader's convenience, we 
begin with a summary of what we consider to be the main issues and 
open questions. 

\begin{itemize}

\item \emph{How do Einstein's equations appear in the classical
limit?}\\[\bulletskip] The space of quantum states used in LQG (not
necessarily a Hilbert space) is very different from the one used in
Fock space quantisation.  As a consequence, it becomes non-trivial to
see how semi-classical `coherent' states can be constructed, and how a
smooth classical spacetime might emerge. In simple toy examples, such
as the harmonic oscillator, it has been shown that the LQG
quantisation method indeed leads to quantum states whose properties
are close to those of the usual Fock space coherent
states~\cite{Ashtekar:2002sn}. In full \mbox{(3+1)-dimensional} LQG,
the classical limit is, however, far from understood (so far only
kinematical coherent states are
known~\cite{Thiemann:2000bw,Thiemann:2000ca,Thiemann:2000bx,Thiemann:2000by,Thiemann:2002vj,Sahlmann:2001nv}). In
particular, we do not know how to describe or approximate classical
spacetimes in this framework that `look' like, say, Minkowski space,
or how to properly derive the classical Einstein equations and their
quantum corrections. A proper understanding of the semi-classical
limit is also indispensable to clarify the connection (or lack
thereof) between conventional perturbation theory in terms of Feynman
diagrams, and the non-perturbative quantisation proposed by LQG.
\bulletsep

\item \emph{Renormalisation vs.~regularisation: where is the 2-loop
divergence?}\\[\bulletskip] No approach to quantum gravity can claim
success that does not explain the ultimate fate of the two-loop
divergence \eqn{e:2loopdiv}.  In a consistent scheme, this divergence
must be either eliminated by cancellation, or disposed of by some
other mechanism. The key question raised by~\eqn{e:2loopdiv} is
therefore what happens when one expands the results of
non-perturbatively quantised Einstein gravity in Newton's
constant. When such an expansion is performed about a semiclassical
state (which remains to be found in LQG, see above), the two-loop
divergence should manifest itself in one form or another. In its
present incarnation, LQG cannot (yet?) `see' and accommodate this
divergence. Its possible `disappearance' is occasionally argued to be
due to the emergence of an effective cut-off (regulator) which might
eventually make the perturbation theory finite. If this is the case,
an obvious question concerns the true origin of this cut-off: is it
generated dynamically, or covertly put in `by hand' (for instance, by
working with the compact group SU(2) instead of the full Lorentz
group)?  There are also questions concerning the meaning of
`regularisation'.  According to conventional (quantum field theory)
wisdom, physics is supposed not to depend on the way in which the
theory is regulated before the cutoff is removed; how can it be that
physics predictions of LQG do depend on the chosen regularisation
prescription? This question is in part answered by the fact that the
notions of `finiteness' and `regulator independence' as currently used
in LQG on the one hand, and in conventional quantum field theory and
perturbative quantum gravity on the other, are \emph{not} the same;
see section~\ref{s:regularisation}.  Let us furthermore stress that in
spite of its perturbative origin, the result \eqn{e:2loopdiv} cannot
be so easily dismissed as a background artifact: while it does require
\emph{some} background for its derivation ({\it i.e.\/}~a spacetime
solving Einstein's equations), the counterterm is actually independent
of the particular background about which one expands, see
\cite{vandeVen:1991gw}, as is also evident from the manifestly
space-time covariant form in which it is written.  \bulletsep

\item \emph{Status of the Hamiltonian constraint?}\\[\bulletskip] In
the current LQG literature, there is surprisingly little discussion of
certain basic aspects concerning the Hamiltonian constraint operator,
which are of central importance for the theory (recent exceptions
are~\cite{Borissov:1997ji}, and the so-called `master constraint
programme' of~\cite{Thiemann:2003zv}). For this reason, we will here
describe the Hamiltonian constraint operator and its action on a given
spin network wave function in rather pedestrian detail, as far as we
have been able to work it out.  In particular, we will exhibit the
numerous choices and ambiguities inherent in this construction, as
well as the extraordinary complexity of the resulting expression for
the constraint operator. The number of ambiguities can be reduced by
invoking independence of the spatial
background~\cite{Thiemann:2001yy}, and indeed, without making such
choices, one would not even obtain sensible expressions, as we shall
see very explicitly. In other words, the formalism is partly
`on-shell' in that the very existence of the (unregulated) Hamiltonian
constraint operator depends very delicately on its `diffeomorphism
covariance', and the choice of a proper `habitat', on which it is
supposed to act in a well defined manner. A further source of
ambiguities, which, for all we know, has not been considered in the
literature so far, consists in possible $\hbar$-dependent `higher
order' modifications of the Hamiltonian~\eqn{e:mess1}, which might
still be compatible with all consistency requirements of LQG.

The attitude often expressed with regard to the remaining ambiguities 
is that they correspond to \emph{different} physics, and therefore 
the choice of the correct Hamiltonian is ultimately a matter of physics
(experiment?), and not mathematics.  We disagree, because we cannot
believe that Nature will allow such a great degree of arbitrariness at
its most fundamental level: recall that it was precisely the infinite
number of parameters and the concomitant ambiguities which killed
perturbative quantum gravity!
\bulletsep

\item \emph{Does the quantum theory possess full spacetime 
covariance?}\\[\bulletskip]
Spacetime covariance is a central property of Einstein's theory.
Although the Hamiltonian formulation is not manifestly covariant, full
covariance is still present in the classical theory, albeit in a
hidden form, via the classical (Poisson or Dirac) algebra of
constraints acting on phase space. However, this is not necessarily so
for the quantised theory. LQG treats the diffeomorphism constraint and
the Hamiltonian constraint in a very different manner. Why and how
then should one expect such a theory to recover full spacetime (as
opposed to purely spatial) covariance?  The crucial issue here is
clearly what LQG has to say about the quantum algebra of constraints. 
Unfortunately, to the best of our knowledge, the `off-shell' calculation of 
the commutator of two Hamiltonian constraints in LQG -- with an explicit 
operatorial expression as the final result -- has never been fully 
carried out. Instead, a survey of the possible terms arising in this 
computation has led to the conclusion that the commutator vanishes on 
a certain restricted `habitat' of 
states~\cite{Thiemann:1996ay,Gambini:1997bc,Lewandowski:1997ba}, 
and therefore the LQG constraint algebra closes without anomalies. By 
contrast, we will here argue that this `on shell closure' is not sufficient
for a full proof of \emph{quantum spacetime covariance}, but that a 
proper theory of quantum gravity requires a constraint algebra 
that closes `off shell', {\it i.e.\/} without prior imposition of a
subset of the constraints. The fallacies that may ensue if one does not
insist on off-shell closure can be illustrated with simple
examples. In our opinion, this requirement may well provide the acid
test on which any proposed theory of canonical quantum
gravity will stand or fail. 
\bulletsep

\item \emph{Matter couplings: anything goes?}\\[\bulletskip]
Because LQG is claimed to be a finite and fully consistent theory of
quantum gravity, it does not appear to impose any restrictions on the
types of matter that are coupled to gravity, nor on their interactions.  
Indeed it is straightforward, though sometimes cumbersome, to extend 
the formalism to include matter: in this perspective, matter appears to 
be a mere accessory that can be added on to pure gravity as one chooses.  
This is in marked contrast to supergravity and superstring theory, which 
are based on the hypothesis that the very \emph{raison d'\^etre} of matter 
is its indispensability for curing the perturbative (and non-perturbative) 
inconsistencies of quantum gravity, and the desire to `geometrise' matter 
in the framework of a (probably supersymmetric) unified theory. It is 
not inconceivable that LQG might eventually encounter new consistency 
requirements when descending from the kinematical to the physical Hilbert 
space, but we see no evidence for this so far. Therefore the question remains
whether and how LQG can recover the consistency requirements that conventional 
perturbative quantum field theory imposes on the matter content of the 
world, in particular those resulting from cancellation of \emph{local 
(gauge) anomalies}. Let us recall that Nature does `care' about such 
consistency requirements, in that it has chosen to put the three known 
generations of fermions into \emph{anomaly free multiplets} of the standard 
model gauge group.  Similar comments apply to global anomalies. Is it 
possible to obtain the correct answer for pion decay when fermions are 
coupled to electromagnetism in the LQG approach, or would LQG predict 
the neutral $\pi$~meson to be a stable particle? \bulletsep

\item \emph{Structure of space(-time) at the smallest
scales?}\\[\bulletskip] There is a general expectation (not only in
the LQG community) that at the very shortest distances, the smooth
geometry of Einstein's theory will be replaced by some quantum space
or spacetime, and hence the continuum will be replaced by some
`discretuum'. LQG does not do away with conventional spacetime
concepts, in that it still relies on a spatial continuum as its
`substrate'. At the kinematical level, it imposes a discrete structure
which is very different from the discreteness of a lattice or naive
discretisation of space ({\it i.e.\/}~of a finite or countable set),
by `polymerising' the continuum via the scalar
product~\eqn{e:scalar_product}.  This is similar to the discrete
topology (`pulverisation') of the real line with countable unions of
points as the open sets. Because the only notion of `closeness'
between two points in this topology is whether or not they are
coincident, whence \emph{any} function is continuous in this topology,
this raises the question as to how one can recover conventional
notions of continuity in this scheme.

However, the truly relevant question here concerns the structure (and
definition!) of \emph{physical} space and time. This, and not the
kinematical `discretuum' on which holonomies and spin networks
`float', is the arena where one should try to recover familiar and
well-established concepts like the Wilsonian renormalisation group,
with its \emph{continuous} `flows'. Because the measurement of lengths
and distances ultimately requires an operational definition in terms
of appropriate matter fields and states obeying the physical state
constraints, `dynamical' discreteness is expected to manifest itself
in the spectra of the relevant physical observables (the associated
\emph{kinematical operators} will be discussed in
section~\ref{s:kinop}). It is not clear whether discreteness of
(physical) space would entail a discrete structure for \emph{time}
also, hence space-\emph{time}, as there is no \emph{a priori} notion
of `time' in quantum gravity. Instead, `time' must also be defined
operationally in terms of a `clock field', see
e.g.~\cite{b_zeh,isha1,Banks:1984cw}. Continuity or discreteness of
physical space and time thus follows from the properties of the
relevant `measuring rod fields' and `clock fields', and their
spectra. \bulletsep

\item \emph{Conceptual issues}\\[\bulletskip] Last but not least:
although we will have nothing new to say here on the grand conceptual
issues of quantum gravity and quantum cosmology we wish to emphasise
that these issues must be addressed and resolved by \emph{all}
approaches to quantum gravity. This comment concerns not only the
difficult interpretational problems but also more technical
aspects. Among the former, we would like to mention the problem of
interpreting the `wave function of the universe' and associated
`matrix elements' between different such wave functions, or the
problem of constructing `observables' and their physical
interpretation; among the latter, there is the question of whether we
have any right to expect the `wave function of the universe' to be
normalisable, or the related question of whether the familiar Hilbert
space formalism of standard quantum mechanics is really the correct
mathematical framework for quantum gravity.\footnote{In ordinary
quantum mechanics, the Hilbert space formalism and the concomitant
notion of `unitarity' are tightly linked to the physical
interpretation of the theory in terms of probabilities and their
conservation in time. In the absence of an \emph{a priori} notion of
time (as in quantum gravity) it is therefore by no means evident
whether these will remain the relevant concepts.} Over the past
decades, there has been slow but steady progress on several fronts
(see
e.g.~\cite{b_zeh,isha1,Hartle:1992as,Kuchar:1993ne,Kiefer:2004gr,Rovelli:1990ph,Rovelli:2001my,Rovelli:2001bz}
and references therein), but it is probably fair to say that we are
still far from fully understanding these issues.

\end{itemize}

\section{Topics omitted}

As we mentioned already, there are several recent developments 
and advances which we cannot cover here for lack of space, or lack 
of expertise on our side. We will list these and briefly comment on 
them below, but otherwise must refer readers to the pertinent 
`inside' reviews~%
\cite{Gambini:1996ik,Thiemann:2001yy,Ashtekar:2004eh,b_rove1,Perez:2004hj} 
and the more recent original references for more information and 
other points of view. 

\subsection{Loop Quantum Cosmology}
Symmetry reduced versions of LQG have recently been studied as models
of quantum cosmology~\cite{Bojowald:2001xe,Ashtekar:2003hd}. Two main
features stand out. The first is a new mechanism for the avoidance of
the big bang singularity in the framework of mini-superspace models of
quantum gravity.\footnote{Let us note that `singularity avoidance
mechanisms' may exist also in more conventional mini-superspace
quantum geometrodynamics~\cite{Russo:2004am}.} This mechanism is
based on the fact that the inverse scale factor is represented by an
operator that stays bounded as the classical radius of the universe
shrinks to zero.  Alternatively, one might say that the effective
discretisation of the Hamiltonian constraint enables the quantum wave
function to `jump over' the singularity, and to continue to evolve
past the singularity. However, it is not clear whether and how these
models can be derived from full-fledged LQG, and whether singularity
avoidance is also a property of the full theory. In fact, a very recent
investigation ~\cite{Brunnemann:2005in} has revealed that the spectrum of 
the operator corresponding to the inverse volume is \emph{not} bounded 
from above in full LQG. In addition, it is by no means excluded (and 
some would even say likely) that inhomogeneities, and possibly other 
degrees of freedom, will essentially alter the nature of the quantum 
state near the singularity. See also~\cite{Coule:2003ig} for other comments.

The second new feature arises in applications of loop quantum
cosmology to inflationary models. It is the possibility that inflation
might be triggered and eventually stopped (gracefully) by gravity
itself, via an intrinsically quantum gravitational
mechanism~\cite{Bojowald:2002nz}.  Even if scalar fields cannot be
avoided altogether, an appealing feature here is that the inflaton
potential engineering characteristic of most current models of
inflation could become unnecessary. Perhaps less attractive (to an
outsider) is that the regularisation ambiguities of LQG feed through
to the symmetry reduced models and lead to physical effects
(see~\cite{Vandersloot:2005kh} for a recent discussion). Concerning
the value of the Barbero-Immirzi parameter, obtained from black hole
entropy calculations (see below), one might arrive at an interesting
`internal' consistency check by matching it with the duration of the
inflationary period, which is also linked to this parameter in these
models. It has furthermore been suggested that measurements of the 
CMB fluctuation spectrum at large angles might provide experimental 
tests of LQG, but it is not clear to us whether the predicted effects, 
if present, might not be explained in many other ways, too.

\subsection{Microscopic origin of black hole entropy}

The explanation of the Bekenstein-Hawking entropy of black
holes~\cite{Bekenstein:1973ur,hawk2} in terms of microstates has been
claimed as a success for both
LQG~\cite{Ashtekar:1997yu,Ashtekar:2000eq} and string
theory~\cite{stro2}. The main achievement of string theory is that it
not only explains the area law, but also predicts the factor~1/4
relating entropy and area; furthermore, the agreement has been shown
to extend to the higher order curvature terms predicted by string
theory, see e.g.~\cite{deWit:2005fz}. However, the argument requires a
huge extrapolation in the string coupling constant, and is essentially
limited to BPS-type extremal black holes.  The LQG explanation, on the
other hand, succeeds in implementing the condition for isolated
horizons~\cite{Ashtekar:2004cn} in the quantum theory, and works for
ordinary (Schwarzschild and Kerr) black holes, but requires a `fit'
for the Barbero-Immirzi parameter to get the prefactor right. Although
the two ans\"atze thus both reproduce the desired result, we are faced
with something of a paradox here, as the two explanations seem almost
impossible to reconcile, given the very different hypotheses
underlying them --- pure gravity on the one hand, and an exponentially
growing spectrum of D-brane states on the other hand.

\subsection{Spin foams}

Attempts to overcome the difficulties with the Hamiltonian constraint
have led to yet another development, \emph{spin foam
models}~\cite{Reisenberger:1994aw,Reisenberger:1997pu,Baez:1997zt}.
These can be regarded as space-time versions of spin networks, to wit,
evolutions of spin networks in `time'. Mathematically, these models
represent a generalisation of spin networks, in the sense that group
theoretical objects (holonomies, representations, intertwiners, etc.)
are attached not only to vertices and edges (links), but also to
higher dimensional faces in a simplicial decomposition of
space-time. In addition, they may pave the way towards the
construction of a physical inner product, which in turn is defined by
the quantum dynamics. Many of the recent advances in this area concern
purely topological theories, so-called ``$BF$ models'', where $F(A)$
is a field strength, and $B$ the Lagrange multiplier (tensor) field
whose variation enforces $F(A)=0$. The formalism can thus be nicely
applied to (2+1) gravity, which \emph{is} topological. In (3+1)
dimensions one needs extra constraints in order to recover the
propagating gravitational degrees of freedom back into the
theory~\cite{Barrett:1997gw}. 
Interesting as they are, however, these developments have so far not
shed much new light on the problems with the Hamiltonian constraint,
or the constraint algebra, in our opinion. A derivation of spin foam
models from the Hamiltonian formulation remains incomplete due to the
complexity of the Hamiltonian constraint~\cite{Rovelli:1998dx}. Hence,
a decisive proof of the connection between spin foam models and 
the full Einstein theory and its canonical formulation
still appears to be lacking. On the contrary, it has even been
suggested that these models may provide a possible `way out' if the
difficulties with the `orthodox' Hamiltonian approach, which we
follow here, should really prove insurmountable.

\section{Old vs.~new variables: from metric to loops}
\subsection{Prelude: the metric (or dreibein) approach}
\label{s:prelude}

There are many introductory texts to which we can refer readers for a
more detailed treatment of canonical gravity (see
e.g.~\cite{Arnowitt:1962hi,b_mtw,b_dira2,isha1,Henneaux:1992ig,Kuchar:1993ne,
Ashtekar:2004eh,Kiefer:2004gr}), but let us
nevertheless briefly review the traditional way of doing canonical
quantum gravity, also called \emph{geometrodynamics}. Our exposition
in this section follows \cite{Nicolai:1992xx}, whose notations and
conventions we adopt in the remainder. We will be using the vierbein
formalism with a vierbein ${E_\mu}^A$ and space-time metric
$G_{\mu\nu} = {E_\mu}^A {E_\nu}^B\eta_{AB}$ with flat (tangent space)
metric $\eta_{AB} = {\rm diag}\, (-1,+1,+1,+1)$ \footnote{Modulo
some dimension dependent factors, the results described in this
subsection are valid in any dimension.}. The vierbein is covariantly
constant under a derivative which is covariant w.r.t.~both spacetime
diffeomorphisms and local Lorentz transformations, viz.
\begin{equation}
D_\mu {E_\nu}^A := \nabla_\mu {E_\nu}^A - {\omega_\mu}^{AB} E_{\nu B} = 0\,,
\end{equation}
where $\nabla_\mu$ is the covariant derivative which involves only the
Christoffel connection,
\begin{equation}
\nabla_\mu V_\nu := \partial_\mu V_\nu - {\Gamma_{\mu\nu}}^\rho V_\rho
\end{equation}
and $\omega_{\mu AB}$ is the spin connection. It is a standard result that
both $ {\Gamma_{\mu\nu}}^\rho$ and $\omega_{\mu AB}$ can be explicitly
solved for the vielbein from the above equation (in the absence of
torsion).

As is customary in canonical gravity, and following the standard ADM
prescription \cite{Arnowitt:1962hi} we assume the space-time manifold 
${\cal M}$ to be globally hyperbolic. It then follows from a Theorem
of Geroch~\cite{geroch3} that ${\cal M}$ can be foliated according to 
${\cal M} = \Sigma \times {\mathbb R}$, with a spatial manifold $\Sigma$ of 
fixed topology (and no boundary, for simplicity). Using letters $m,n,...$ 
and $a,b,...$ for curved and flat \emph{spatial} indices, respectively, we 
choose the triangular gauge for the vierbein  
\begin{equation}
E_\mu{}^A =
\left(\begin{array}{ccc}
          N &  N^a\\
          0 & e_m{}^a
\end{array}\right) \quad \Longrightarrow \quad
G_{\mu\nu} =
\left(\begin{array}{ccc}
          -N^2 + N_a N_a &  N_n\\
          N_m & g_{mn}
\end{array}\right)\,,
\end{equation}
where $g_{mn} \equiv {e_m}^a e_{na}$ is the spatial metric. The vierbein
components $N$ and $N_a$ are Lagrange multipliers (`lapse' and `shift');
their variation yields the canonical constraints. In the remainder 
we will freely convert spatial indices by means of the spatial dreibein 
and its inverse.
 
The canonical momenta are obtained in the standard fashion from
the Einstein-Hilbert action
\begin{equation}
{\Pi_a}^m := \frac{\delta {\cal L}}{\delta \partial_t {{ e}_m}^a} 
          = \frac12  e {e_b}^m (K_{ab} - \delta_{ab} K)\,,
\end{equation}
where $e\equiv \det({e_m}^a)$, and
\begin{equation}
\label{e:K_def}
K_{ab} := {E_a}^\mu {E_b}^\nu \nabla_\mu E_{\nu 0} = \omega_{ab0}
\end{equation}
is the extrinsic curvature of $\Sigma$ in ${\cal M}$ (expressed in terms 
of flat spatial indices, with $\omega_{ab0}\equiv {E_a}^\mu \omega_{\mu b0}$),
and $K\equiv K_{aa}$. The inverse formula reads
\begin{equation}
K_{ab} = \frac{2}{e} \big(\Pi_{ab} - 
\frac12 \delta_{ab} \Pi \big) \quad , \qquad
(\Pi\equiv \Pi_{aa})\,.
\end{equation}
From the symmetry $K_{ab} = K_{ba}$ we immediately deduce the Lorentz 
constraint
\begin{equation}
L_{ab} = e_{m[a} {\Pi_{b]}}^m \approx 0\,,
\end{equation}
which is also the canonical generator of spatial rotations on the dreibein
($\approx$ means `weakly zero'~\cite{b_dira2}).
The canonical Hamiltonian (really Hamiltonian density) is
\begin{equation}
H = \partial_t {e_a}^m {\Pi_m}^a - {\cal L} = N H_0 + N_a H_a \,,
\end{equation}
with the diffeomorphism constraint
\begin{equation}
H_a \equiv D_m {\Pi_a}^m \approx 0\,,
\end{equation}
and the Hamiltonian constraint (\emph{alias} the scalar constraint)
\begin{equation}
H_0 \equiv e^{-1} \Big(\Pi_{ab} \Pi_{ab} - \frac12 \Pi^2 \Big)
      - e R^{(3)}
= \frac14 e \big( K_{ab} K_{ab} - K^2 \big) - e R^{(3)} \approx 0\,,
\end{equation}
where $R^{(3)}$ is the spatial Ricci scalar.
The canonical equal time (Poisson) brackets are
\begin{equation}
\{ e_{ma} ({\bf x}), {\Pi_b}^n ({\bf y}) \} = \delta_{ab} \delta^n_m
     \delta^{(3)}({\bf x}, {\bf y})\,,
\end{equation}
with the other brackets vanishing (for the canonical variables with the 
indices in the indicated positions). Canonical quantisation in the `position 
space representation' now proceeds by representing the dreibein as a
multiplication operator, and the canonical momentum by the functional 
differential operator
\begin{equation}\label{quant1}
{\Pi_a}^m ({\bf x})= \frac{\hbar}{i} \frac{\delta}{\delta {e_m}^a ({\bf x})}\,.
\end{equation}
With these replacements, the classical constraints are converted to 
quantum constraint operators which  act on suitable wave functionals. 
The diffeomorphism and Lorentz constraints become
\begin{equation}
H_a ({\bf x}) \Psi[e] = 0 \quad ,\qquad
L_{ab} ({\bf x})\Psi[e] = 0\,.
\end{equation}
They will be referred to as `kinematical constraints' throughout.
Dynamics is generated via the Hamiltonian constraint, the 
\emph{Wheeler-DeWitt (WDW) equation}~\cite{DeWitt:1967yk,DeWitt:1967ub,DeWitt:1967uc,whee1}:
\begin{equation}\label{WDW}
H_0 ({\bf x}) \Psi[e] = 0\,.
\end{equation}
It is straightforward to include matter degrees of freedom, in which 
case the constraint operators and the wave functional $\Psi [e,\dots]$ 
depend on further variables (indicated by dots). The functional 
$\Psi[e,\dots]$ is sometimes referred to as the `wave function 
of the universe', and is supposed to contain the complete information
about the universe `from beginning to end'. A good way to visualise 
$\Psi$ is to think of it as a film reel; `time' and the illusion that 
`something happens' emerge only when the film is played.

The substitution (\ref{quant1}) turns the Hamiltonian constraint 
into a highly singular functional differential equation, which most 
likely cannot be made mathematically well defined in this form, even 
allowing for certain `renormalisations'. Here we do not wish to belabour 
the well known difficulties, both mathematical and conceptual, which 
have stymied progress with the WDW equation for over forty years, 
see e.g.~\cite{b_mtw,isha1,Kiefer:2004gr} \footnote{One of the 
inventors of (\ref{WDW}) has been overheard aptly referring to it as 
{\it ``that damned equation...''}.}. We should point out the (surely 
well known) fact that, at least formally and ignoring subtle points 
of functional analysis, solutions to both the Lorentz constraint 
and the diffeomorphism constraint can be rather easily obtained by 
integrating suitably densitised invariant combinations of the spatial 
dreibein (or metric) and curvature components, and the matter `position 
variables' over the spatial three-manifold $\Sigma$. The associated wave 
functionals $\Psi$ are then automatically functionals of diffeomorphism 
classes of spatial metrics. In saying this, it must be stressed, however, 
that geometrodynamics has so far not succeeded in constructing a suitable 
scalar product and an appropriate Hilbert spaces of wave functionals.

The absence of a suitable Hilbert space is often invoked by LQG
proponents as an argument against the geometrodynamics approach.
While LQG is certainly `ahead' in that it does succeed in constructing
a Hilbert space (more on this later), we should like to emphasise that
in all approaches there looms the larger conceptual problem of whether
conventional quantum mechanical concepts are indeed sufficient for
quantum gravity~\cite{Hartle:1992as,isha1}. For instance, even if we
can compute matrix elements of wave function(al)s, we still have no
idea what their correct physical interpretation is in the context of
quantum cosmology, or whether the normalisability of the wave function
of the universe is a necessary condition. In other words, we do not 
know whether these cherished concepts may not have to be amended 
or abandoned in the final theory. For this reason, we believe that 
besides the emphasis on mathematical rigour, it is equally important 
to develop some physical intuition for the states one is dealing with, 
and in this regard, we do not think that geometrodynamics lags behind. 
From this point of view, it appears to us that, beyond the technical 
subtleties, the kinematical constraints are not the real problem of 
quantum gravity. The core difficulties of canonical quantum gravity 
are all connected in one way or another to the Hamiltonian constraint 
-- irrespective of which canonical variables are used.

\subsection{Ashtekar's new variables}
\label{s:ashtekar_variables}

Much of the initial 
excitement over Ashtekar's discovery \cite{Ashtekar:1987gu} of new 
canonical variables was due to the change of perspective they bring 
about, which fuelled hopes that they might alleviate some of the 
longstanding unsolved problems  of quantum gravity. Let us therefore 
first describe what they are, and how they are obtained. There is
an {\it ab initio\/} derivation  based on the addition of a term 
$\propto \int E\wedge E \wedge R$ to the Einstein-Hilbert action 
\cite{Holst:1995pc,Samuel:2001qy} (this term vanishes upon use of 
the Bianchi identity and the equations of motion for the spin connection
with vanishing torsion), but we will skip this step here. Instead consider 
the combination~\footnote{We alert readers that our notational conventions 
differ from the ones used in most of the LQG literature, where $a,b,\dots $ 
denote curved and $i,j,\dots$ flat spatial indices.} 
\begin{equation}\label{Ashtekar1}
A_{ma} := -\frac12 \epsilon_{abc} \omega_{mbc} + 
     \gamma e^{-1} (\Pi_{ma} - \frac12 e_{ma} \Pi) 
   = -\frac12 \epsilon_{abc} \omega_{mbc} + \gamma K_{ma} \,,
\end{equation}
where $\omega_{mbc}$ is the \emph{spatial} spin connection. The
parameter $\gamma \neq 0$ is referred to as the `Barbero-Immirzi
parameter' in the LQG literature~\cite{Barbero:1994ap,Immirzi:1996di}.
Classically, $\gamma$ has no physical significance, but is believed to
become physically relevant upon quantisation, for instance, by setting
the scale for the fundamental areas and volumes (in this sense it is
somewhat analogous~\cite{Gambini:1998ip} to the $\theta$ parameter of
QCD).  One can now show that~\cite{Ashtekar:1987gu}
\begin{equation}
\label{Poisson1}
\begin{aligned}
\{ A_{ma} ({\bf x}),\; A_{nb} ({\bf y}) \} &= 0\,, \\[1ex]
\{ {\tilde E_a}^m ({\bf x}),\; {\tilde E_b}^n ({\bf y}) \} &= 0\,, \\[1ex]
\{ {A_m}^a (\x) \, , \,  {\tilde E_b}^n ({\bf y})\} &= 
\gamma \delta^n_m \delta_b^a \delta^{(3)}({\bf x}, {\bf y}) \,,
\end{aligned}
\end{equation}
where the canonically conjugate variable to $A_{ma}$ is the \emph{inverse 
densitised spatial dreibein}
\begin{equation}\label{Ashtekar2}
{\tilde E_a}^m := e {e_a}^m \qquad \Longrightarrow \quad \tilde E = e^2 \,,
\end{equation}
with $e\equiv \det {e_m}^a$ and $\tilde{E}\equiv \det {\tilde{E}_a}^m$.
The parameter $\gamma$ is often eliminated from these brackets by 
absorbing it into the definition of ${\tilde E_a}^m$. 

To rewrite the constraints in terms of the new variables, we first observe 
that the covariant constancy of the spatial dreibein and the Lorentz 
constraint imply the Gauss constraint:
\begin{equation}
D_m {\tilde E_a}^m \equiv \partial_m {\tilde E_a}^m + 
    \epsilon_{abc} {A_m}^b {\tilde E}^{cm} \approx 0\,.
\end{equation}
Defining the field strength
\begin{equation}
\begin{aligned}
F_{mna} :&= \partial_m A_{na} - \partial_n A_{ma}
         + \epsilon_{abc} A_{mb} A_{nc} \\
&=  -\frac12 \epsilon_{abc} R_{mnbc} + 
     \gamma \big( D_m K_{na} - D_n K_{ma}\big) +
     \gamma^2 \epsilon_{abc} {K_m}^b {K_n}^c   \,,
\end{aligned}
\end{equation}
it follows that the diffeomorphism constraint takes the form
\begin{equation}
{\tilde E_a}^m F_{mna} \approx 0\,.
\end{equation}
Furthermore, 
\begin{equation}
\label{e:EEF}
\begin{aligned}
\epsilon_{abc}{\tilde E_a}^m {\tilde E_b}^n F_{mnc} &=
 - \gamma^2 \big(\Pi_{ab} \Pi_{ab} - \frac12 \Pi^2 \big)
  - e^2 R^{(3)}\\
&= - \gamma^2 e{\cal H}_0 - \frac14 (1+\gamma^2)e^2 (K_{ab} K_{ab} -
  K^2) \,.
\end{aligned}
\end{equation}
These relations immediately suggest interpreting ${A_m}^a$ as a 
\emph{gauge connection} for the gauge group SO(3) of spatial rotations 
(for $D=2+1$ gravity, this group is replaced by its non-compact form 
SO(1,2)~\cite{Bengtsson:1989fc}). Accordingly, the new variables
are conveniently rewritten as~\cite{Ashtekar:1987gu}
\begin{equation}
A_{m\alpha\beta} \equiv A_{ma} \tau^a_{\alpha\beta}\,,
\end{equation}
where $\tau^a$ are the Pauli matrices. 

For the special choice $\gamma = \pm i$ \cite{Ashtekar:1987gu} the
second term on the r.h.s. of \eqn{e:EEF} drops out, and --- save for
an extra density factor of $e$ --- the Hamiltonian constraint is
expressed entirely in terms of the new canonical variables, and
furthermore depends on them \emph{polynomially}.\footnote{More
recently it has, however, been appreciated that the extra density
factor in~\eqn{e:EEF} is a major source of problems for a
background-independent quantisation, since it makes the Hamiltonian an
object of density weight two; for details
see~\cite{Thiemann:1996aw}.\label{densityfootnote}} In other words,
this particular choice allows us to combine the diffeomorphism and
Hamiltonian constraints, which are schematically of the form
`$(\partial + \omega)\Pi \approx 0$' and `$(\partial\omega + \omega^2
+ \Pi^2) \approx 0$', respectively, into a single expression in terms
of the connection $A = \omega \pm i \Pi$, and its canonically
conjugate variable. Moreover, for this choice of $\gamma$ the
connection $A_{ma}^{(\pm)}$ is nothing but the pullback of the
four-dimensional spin connection to the spatial hypersurface $\Sigma$,
with the two signs corresponding to the two chiralities (indicated by
superscripts~$^{(\pm)}$)
\begin{equation}\label{omega}
\omega_{mAB} \gamma^{AB} = \omega_{mab} \gamma^{ab} 
                      + 2 \omega_{ma0} \gamma^a \gamma^0
 = A^{(+)}_{ma} \gamma^0 \gamma^a (1 + \gamma^5) +
   A^{(-)}_{ma} \gamma^0 \gamma^a (1 - \gamma^5) \,,
\end{equation}
(cf.~eqn.~\eqn{e:K_def}). Equivalently, the variables $A^{(\pm)}_{ma}$ 
are associated with what, in a Euclidean formulation, would be the 
selfdual and anti-selfdual parts of the spin connection, respectively,
\begin{equation}
A^{(\pm)}_{ma} = \omega_{ma0}^{(\pm)} \qquad \mbox{with} \qquad
\omega_{mAB}^{(\pm)} := \frac12 \left(\omega_{mAB} \pm 
\frac{i}2 \epsilon_{ABCD} {\omega_m}^{CD} \right)\,.
\end{equation}
This is therefore also the natural choice for $\gamma$ when one considers
coupling gravity to chiral fermions. In fact, one of the authors (H.N.) 
was first enticed into learning about Ashtekar's variables when realising 
that they simplify the calculation of the constraint algebra of 
supergravity considerably \cite{Nicolai:1990vb,Matschull:1993hy}. 
This is because the local supersymmetry constraint ({\it i.e.\/}~the time
component of the Rarita-Schwinger equation) always contains a factor
$D_{[m}(A)\psi_{n]}$, where $\psi_n$ is the gravitino, and $A$ is just 
the Ashtekar connection (of course with $\gamma = \pm i$). More
succinctly, in supergravity, the commutation relations~\eqn{Poisson1}, 
as well as the polynomiality of the constraints must necessarily hold,
if the commutator of two local supersymmetry constraints is to close 
into the scalar and diffeomorphism (and possibly other) constraints.

From the esthetical point of view $\gamma=\pm i$ is therefore clearly
the preferred choice. Nevertheless, this value has been abandoned 
in most of the recent LQG literature, because there is a major difficulty
with it: the phase space of general relativity must be \emph{complexified}
with imaginary or complex $\gamma$. To recover the \emph{real} phase space 
of general relativity and to ensure that real initial data evolve into 
real solutions, suitable reality conditions must be imposed. This is 
straightforward to achieve for the classical theory --- after all,
we have merely changed the variables, not the theory itself --- but not 
so for the quantum theory. There, the complexification poses subtle
problems concerning the definition and imposition of appropriate hermiticity 
conditions on the states and operators, and no consensus has been reached 
so far on how to circumvent these difficulties (or on whether they can
be circumvented at all).

For this reason one now usually takes $\gamma$ to be \emph{real}.  In
this case, no problem arises with reality of solutions in either the
classical or the quantum theory, but the polynomiality of the
constraints, and hence one of the most attractive features of the new
variables, is lost. This is because the extra term in~(\ref{e:EEF}) no
longer vanishes, but must be subtracted from both sides to recover the
correct WDW Hamiltonian. Accordingly, one must express the extra
contributions in terms of the new canonical variables ${A_m}^a$ and
${\tilde{E}_a}^m$ via (\ref{Ashtekar1}) and (\ref{Ashtekar2}), and
this in turn requires expressing the original dreibein as well as the
extrinsic curvatures in terms of the new canonical variables. For a
while this was regarded as a chief obstacle, until a way to solve it
was discovered by Thiemann~\cite{Thiemann:1996aw}. To this aim let us
first introduce the volume associated with a region
$\Omega\subset\Sigma$ (considered as a phase space variable)
\footnote{The $\epsilon$-symbol is always the invariant \emph{tensor
density}: $\epsilon_{mnp} = e^{-1} {e_m}^a {e_n}^b {e_p}^c
\epsilon_{abc}$, {\it i.e.\/}~assumes the values $0,\pm 1$.}
\begin{equation}
V(\Omega) = \int_\Omega\! {\rm d}^3x \, e = \int_\Omega\! {\rm d}^3 x \, \sqrt{\tilde{E}}
\equiv \int_\Omega\! {\rm d}^3x \, \sqrt{\frac1{3!} \epsilon_{mnp} \epsilon^{abc}
{\tilde{E}_a}^m {\tilde{E}_b}^n {\tilde{E}_c}^p }\,.
\end{equation}
Writing $V\equiv V(\Sigma)$, we first use the substitution
\begin{equation}
\label{e:trick1}
e_m{}^a (\x )=  \epsilon_{mnp} \epsilon^{abc} 
  \tilde{E}^{-1/2} \tilde{E}_b^n \tilde{E}_c^p (\x )
= \frac1{4\gamma} \Big\{ A_m{}^a (\x) , V \Big\}
\end{equation}
to recover the spatial dreibein. The second trick is to eliminate the 
extrinsic curvature using a doubly nested bracket. The first bracket is 
introduced by rewriting
\begin{equation}
\label{e:trick2}
K_m{}^a (\x )= \frac{1}{\gamma}\big\{ A_m{}^a (\x)\, ,\,\bar{K} \big\} \qquad\text{where}\qquad
\bar{K} \equiv \bar{K}(\Sigma)
:= \int_\Sigma\!  {\rm d}^3x \, K_m{}^a \tilde{E}_a{}^m\,.
\end{equation}
The second bracket comes in through identity 
\begin{equation}
\label{e:trick3}
\bar{K} (\x) = \frac{1}{\gamma^{3/2}}\Big\{ \frac{\tilde{E}_a{}^m
        \tilde{E}_b{}^n}{\sqrt{\tilde{E}}} \epsilon^{abc} F_{mnc}(\x) \,,\, V \Big\}\,.
\end{equation}
Last, we need another dreibein factor to convert the curved index $m$
on ${K_m}^a$ to a flat one, for which we must use~\eqn{e:trick1} once
again. We will postpone writing out the Hamiltonian constraint in
terms of these multiple brackets until
section~\ref{s:hamiltonian_constraint}. But let us note already here
that the above relations supply the `tool kit' also for transcribing 
any given matter Hamiltonian in terms of the new variables.

For $\gamma\neq \pm i$ the connection is no longer the pullback of a 
four-dimensional spin connection~\cite{Samuel:2000ue}. Although this 
does not immediately lead to problems with spacetime covariance, we will 
see in section~\ref{s:matter_couplings} that the problem comes back  
when one considers coupling the theory to fermionic matter degrees of 
freedom.

\subsection{The connection representation}

Independently of the choice of $\gamma$, the reformulation of canonical
gravity in terms of connection variables opens many new avenues,
in particular the use of concepts, tools and techniques from 
Yang-Mills theory. Early attempts at quantisation were based on 
the \emph{connection representation} with the original Ashtekar 
connection, {\it i.e.\/}~$\gamma=\pm i$. Although these were ultimately 
not successful, let us nonetheless briefly summarise them here.
In this scheme, one represents the connection ${A_m}^a$ by a 
multiplication operator, and sets
\begin{equation}\label{quant2}
{\tilde E_a}^m ({\bf x})= 
  \frac{\hbar}{i}\frac{\delta}{\delta {A_m}^a ({\bf x})}\,.
\end{equation}
The WDW functional depending on the spatial metric (or dreibein)
is accordingly replaced by a functional $\Psi[A]$ living on the space 
of connections (modulo gauge transformations). The price one pays
is that this representation is much harder to `visualise' because
the spatial metric is no longer represented by a simple multiplication
operator, but must now be determined from the operator for the 
inverse densitised metric
\begin{equation}\label{gmn}
g g^{mn} ({\bf x})= - {\hbar^2}\frac{\delta}{\delta {A_m}^a ({\bf x})}
 \frac{\delta}{\delta {A_n}^a ({\bf x})} \;\; .
\end{equation}
Even if one ignores the clash of functional differential operators at 
coincident points, finding suitable states and computing their expectation 
values is obviously not an easy task (and has not been accomplished so far). 
Similarly, the spatial volume density is obtained from
\begin{equation}\label{g}
g ({\bf x})= \tilde{E} ({\bf x})=  
\frac{i\hbar^3}6 \epsilon^{abc} \epsilon_{mnp}
\frac{\delta}{\delta {A_m}^a ({\bf x})}
\frac{\delta}{\delta {A_n}^b ({\bf x})} 
\frac{\delta}{\delta {A_p}^c ({\bf x})} \quad .
\end{equation}
Again this operator is very singular. Equations \eqn{gmn} and \eqn{g}
provide a first glimpse of the difficulties that the formalism has in 
finding semi-classical states and thereby establishing a link between 
the quantum theory and classical smooth spacetime geometry.

For the quantum constraints the replacement of the metric by
connection variables leads to a Hamiltonian which is simpler 
than the original WDW Hamiltonian, but still very singular. 
Allowing for an extra factor of $e$ (and assuming $e\neq 0, \infty$) 
the WDW equation becomes 
\begin{equation}\label{HamA}
\epsilon^{abc} F_{mna} \big(A({\bf x})\big)
\frac{\delta}{\delta {A_m}^b ({\bf x})}\frac{\delta}{\delta {A_n}^c ({\bf x})}
\Psi[A] = 0\,.
\end{equation}
Here we have adopted a particular ordering, which however is by no
means singled out. No viable solutions to this constraint have been 
found, but there is at least one interesting solution if one 
allows for a non-vanishing cosmological constant $\Lambda$. Namely,
using an ordering opposite to the one above, and including a term 
$\Lambda g$ with the volume density~(\ref{g}), the WDW equation reads
\begin{equation}\label{Kodama}
\epsilon_{abc} \frac{\delta}{\delta {A_m}^a ({\bf x})}
\frac{\delta}{\delta {A_n}^b ({\bf x})} \left( F_{mnc}\big( A (\x)\big)
- \frac{i\hbar\Lambda}6 \epsilon_{mnp}\frac{\delta}{\delta {A_p}^c (\x)} 
\right) \Psi_\Lambda[A] = 0\,.
\end{equation}
This is solved by
\begin{equation}
\Psi_\Lambda [A] = 
\exp \left( \frac{i}{\hbar\Lambda} 
\int_\Sigma\! {\rm d}^3 {\bf x}\, {\cal L}_{CS} (A) \right) \,,
\end{equation}
with the Chern-Simons Lagrangian ${\cal L}_{\text{CS}} = A\wedge {\rm
d}A + i A\wedge A \wedge A$ (actually, $\Psi_\Lambda$ is already
annihilated by the operator in parentheses, so the first factor in
the Hamiltonian constraint operator~\eqn{Kodama} is not `needed' for 
this result). In the literature this state is known as the Kodama 
state~\cite{Kodama:1990sc}, but the solution had been 
known for a long time in Yang-Mills theory~\cite{Jackiw:1985k1}, 
where however it has rather unusual physical
properties~\cite{Witten:2003mb}. The difficulties with this solution
have been much discussed recently~\cite{Freidel:2003pu,Witten:2003mb};
an obvious one concerns the flat space limit $\Lambda\rightarrow 0$
(idem for the `semiclassical' limit $\hbar\rightarrow 0$).

What happens when we choose $\gamma$ to be different from~$\pm i$, and
real in particular? As we explained already, the extra term
in~\eqn{e:EEF} then no longer vanishes, must be dealt with
separately~\cite{Thiemann:1996aw}. The nice polynomial form of the
Hamiltonian constraint operator~\eqn{HamA} is lost. When implementing
the translation rules at the end of the foregoing subsection in the
connection formulation, one finds that the new Hamiltonian is not
significantly simpler any more than the original one of geometrodynamics 
in terms of metric or dreibein variables.

\subsection{From connections to holonomies}
\label{s:conn2hol}

The \emph{loop representation} is an attempt to overcome the
difficulties with the connection representation which we sketched
above.  The transition between the connection and the loop
representation was originally obtained via the \emph{loop transform},
which can be thought of as a kind of functional Fourier
transform~\cite{Rovelli:1989za}. We will not describe that 
construction here, but turn immediately to the formulation in
terms of holonomies, on which the modern formulation of LQG -- spin
networks and spin foams -- are based.

Whereas in the connection representation one works with functionals
$\Psi[A]$ which are supported `on all of $\Sigma$', one now switches
to the \emph{holonomies} as the basic variables. These are gauge covariant 
functionals supported on one-dimensional links, or `edges', which we
will designate by~$e$ (following established LQG notation). For a given
edge, {\it i.e.\/}~some (open) curve embedded in $\Sigma$, we set
\begin{equation}
\label{e:holonomy_def}
h_e [A] = {\cal P} \exp \int_e A_m {\rm d}x^m\,,\quad\text{with
  $A\equiv A^a \tau_a$}\,.
\end{equation}
Hence, $h_{e}[A]$ is a matrix valued functional. The holonomy
transforms under the action of SU(2) at each end of the edge~$e$:
\begin{equation}
\label{gau-rot}
h_e[A] \rightarrow h^g_e[A] = g\big(e(0)\big)\;
h_e[A]\; g^{-1}\big(e(1)\big) \; , 
\qquad\text{with~~ $g\big(e(0)\big)\;, \; g\big(e(1)\big)\; \in$  SU(2)}\,.
\end{equation}
For the remainder it is important that the holonomies are to be
regarded as variables \emph{in their own right}, subject to these
transformation properties (so in some sense one can `forget' about
the original connection $A$ defining the holonomy).
The \emph{distributional nature} of the holonomy is not only evident
from its singular support (on a line rather than all of $\Sigma$), but
also from the fact that we do not assume~$h_e[A]$ to be close to the
identity if the edge~$e$ is `small' (this terminology has to be used
with due care, as there is no a priori measure that tells us when~$e$
is `small', but we can still imagine making it `small' by chopping~$e$
into as many `subedges' as we like). The fact that the typical field
configuration is generically a distribution rather than some smooth
function is well known from constructive quantum field
theory~\cite{Simon:1974dg}.

The holonomies are taken to transform in SU(2) representations 
$\rho_{j_e}$ of \emph{arbitrary spin} $j_e = \frac12, 1, \frac32, \dots$ 
for each link $e$ (with the convention that $j_e=0$
means that there is no edge). We will denote such a spin-$j_e$ valued
holonomy by
\begin{equation}
\big(\rho_{j_e}( h_e[A]) \big)_{\alpha\beta}\,,
\end{equation}
with indices $\alpha, \beta, \dots$ as appropriate for the representation 
at hand. To make the notation less cumbersome, we will occasionally
suppress $\rho_{j_e}$ and the representation labels, and simply denote 
the above matrix as $(h_e[A])_{\alpha\beta}$.

To define the conjugate variable, we recall that the area element for 
the spatial manifold $\Sigma$ can be expressed as a Lorentz vector 
({\it i.e.\/}~with flat SO(3) indices) via
\begin{equation}\label{dFa}
{\rm d}F_a := \epsilon_{abc} \theta^b \wedge \theta^c \qquad
\mbox{with} \quad \theta^a \equiv \d x^m {e_m}^a \, .
\end{equation}
Happily, this can be nicely rewritten in terms of the new canonical variables
\begin{equation}
{\rm d}F_a = \epsilon_{abc} {e_n}^b {e_p}^c {\rm d}x^n\wedge {\rm d}x^p =
  \epsilon_{mnp} {\tilde E}^{m}_a {\rm d}x^n\wedge {\rm d}x^p \,.
\end{equation}
As the conjugate variable to $h_e[A]$ one takes the `flux' vector
\begin{equation}
\label{e:flux_def1}
F^a_S[\tilde E] := \int_S \, {\rm d}F^a 
\end{equation}
through any two-dimensional surface $S$ embedded in $\Sigma$. There is 
also a smeared version of this variable, with a test function $f_a$ to 
soak up the free index $a$, which reads
\begin{equation}
\label{e:flux_def}
F_S[\tilde E,f] := \int_S \, f^a \epsilon_{abc}
{e_n}^b {e_p}^c {\rm d}x^n\wedge {\rm d}x^p  = 
\int_S\, f_a \epsilon_{mnp} {\tilde E}^{ma} {\rm d}x^n\wedge {\rm d}x^p \, .
\end{equation}
We note that the standard notation for this variable in the LQG 
literature is $E[S,f]$, but we prefer the one above because
it is in parallel with the notation for the holonomy itself. 
Both $F_S[\tilde{E},f]$ and $F_S^a[\tilde{E}]$ are \emph{distributional} 
in the sense that they are supported on a two-dimensional submanifold
of~$\Sigma$.

To compute the Poisson brackets between the new canonical variables
introduced above, we consider a surface $S$ and an edge $e$ that 
`pierces' $S$ at the point $P$ (if $e$ does not intersect $S$, the
bracket simply vanishes). 
\begin{figure}[t]
\psfrag{e1}{$e_1$}
\psfrag{e2}{$e_2$}
\psfrag{eeps}{$e(\epsilon)$}
\begin{center}
\includegraphics*[height=4cm]{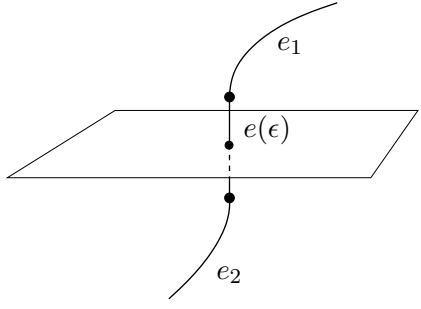}
\caption{Setup used for the computation of the
  bracket~\eqn{e:basic_bracket}. In the limit $\epsilon\rightarrow 0$
  the edge $e(\epsilon)$ shrinks to zero and
  the two nodes just above and below the surface coincide.}
\label{f:compute_basic_bracket}
\end{center}
\end{figure}
We next subdivide this edge into three pieces as shown in
figure~\ref{f:compute_basic_bracket}: two subedges $e_1$ and $e_2$,
with associated holonomies $h_{e_1}^\epsilon$ and $h_{e_2}^\epsilon$,
which touch $S$ only in the limit $\epsilon\rightarrow 0$, and a third
`infinitesimal' edge $e(\epsilon)$ intersecting $S$, for which the
path ordered exponential can be approximated by the linear term. Then
\begin{multline}
\label{e:basic_bracket}
\Big\{ (h_e[A])_{\alpha\beta}, \, F_S[\tilde{E},f] \Big\}  \\
\begin{aligned}[t]
& = \lim_{\epsilon\rightarrow 0} \left[\big(h^\epsilon_{e_1}[A]\big)_{\alpha\gamma}
\left\{ \int_{e(\epsilon)} {\rm d}x^m {A_m}^a (\x ) \tau^a_{\gamma\delta} \;,\;\,
\int_S {\rm d}y^n \wedge {\rm d}y^p \epsilon_{npq} 
f_b(\y) \tilde{E}^{bq} (\y ) \right\}
 \big(h^\epsilon_{e_2}[A]\big)_{\delta\beta} \right] \\[1ex]
& = \N (e,S) \gamma\, f_a(P) \big(h_{e_1}[A]\,\tau^a\, h_{e_2}[A]\big)_{\alpha\beta} \,.
\end{aligned}
\end{multline}
Here, the intersection number
\begin{equation}
\N(e,S) := \int_{e} {\rm d}x^m \int_S {\rm d}y^n {\rm d}y^p \,\epsilon_{mnp}\,
\delta^{(3)} (\x , \y ) = \pm 1 \;\; \text{or} \;\; 0\,,
\end{equation}
encodes the information on how $e$ intersects $S$ in a coordinate independent 
way. The integral is equal to $\pm 1$, depending on the orientation
of $e$ and $S$, if $e$ intersects $S$ transversally. When the edge 
$e$ intersects or touches $S$ tangentially, a little care must be 
exercised; one then finds that $\N(e,S)=0$, and the above bracket 
vanishes~\cite{Ashtekar:1998ak}. The fact that $h_e[A]$ and $F_S[A,f]$ 
are, respectively, supported on one-dimensional and two-dimensional 
subsets of $\Sigma$ is thus precisely what is required to perform the 
integral over the three-dimensional $\delta$-function. Evidently, the 
integral is ill-defined when an entire segment of $e$ lies within $S$; 
for a discussion of how to deal with this difficulty, 
see~\cite{Ashtekar:1998ak}.

\section{Quantisation: kinematics}
\label{s:kinematics}

Having determined the classical canonical variables, one would now
like to promote them to quantum operators obeying the appropriate
commutation relations.  The essential assumption of LQG is that this
quantisation should take place at the level of the bounded hermitean
operators~$h_e[A]$ rather than the connection~$A$ itself. This is
analogous to ordinary quantum mechanics, when one replaces the
Heisenberg operators~$x$ and~$p$ by Weyl operators~$e^{i x}$ and~$e^{i
p}$; the spin network representation actually uses the analog of a
hybrid formulation with~$x$ and~$e^{i p}$. The Stone-von Neumann
theorem~\cite{stone1,neum1,rose1} is usually invoked to argue that it
makes no difference whether one quantises the Heisenberg or the Weyl
algebra, {\it i.e.\/}~that these quantisations are equivalent. The
theorem does require, however, that the representations which are used
are `weakly continuous'. In the case of ordinary quantum mechanics,
for example, this means that matrix elements of the operators
corresponding to~$e^{i\alpha x}$ and~$e^{i\beta p}$ are smooth
functions of the parameters~$\alpha$ and~$\beta$. In LQG the
representations of operators do not satisfy this
requirement.\footnote{This is also the reason why the kinematical
Hilbert space employed in loop quantum cosmology is different from
the standard one already for a \emph{finite} number of degrees of 
freedom~\cite{Ashtekar:2003hd}. When the number of degrees of freedom
is infinite (as in quantum field theory), the Stone-von Neumann theorem 
anyhow does not apply.}

The failure of operators to be weakly continuous can, as we will see,
be traced back to the very special choice of the scalar
product~\eqn{e:scalar_product} below, which LQG employs to define its
kinematical Hilbert space~$\Hk$. This Hilbert space does not admit a
countable basis, hence is \emph{non-separable}, because the set of all
spin network graphs in~$\Sigma$ is uncountable, and non-coincident
spin networks are orthogonal w.r.t.~\eqn{e:scalar_product}.
Therefore, any operation (such as a diffeomorphism) which moves around
graphs continuously corresponds to an uncountable sequence of mutually 
orthogonal states in~$\Hk$. That is, no matter how `small' the deformation 
of the graph in~$\Sigma$, the associated elements of~$\Hk$ always 
remain a finite distance
apart, and consequently, the continuous motion in `real space' gets
mapped to a highly discontinuous one in~$\Hk$. Although unusual, and
perhaps counter-intuitive, as they are, these properties constitute a
cornerstone for the hopes that LQG can overcome the seemingly
unsurmountable problems of conventional geometrodynamics: if the
representations used in LQG were equivalent to the ones of
geometrodynamics, there would be no reason to expect LQG not to end up
in the same quandary.

It is perhaps also instructive to contrast the LQG approach with the
standard lattice approach to field theory. In the lattice approach,
all quantities depend explicitly on the lattice spacing ({\it
i.e.\/}~the regulating parameter). In the limit in which the lattice
spacing is taken to zero, one recovers the continuum results (all
expectation values are smooth functions of the regulating
parameter). In the LQG approach the `discretuum' is instead built in
by the very construction of the scalar product, rather than by
introducing a regulating parameter. LQG is therefore quite different
from `quantum gravity on the lattice'.  While the radical modification
underlying LQG has certain appealing properties, it makes it hard to
recover long-distance physics and the usual notion of continuity.

In short, one thus implements the quantisation not by the
replacement~\eqn{quant2} but rather by promoting the above Poisson
bracket~\eqn{e:basic_bracket} to a quantum commutator:
\begin{equation}
\label{e:basic_bracket1}
\Big[ (h_e[A])_{\alpha\beta} \,, \, \hat{F}_S[\tilde{E},f] \Big] = i\hbar l_P^2 \gamma\,
\N(e,S) \, f^a(P) \big(h_{e_1}[A]\,\tau_a\, h_{e_2}[A]\big)_{\alpha\beta} 
\end{equation}
or equivalently,
\begin{equation}
\label{e:basic_bracket2}
\Big[ (h_e[A])_{\alpha\beta} \, , \, \hat{F}^a_S[\tilde{E}] \Big] = i \hbar l_P^2 \gamma\,
\N(e,S) \, \big( h_{e_1}[A]\tau^a\, h_{e_2}[A] \big)_{\alpha\beta} \,.
\end{equation}
On the spin network representation the holonomies~$h_e[A]$ will be
represented as multiplication operators; the action of the canonically
conjugate operators will be explained below. The inequivalence of LQG
quantisation with Fock-space quantisation arises through a special
scalar product, to be discussed below.

\subsection{The Hilbert space of spin networks}
\label{s:spin_networks}

After defining the basic variables in which the theory should be
quantised, the next step is to choose a Hilbert space in which the
operators act.  Starting from this space, one should construct a
Hilbert space of physical states, {\it i.e.\/}~space of states for which all
constraints hold. The initial Hilbert space of LQG is the space of
spin networks. While the Gauss constraint is easily solved in this
space it turns out that a solution of the diffeomorphism constraints 
lies \emph{outside} this `naive' initial space, and one is forced to 
introduce a larger space.

The intuitive idea behind spin networks is that the geometry at the
Planck scale is foam-like. Physical gravitational and matter degrees
degrees of freedom are excited only on so-called spin networks, {\it
i.e.\/}~one-dimensional `edges' or `links', and the vertices
connecting them.  Geometries which look smooth at large scales are
supposed to arise only from complicated spin network states with many
edges. In order to find the Hilbert space of these objects, one has to
find a basis of wave functions over the configuration space, which
associate a complex number to each and every configuration of the
gauge connection. LQG makes use of wave functions which have singular
support in the sense that they only probe the gauge connection on
one-dimensional networks embedded in the three-dimensional spatial
hypersurface~$\Sigma$, which is a (not necessarily differentiable)
manifold, {\it i.e.\/}~can be mapped out by local charts. This three
dimensional `reference space', or `carrier space' of the spin
networks, does not carry any physical metric. LQG makes occasional use
of local coordinates, or fiducial background metrics for certain
intermediate steps in the construction, but physical quantities must
not depend on such background data. We will here avoid their use as
far as possible.  Let us emphasise again that the `discreteness' of
the spin networks does not correspond to a naive discretisation of
space. Rather, the underlying continuum, on which the spin networks
`float', the spatial manifold~$\Sigma$, is still present. As we will
see, the setup furthermore requires the \emph{a priori} exclusion of
infinite spin networks, that might contain Cantor-like or `fractal'
sets.

\begin{figure}[t]
\psfrag{e1}{\small $e_1$} \psfrag{e2}{\small $e_2$}
\psfrag{e3}{\small $e_3$} \psfrag{e4}{\small $e_4$} \psfrag{e5}{\small
$e_5$} \psfrag{e6}{\small $e_6$} \psfrag{e7}{\small $e_7$}
\psfrag{e8}{\small $e_8$} \psfrag{n1}{\small $v_1$} \psfrag{n2}{\small
$v_2$} \psfrag{n3}{\small $v_3$} \psfrag{n4}{\small $v_4$}
\psfrag{n5}{\small $v_5$}
\begin{center}
\includegraphics*[width=0.7\textwidth]{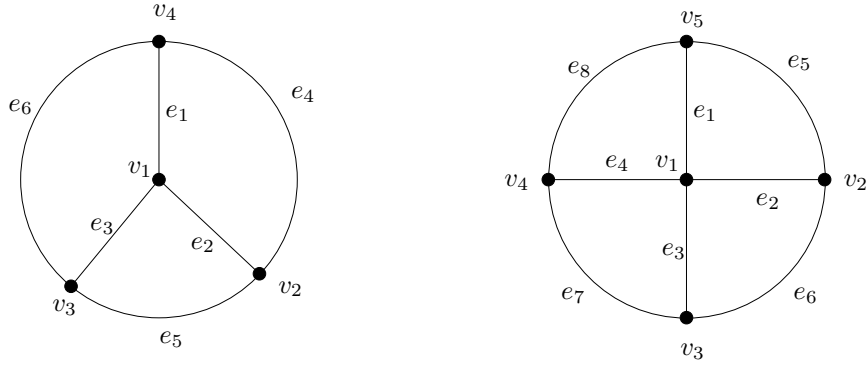}
\end{center}
\caption{Examples of spin network states. For the 3-valent vertices on
  the left, the three incoming edges at each vertex are connected by a
  Clebsch-Gordan coefficient. For the 4-valent vertex on the right,
  one has to decide on a given way to construct a higher-order
  invariant tensor from two Clebsch-Gordan coefficients.}
\label{f:network1}
\end{figure}

\begin{figure}[t]
\psfrag{j1}{\small $j_1$} \psfrag{j2}{\small $j_2$}
\psfrag{j3}{\small $j_3$} \psfrag{j4}{\small $j_4$} 
\psfrag{j5}{\small $j_5$}
\psfrag{k1}{\small $k_1$}
\psfrag{k2}{\small $k_2$}
\psfrag{a}{(a)}
\psfrag{b}{(b)}
\psfrag{c}{(c)}
\psfrag{k}{\small $k$}
\psfrag{Cj1j2}{}
\psfrag{Ck1k2}{}
\psfrag{Cj4j5}{}
\begin{center}
\includegraphics*[width=0.6\textwidth]{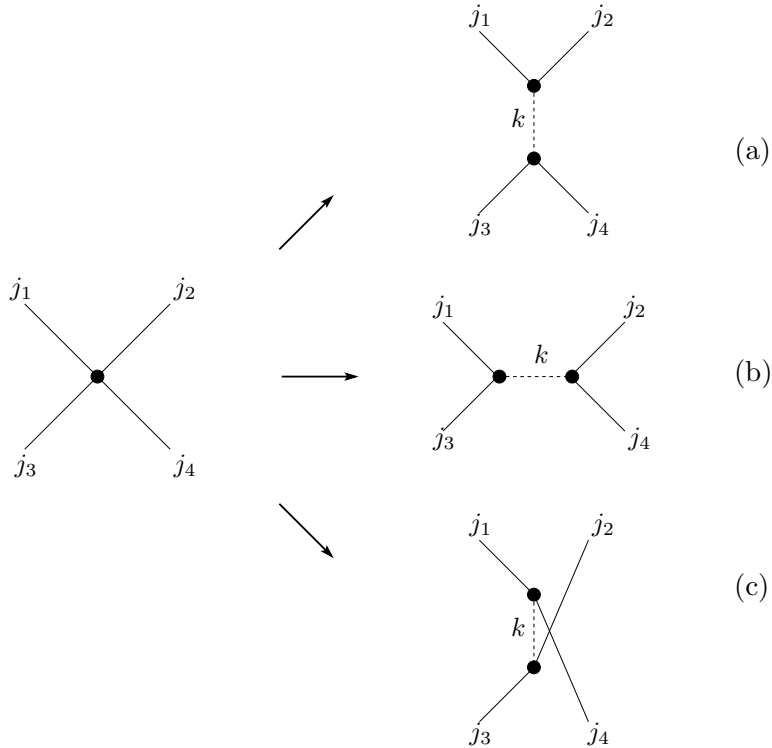}
\end{center}
\caption{A four-valent vertex is defined by a particular way of
  connecting three-valent vertices. The spin $k$ has to satisfy the
  triangle inequalities, but is otherwise
  arbitrary.\label{f:fourvalent}}
\end{figure}

By definition, each network is a (not necessarily connected) graph
$\Gamma$ embedded in $\Sigma$ and consisting of \emph{finitely many}
edges~$e_i\in\Gamma$ and vertices $v\in\Gamma$. The edges are
connected at the vertices. Each edge $e$ carries a holonomy $h_{e}[A]$
of the gauge connection $A$ (this connection does not have to be
smooth on the edge). The wave function on the spin network over the
graph $\Gamma$ can be written as
\begin{equation}
\label{psi-gamma}
\Psi_{\Gamma,\psi}[A] = \psi\big( h_{e_1}[A],\, h_{e_2}[A],\, \ldots \big)\,,
\end{equation}
where the $\psi$ is some function of the basic holonomies associated
to the edges $e\in\Gamma$. If, in addition, the wave function $\psi$
is invariant under arbitrary SU(2) gauge transformations it satisfies
the \emph{Gauss constraint}, and vice versa. A gauge invariant
function $\psi$ thus takes care of joining the collection of
holonomies into an SU(2)-invariant complex number by contracting all
SU(2)-indices of the holonomies with invariant tensors `located' at
the vertices $v$, see figure~\ref{f:network1}. The basic building
blocks of the spin network wave functions are therefore expressions of
the following type.  A three-valent vertex connects three edges
according to
\begin{equation}\label{e:threevalent}
\psi [{\text{fig.\ref{f:network1}a}}] = 
\Big( \rho_{j_1}(h_{e_1}[A])\Big)_{\alpha_1 \beta_1}\; 
\Big( \rho_{j_2}(h_{e_2}[A])\Big)_{\alpha_2 \beta_2}\;
\Big(\rho_{j_3}(h_{e_3}[A])\Big)_{\alpha_3 \beta_3}\, 
C^{j_1 j_2 j_3}_{\beta_1 \beta_2 \beta_3}  \ldots \,,
\end{equation}
where dots represent the remainder of the graph. While the contraction 
is obviously unique if only two or three edges meet at a vertex, there 
may be more and independent choices for vertices of valence four and 
higher, depending on the way in which the edges are connected with the 
Clebsch-Gordan coefficients. For this reason, any given spin network 
will in general admit several independent wave functions of the above 
type. As an example, consider a four-valent vertex: one first has to 
decide on how to pair the edges into two groups of two. One such choice 
leads to {\it e.g.} 
\begin{multline}
\label{e:fourvalent}
\psi [{\text{fig.\ref{f:network1}b}}] = 
\Big(\rho_{j_1}(h_{e_1}[A]\Big)_{\alpha_1 \beta_1} \;
\Big(\rho_{j_2}(h_{e_2}[A])\Big)_{\alpha_2 \beta_2} \;
\Big(\rho_{j_3}(h_{e_3}[A])\Big)_{\alpha_3 \beta_3} \;
\Big(\rho_{j_4}(h_{e_4}[A])\Big)_{\alpha_4 \beta_4} \\[1ex]
\times C^{j_1 j_2 k}_{\beta_1 \beta_2 \beta} 
C^{j_3 j_4 k}_{\beta_3 \beta_4 \beta} \ldots \,.
\end{multline}
In these equations $C^{j_1 j_2 k}_{\beta_1 \beta_2 \beta}$ are
intertwiners (Clebsch-Gordan coefficients). In~\eqn{e:fourvalent} the
intermediate spin~$k$ can be freely chosen in accordance with the
standard rules for the vector addition of angular momenta:
$|j_1-j_2|\leq k \leq j_1 + j_2$.  In other words, we can graphically
represent the 4-valent vertex by splitting it into two `virtual'
3-valent vertices and adding a `virtual' edge, carrying angular
momentum $k$ (see figure~\ref{f:fourvalent}). The same wave function
can be re-expressed by performing this split in a different `channel'
by means of recoupling relations for the Clebsch Gordan coefficients
(see for instance~\cite{Perez:2003vx})
\begin{equation}
\label{e:recoupling}
C^{j_1 j_2 k}_{\alpha_1\alpha_2\beta} 
C^{j_3 j_4 k}_{\alpha_3\alpha_4\beta} 
= \sum_{m} \sqrt{2k+1}\sqrt{2m+1} 
   \left\{\begin{matrix} j_1 & j_2 & k \\
                         j_3 & j_4 & m \end{matrix}\right\}
 C^{j_1 j_3 m}_{\alpha_1\alpha_2\beta} 
 C^{j_2 j_4 m}_{\alpha_3\alpha_4\beta} \,,
\end{equation}
where the object with curly brackets is the Wigner $6j$-symbol. 
To make the dependence of a general spin network wave functional
on the spins~$j_e$ associated with the edges and the
intertwiners~$C_v$ associated with the vertices explicit, one
occasionally designates functionals such as~(\ref{psi-gamma}) also
by $|\Gamma, \{j\}, \{ C\}\rangle $. The space of finite linear
combinations of such states is denoted by $\S$ (for `spin networks').
Although three-valent spin networks are obviously simplest, we will
see later (see section~\ref{s:volume}) that higher valence is even
generic because three-valent networks correspond to `zero volume', and
hence are deemed to be not of much interest.

The wave functionals (\ref{psi-gamma}) are called \emph{cylindrical}, 
because they probe the connection $A$ only `on a set of measure 
zero' (like the $\delta$-function does for ordinary functions). Similar 
`cylindrical functionals' were formerly used in constructive quantum 
field theory in order to rigorously define the functional measures for 
free and certain interacting models of quantum field theory as limits of 
finite dimensional integration measures~\cite{Simon:1974dg}. The 
space $\S$ spanned by finite linear combinations of such cylindrical 
functions over all possible graphs is the starting point for the 
construction of the Hilbert space of spin networks. Obviously, the 
product of two cylindrical functions supported on the same or 
different spin networks is again a cylindrical function.

To complete the definition of the space of spin network states, we must
introduce a suitable scalar product. In LQG this is not the standard scalar 
product induced by a Fock space representation (see the following section 
for more on this); instead, the scalar product of two cylindrical functions
$\Psi_{\Gamma,\{j\},\{C\}}[A]$ and $\Psi_{\Gamma',\{ j'\},\{C'\}}[A]$ 
is defined as
\begin{multline}
\label{e:scalar_product}
\big\langle \Psi_{\Gamma,\{j\},\{C\}} \,\big|\, 
\Psi'_{\Gamma',\{j'\},\{C'\}} \big\rangle \\[1ex]
=\begin{cases}
\;\;0 &\text{if $\Gamma \neq \Gamma'$}\,, \\[1ex]
\displaystyle\int \prod_{e_i\in\Gamma} {\rm d}h_{e_i}\;
\, \bar\psi_{\Gamma,\{j\}, \{C\}}\Big(h_{e_1},\ldots\Big)\,
\psi'_{\Gamma',\{j'\},\{C'\}}\Big(h_{e_1},\ldots\Big) & 
\text{if $\Gamma = \Gamma'$} \,,
\end{cases}
\end{multline}
where the integrals~$\int\!{\rm d}h_e$ are to be performed with the
SU(2) Haar measure. The above scalar product may look artificial, but
given a few reasonable assumptions, there in fact exists a strong
uniqueness theorem; the reader is referred
to~\cite{Sahlmann:2002xu,Sahlmann:2002xv,Sahlmann:2003in,Okolow:2003pk,Sahlmann:2003qa,Okolow:2004th}
for details.

From \eqn{e:scalar_product} we see that the inner product vanishes if the 
graphs $\Gamma$ and $\Gamma'$ do not coincide (even if they are `very close
to each other' in any given fiducial background metric). If $\Gamma$
and $\Gamma'$ coincide, the product may still vanish, depending on the
choice of spins and intertwiners. At any rate, for $\Gamma= \Gamma'$,
the inner product is given by integrating the product of $\bar\psi$
and $\psi'$ over the holonomies at all edges, using the standard Haar
measure.  Because $\Gamma$ was assumed to consist of \emph{finitely}
many edges, this is always a \emph{finite-dimensional} integral, with
one SU(2) integral for every edge in the graph. It is noteworthy that
the above scalar product is \emph{invariant under spatial
diffeomorphisms}, even if the states $\Psi_1$ and $\Psi_2$ themselves
are not, because the statement whether two graphs coincide or not is
diffeomorphism invariant.  Whereas the original wave functionals
$\Psi[A]$ probe the value of the connection and therefore also depend
on the position of the graph, this information is `lost' in the scalar
product \eqn{e:scalar_product} which makes no more reference to the
underlying space of connections or the `shape' of the spin network
graph.

One now defines the \emph{kinematical Hilbert space} $\Hk$ as the completion 
of the space of gauge invariant spin network states $\S$ w.r.t.~to this norm. 
$\Hk$~thus consists of all linear superpositions of spin network states 
$\Psi_n\equiv | \Gamma_n,\{j_n\},\{C_n\}\rangle$ such that they have
finite norm,
\begin{equation}\label{Norm1}
\Psi = \sum_{n=1}^\infty a_n \Psi_n \quad , \qquad
||\Psi||^2 =\sum_{n=1}^\infty |a_n|^2 \, ||\Psi_n ||^2 < \infty 
\end{equation}
where the norm $||\Psi_n||^2$ is defined by \eqn{e:scalar_product}.
One of the distinctive features of the Hilbert space $\Hk$, which will 
play a key role in the further development of the theory, is its 
\emph{non-separability}. This non-separability can be traced back to 
the existence of an underlying continuum, the spatial manifold $\Sigma$,
and accounts for the difference between this scheme and the standard
(Fock space) quantisation of gauge theories, see the following subsection.
Although each $\Psi_n$ in the above sum is associated to a \emph{finite} 
graph $\Gamma_n$ the expected number of edges need not be finite, 
because the sum
\begin{equation}
\langle \#(\text{edges})\rangle \; \propto  \;
\sum_{n=1}^\infty |a_n|^2 \, L(\Gamma_n)||\Psi_n ||^2 \; ,
\end{equation}
with $L(\Gamma_n)$ the number of edges in $\Gamma_n$, can be made to
diverge if $L(\Gamma_n)$ increases sufficiently rapidly with $n$, even
if $\sum_n |a_n|^2 \, ||\Psi_n||^2 < \infty$. Idem for the expectation
value for the number of vertices. Let us, however, caution readers
already at this point that $\Hk$ is \emph{not} the relevant Hilbert
space for solving the quantum constraints in LQG. The physical Hilbert
space, consisting of those states which satisfy the constraints,
\emph{is} expected to be separable; we will have more to say about
this in the following sections.

A second unusual feature, at least for a traditional quantum field
theorist, is the \emph{ab initio} absence of negative norm states,
despite the fact that at this stage the constraints have not even
appeared yet. This is in stark contrast to the usual covariant Fock
space quantisation of gauge theories, where negative norm states are
unavoidable, and can only be eliminated (if they can be eliminated at
all) by restricting the Hilbert space to the subspace of
\emph{physical states}, which by definition are annihilated by the
constraints. Curiously, there does exist a reformulation of (free)
second quantised Maxwell theory with similar
properties~\cite{Thirring:1992hu}. Roughly speaking, this consists in
($i$) defining the scalar product to be identical with the standard
Fock space one for gauge invariant expectation values (as is well
known, in second quantised Maxwell theory this restriction indeed
eliminates negative norm states), and ($ii$) by declaring
gauge-variant scalar products to be zero. In this way one preserves
Lorentz covariance (because a gauge variant state is never transformed
into a gauge invariant one by a Lorentz transformation, and vice
versa), and eliminates negative norm states, in such a way that
\emph{for observable quantities the results are identical with the
usual Fock space quantisation}. The price one pays is the loss of weak
continuity (because the expectation values for gauge variant
expressions always vanish, no matter how close they are to gauge
invariant ones) and the loss of separability of the Hilbert
space. These are precisely the features encountered in the LQG
quantisation procedure above. However, the crucial difference is that
the prescription~\cite{Thirring:1992hu} is tailored to reproduce the
standard Fock space quantisation in the observable sector, whereas LQG
does not. Also, it is not known how to extend the prescription of
\cite{Thirring:1992hu} to full QED or Yang-Mills theory, or how to
implement the procedure in a perturbative approach (with unphysical
intermediate states in Feynman diagrams).
\bigskip

\subsection{Comparison with the loop approach in gauge theories}

The (Wilson) loop approach has been advocated in the past in the
context of gauge theories, as an attempt to non-perturbative
quantisation.  This approach eventually did not lead to the expected
success, see \cite{Migdal:1984gj,Polyakov:1987ez} for detailed reviews
of the subject.  Nevertheless LQG practitioners argue that their
approach, despite its similarity to the loop approach in gauge
theories, should work for gravity.  Because this relates to one of the
frequently asked questions about the LQG programme, we here briefly
contrast some key features of the Wilson loops in gauge theory with
the behaviour of spin networks in LQG.  Not unexpectedly, the
essential difference turns out to lie in the special scalar
product~\eqn{e:scalar_product} used in LQG.

The first difference between Wilson loops and spin networks is that,
due to the product~\eqn{e:scalar_product}, any spin network is orthogonal 
to any other spin network except itself, including the trivial empty one
({\it i.e.\/}~the vacuum). Hence,
\begin{equation}
\label{e:LQG_single_loop}
\langle\, \Psi_{\Gamma, \{C\}}[A] \,\rangle = 
\langle\, {\bf 1} \,|\, \Psi_{\Gamma, \{C\}}[A] \,\rangle = 0 \, .
\end{equation}
By contrast, in gauge theory, the expectation value of a single Wilson loop 
encodes key physical information: it determines the behaviour and
properties of quark\,-\,anti-quark potentials, and can signal
confinement of quarks. More specifically, in a perturbative approach we have
\begin{equation}
\label{e:W-vev}
\langle\, W_{\cal C}\, \rangle \equiv \left\langle\, \exp\left[i e  
\oint_{\cal C}  A_{\mu} {\rm d}x^{\mu}\right] \right\rangle 
= \exp\left[-\frac{i e}{2} \oint_{\cal C}\oint_{\cal C} 
{\rm d}x^{\mu}{\rm d}x^{\nu}\,\Delta_{\mu\nu}(x-y)\right] \, ,
\end{equation}
where $\Delta_{\mu\nu}$ is the gluon propagator. In conventional 
quantum field theory (QCD), this expression is divergent, in contrast 
to~\eqn{e:LQG_single_loop}) because of the self-interaction of the quarks 
and gluons propagating in the loop. The way in which it diverges generically 
depends on the shape of the loop. Nevertheless, the~UV divergences 
in~\eqn{e:W-vev} can be consistently removed, for example by point 
splitting regularisation, and a well defined answer obtained. All loops 
can be regularised in this way (while it is only known how to renormalise, 
\emph{smooth} and \emph{non-self-intersecting} loops~\cite{Migdal:1984gj}).
A better, and entirely non-perturbative scheme to calculate the Wilson 
loops and other quantities of interest is, of course, provided by 
lattice gauge theory~\cite{Montvay:1994cy,roth1,Creutz:1984mg}.

In the case of correlation functions of several Wilson loops, no new
types of divergences appear. For two non-intersecting loops $\cal C$ 
and ${\cal C}'$ `self energy divergences' can be removed by defining
the `connected correlator'
\begin{equation}
\langle W_{\cal C} W_{{\cal C}'} \rangle_{\text{conn}} :=
\langle W_{\cal C} W_{{\cal C}'} \rangle -
\langle W_{\cal C} \rangle \langle W_{{\cal C}'}\rangle  = \text{finite}  \, ,
\end{equation}
which can be thought of as the analog of the scalar product of two 
single loops in LQG. In first approximation we then obtain
\begin{equation}
\big\langle\, W_{{\cal C}_1}\,| W_{{\cal C}_2}\big\rangle_{\text{conn}} 
= \exp\left[-\frac{i e}{2} \oint_{{\cal C}_1} {\rm d}x^{\mu}
\oint_{{\cal C}_2} {\rm d}y^{\nu}\,\Delta_{\mu\nu}(x-y)\right] \, .
\end{equation}
The main point about this result is that it depends continuously on the 
shape of the loops ${\cal C}_1$ and ${\cal C}_2$ -- unlike the scalar 
product or correlator of two spin networks~\eqn{e:scalar_product}, which 
is non-zero if and only if the two networks completely overlap. Indeed,
different shapes of Wilson loops in gauge theory are inequivalent, 
such that the relative change can be expressed via the Yang-Mills fields 
strength and the `non-abelian Stokes' theorem'. The value of the Wilson 
loop is invariant under continuous deformations only for vanishing field 
strength (as in a topological theory, such as Chern-Simons gauge theory). 
In LQG on the other hand, the physical states (for which we have to enlarge 
the space~$\S$, as we said) are supposed to be diffeomorphism invariant. 
Hence there is no physical information whatsoever in the shape of the 
loop, since two networks of different shape but identical topology 
can always be related by a suitable diffeomorphism. This equivalence 
holds independently of the `value' of the Ashtekar field strength 
in this state.

When matter is included in the LQG formalism, consistency would seem
to require that standard gauge fields must be treated analogously to
the Ashtekar connection, {\it i.e.\/}~with ho\-lo\-nomies associated to
edges, and a scalar product similar to \eqn{e:scalar_product}, leading
to a non-separable kinematical Hilbert space for the matter sector as
well. It is not known how (and whether) this formalism can recover
known results such as the ones sketched above, and in particular the
shape dependence of the Wilson loops, and we do not know of any
attempts in this direction.  At the very least, this will require a
sophisticated analog of the `shadow states' introduced
in~\cite{Ashtekar:2002sn}. For kinematical states (which do not
satisfy all the constraints) progress along these lines has been made
recently
in~\cite{Sahlmann:2002qj,Sahlmann:2002qk,Thiemann:2000bw,Thiemann:2000ca,Thiemann:2000bx,Thiemann:2000by,Sahlmann:2001nv,Thiemann:2002vj}.

\subsection{The action of elementary operators on a spin network}

We would now like to implement the canonical variables as operators on
spin network wave functions in such a way that the canonical
commutation relations~\eqn{e:basic_bracket1} and
\eqn{e:basic_bracket2} are satisfied. This will provide us with the
`building blocks' that will allow us to define the action of the basic
kinematical operators (area and volume operators) and the Hamiltonian
constraint operator on any spin network wave function.  The elementary
holonomy and flux operators defined in~\eqn{e:holonomy_def}
and~\eqn{e:flux_def} act in a rather simple way on the spin network
states.  Namely, the holonomy $(h_e[A])_{\alpha\beta}$ simply acts as
a multiplication operator, and thus adds the edge $e$ to an existing
network $\Gamma$.  This edge may be disconnected from or `overlaid'
on~$\Gamma$; usually it will appear in some gauge invariant
combination, whose action on the spin network wave function produces
another such wave function. The action of the flux
$\hat{F}_S[\tilde{E},f]$ on a network state is either zero (if the
surface element $S$ does not intersect the graph $\Gamma$ anywhere),
or otherwise cuts in two any edge which pierces~$S$, and inserts a new
intertwiner vertex $\tau_a$ into the spin network at the point of
intersection. Consequently, on such a single edge (which is part of
some spin network wave function $\Psi_{\Gamma,\{C\}}$) one has
\begin{equation}
\label{smeaE}
\hat{F}^a_S[\tilde E]\, \Big( \dots \big(h_{e}[A]\big)_{\alpha\beta} \dots \Big)
= 8 \pi i l_P^2\hbar \gamma\, \N(e,S) \; \Big( \dots  
\big(h_{e_1}[A]\,\tau^a\, h_{e_2}[A]\big)_{\alpha\beta} \dots \Big) \, .
\end{equation}
where $e= e_1 \cup e_2$ and we have now omitted the symbol
$\rho_{j_e}$. Dots indicate the remaining part of the spin network
wave function.

\begin{figure}[t]
\psfrag{1}{1}
\psfrag{2}{2}
\psfrag{3}{3}
\begin{center}
\includegraphics*[width=.3\textwidth]{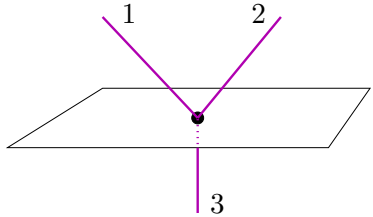}
\end{center}
\caption{A flux operator intersecting a three-valent node. The action
  is given by equation~\eqn{e:flux_3val_node}.}
\label{f:flux_3val_node}
\end{figure}

If $\hat{F}^a_S[\tilde{E}]$ cuts the network at a vertex, the result
depends on the choice for the intertwiner as well as the position and
orientation of the surface~$S$ with respect to the edges. Let us illustrate 
this by considering the 3-valent vertex depicted in
figure~\ref{f:flux_3val_node}.
The flux acts at the end of every edge that emanates from the node. 
Edges which are located above the surface~$S$ contribute with opposite
sign from the ones that are located below, and for the example of the
3-valent vertex the net result is
\begin{multline}
\label{e:flux_3val_node}
\hat{F}_S^a[\tilde{E}] \Big( 
C^{j_1 j_2 j_3}_{\alpha_1\alpha_2\alpha_3} 
   \big( h_{e_1}[A]\big)_{\alpha_1\beta_1}
   \big( h_{e_2}[A]\big)_{\alpha_2\beta_2}
   \big( h_{e_3}[A]\big)_{\alpha_3\beta_3}\Big) \\[1ex]
\begin{aligned}
&= 8\pi l_P^2\hbar\gamma \Big( \big(\rho_{j_1}(\tau^a)\big)_{\alpha_1\gamma_1} 
C^{j_1 j_2 j_3}_{\gamma_1\alpha_2\alpha_3}
+ \big(\rho_{j_2}(\tau^a)\big)_{\alpha_2\gamma_2} 
C^{j_1 j_2 j_3}_{\alpha_1\gamma_2\alpha_3}
- \big(\rho_{j_3}(\tau^a)\big)_{\alpha_3\gamma_3} 
C^{j_1 j_2 j_3}_{\alpha_1\alpha_2\gamma_3}\Big)\\[1ex]
&\quad\quad\quad\quad\quad\times \Big(\big( h_{e_1}[A]\big)_{\alpha_1\beta_1}
   \big( h_{e_2}[A]\big)_{\alpha_2\beta_2}
   \big( h_{e_3}[A]\big)_{\alpha_3\beta_3}\Big)\\[1ex]
&= -16\pi l_P^2\hbar\gamma\, \big(\rho_{j_3}(\tau^a)\big)_{\alpha_3\beta_3} 
  C^{j_1 j_2 j_3}_{\alpha_1\alpha_2\beta_3}
 \Big(\big( h_{e_1}[A]\big)_{\alpha_1\beta_1}
   \big( h_{e_2}[A]\big)_{\alpha_2\beta_2}
   \big( h_{e_3}[A]\big)_{\alpha_3\beta_3}\Big)
\end{aligned}
\end{multline}
(the uncontracted indices $\beta_1, \beta_2, \beta_3$ connect to the
remainder of the spin network wave function). In the last step, we have 
used the invariance property of the Clebsch-Gordan coefficients,
\begin{equation}
\label{e:CG_invariance}
  \Big(\rho_{j_{1}}(\tau^a)\Big)_{\alpha_1\beta_1} C^{j_{1}\,j_{2}\,j_{3}}_{\beta_1\alpha_2\alpha_3}
+ \Big(\rho_{j_{2}}(\tau^a)\Big)_{\alpha_2\beta_2} C^{j_{1}\,j_{2}\,j_{3}}_{\alpha_1\beta_2\alpha_3}
+ \Big(\rho_{j_{3}}(\tau^a)\Big)_{\alpha_3\beta_3} C^{j_{1}\,j_{2}\,j_{3}}_{\alpha_1\alpha_2\beta_3}=0\,.
\end{equation}

Likewise, when the node has higher valence, the action of the flux operator
$\hat{F}^a_S[\tilde{E}]$ inserts factors $\pm \rho_{j_e} (\tau^a)$ in the 
appropriate place in the spin network wave function, with a `+' for edges 
$e$ `above' $S$, and a `$-$' for the edges `below' $S$. This action 
is sometimes symbolically represented in terms of angular momentum
operators $\hat{J}^a_{(e,v)}$. Consequently, the flux operator can be re-expressed 
as a sum of angular momentum operators, viz.
\begin{equation}\label{F=J}
\hat{F}^a_S[\tilde{E}] = \sum_e \sigma(e,S) \hat{J}^a_{(e,v)}\,,
\end{equation}
where the sum runs over all edges entering the vertex $v$, and 
the signs $\sigma(e,S)= \pm 1$ are determined by the position
of $e$ in relation to $S$. When the signs are all `+', the sum 
vanishes for gauge invariant wave functionals by angular 
momentum conservation, {\it i.e.\/}~Gauss' law.

\subsection{The kinematical operators: area and volume operators}
\label{s:kinop}

Two important \emph{kinematical operators} in quantum gravity are the
volume and the area operators. The area operator is relevant in the
application of LQG to black hole physics, while the volume operator is
essential for the very definition and the implementation of the scalar
(Hamiltonian) constraint. It is important to stress that neither of
these operators is an observable \emph{\`a la} Dirac, in the sense
that neither commutes with all the canonical constraints (classically
or quantum mechanically) \footnote{As pointed out
in~\cite{Kuchar:1993ne}, such observables might perhaps better be
called \emph{`perennials'}, to distinguish them from the kinematical
`observables' which commute only with the spatial diffeomorphism
generators.}. Their supposed physical relevance is based on the
general belief that proper Dirac observables corresponding to area and
volume should exist, provided suitable matter degrees of freedom are
included~\cite{Brown:1994py,Kuchar:1993jw}. The reason is that, for an
operational definition of area and volume, one needs `measuring rod
fields' in the same way that the operational definition of `time' in
quantum gravity requires a `clock field', and that the true
observables are appropriate combinations of gravitational and matter
fields (which then would commute with the constraints). Although there
has been much discussion in the literature about these `relational'
aspects (see \cite{isha1,Kuchar:1993ne,b_zeh,Kiefer:2004gr} and
references therein, and \cite{Dittrich:2004cb,Thiemann:2004wk} for a
more recent discussion in the context of LQG), these issues are by no
means conclusively settled, and we are not aware of any fully concrete
realisation of this intuitive idea (see also~\cite{Kuchar:1991qf}).

Setting aside such conceptual worries, the main strategy in constructing 
the area and volume operators is to first write the classical expressions 
for these quantities using the basic variables (holonomies and electric 
field), and then promote those into operators. To do this one first has to 
regularise the classical expressions by a discretisation of the 
integrals~\cite{Ashtekar:1996eg,Ashtekar:1997fb}. It will turn out that 
the ambiguity in the way one does this is irrelevant for the area operator, 
while it does affect the result for the volume operator.

We should point out that the area and volume operators appear to be of
no particular relevance in conventional geometrodynamics. There, they
would simply be represented by multiplication operators (defining a
`length operator' would also be straightforward, as it is in
LQG~\cite{Thiemann:1996at}). As long as these act on smooth wave
functionals, there would be no problem, and issues of regularisation
and renormalisation, like the ones we are about to discuss, would not
even arise. On the other hand, on more singular `distributional'
functionals, these operators would not be well defined, as they depend
non-linearly on the basic metric variables. Whether it is possible or
not to properly define these operators in geometrodynamics therefore
depends on what the `good' and the `bad' wave functionals are. But
this question is impossible to answer in the absence of a suitable
scalar product and a Hilbert space of wave functionals (see
e.g.~\cite{isha1} for a discussion of these and other difficulties in
geometrodynamics).

\subsubsection{The area operator}

A two-dimensional surface $S \subset \Sigma$ possesses a unit normal
vector field $\vec{N}$ and its classical area is given by the
determinant of the induced metric $h_{mn} = g_{mn} - N_m N_n$.
To compute the area, it is however simpler to work directly with
the area element~\eqn{dFa} (cf.~subsection~\ref{s:conn2hol}), so that
\begin{equation}
A_{S}[g] = \int_{S} \, \sqrt{{\rm d}F^a \cdot {\rm d}F^a}\,.
\end{equation}
To convert this into an expression that can act on the spin network
wave function, we subdivide the surface into infinitesimally small 
surfaces $S_I$ (with $I=1,\dots,N$), such that $S= \cup_I S_I$. Next,
we approximate the area by a Riemann sum (which, of course, converges 
for well-behaved surfaces~$S$), using
\begin{equation}
\int_{S_I} \, \sqrt{{\rm d}F^a \cdot {\rm d}F^a} \approx
\sqrt{{F^a_{S_I}[\tilde{E}]\, F^a_{S_I}[\tilde{E}] }}\,.
\end{equation}
Therefore
\begin{equation}
\label{Area}
A_{S}[\tilde{E}_m^a] 
= \lim_{N \rightarrow \infty} \sum_{I=1}^N
\sqrt{{F^a_{S_I}[\tilde{E}]\, F^a_{S_I}[\tilde{E}] }} \, . 
\end{equation}

The Riemann sum on the right-hand side of~\eqn{Area} can now, for fixed~$N$, 
be promoted unambiguously to an operator. Namely, when acting on an
arbitrary spin network wave function, the area operator will receive 
a contribution each time an edge of the spin network pierces
an \emph{elementary} surface~$S_I$ in the sum~\eqn{Area}. More
precisely, one can use equation~\eqn{smeaE} to derive (always
in the appropriate representations $\rho_{j_{e_i}}$ and assuming
$\N(e,S) = \pm 1$) 
\begin{equation}
\label{E2}
\begin{aligned}
\hat{F}^a_{S_I}[\tilde{E}]\,\hat{F}^a_{S_I}[\tilde{E}]\;\; 
\big(h_e[A]\big)_{\alpha\beta} &= i^2 (8 \pi l_P^2 \hbar \gamma)^2\, 
\big(h_{e_1}[A]\,\tau_a\,\tau^a\, h_{e_2}[A]\big)_{\alpha\beta} \\[1ex]
&= (8 \pi l_P^2 \hbar \gamma)^2\,\big( j_e(j_e+1) \big)\, 
\big(h_{e}[A]\big)_{\alpha\beta} \, .
\end{aligned}
\end{equation}
Note also that if the the surface~$S_I$ had been pierced by two edges
(carrying representations $j_1$ and $j_2$) instead of just one, the
previous expression would generalise to
\begin{multline}
\label{E22}
\hat{F}^a_{S_I}[\tilde{E}]\,\hat{F}^a_{S_I}[\tilde{E}]\;\;
\big(h_{e_1}[A]\big)_{\alpha\beta} 
\big(h_{e_2}[A]\big)_{\gamma\delta} \\[1ex]
= (8 \pi l_P^2 \hbar \gamma)^2\, \Big(j_{e_1}(j_{e_1}+1) + j_{e_2}(j_{e_2}+1)\Big)\;
\big(h_{e_1}[A]\big)_{\alpha\beta} \big(h_{e_2}[A]\big)_{\gamma\delta} \, .
\end{multline}
Hence if one applies the operator~(\ref{Area}) to wave function associated
with a fixed graph $\Gamma$ with $L(\Gamma)$ edges, and refines it in such 
a way that each elementary surface~$S_I$ is pierced by only \emph{one} 
edge of the network, one obtains
\begin{equation} 
\label{Area-net}
\hat{A}_S  \Psi  = 8 \pi l_p^2\hbar \gamma \sum_{p=1}^{L(\Gamma)} 
\sqrt{j_p(j_p + 1)}\, \Psi  \, .
\end{equation}
It is obvious that further refinement of the area operator then does
not change the result. It is also clear that the final result,
therefore, does not depend on the shape of the elementary area cells.
From (\ref{Area-net}), one sees that all spin networks are eigenstates
of the area operator (when the~$S_I$ intersect only edges, see
below). Due to the `discreteness' of the spin network states, the
spectrum of the area operator is also discrete.  The quantisation
scale is set not only by the Planck length~$l_p^2$, but also depends
on the parameter~$\gamma$.\footnote{We emphasise that this
discreteness, which is often invoked as the underlying reason for the
absence of divergences in LQG, hinges on the compactness of the
group~SU(2), and would not hold if this group were replaced by a
non-compact or complex group (as would be the case for the original
Ashtekar variables with imaginary~$\gamma$).}

\begin{figure}[t]
\psfrag{1}{\smaller 1}
\psfrag{2}{\smaller 2}
\psfrag{3}{\smaller 3}
\psfrag{4}{\smaller 4}
\begin{center}
\includegraphics*[width=.9\textwidth]{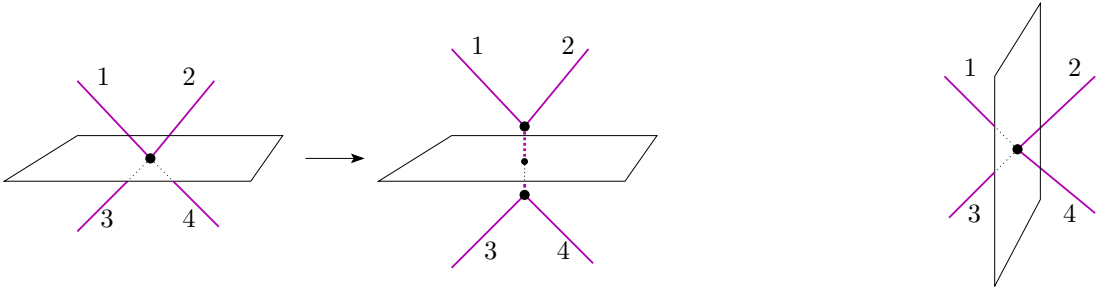}
\caption{The action of the area operator on a node with intertwiner
$C^{j_1 j_2 k}_{\alpha_1 \alpha_2 \beta} C^{j_3 j_4 k}_{\alpha_3
\alpha_4 \beta}$. In the figure on the left, the location of the edges
with respect to the surface is such that the invariance of the
Clebsch-Gordan coefficients~\eqn{e:CG_invariance} can be used to
evaluate the action of the area operator. The result can be written in
terms of a ``virtual'' edge. In the figure on the right, however, the
edges are located in a different way with respect to the surface. The
invariance property~\eqn{e:CG_invariance} does not apply, one has to
use the recoupling relation~\eqn{e:recoupling}, and the spin network state is
therefore not an eigenstate of the corresponding area operator.}
\label{f:area2}
\end{center}
\end{figure}

We have so far assumed that the surface elements $S_I$ in the above 
computation `meet' only the edges (hence insert a bivalent vertex 
into the spin network), but not the vertices of the spin network.
Making use of the general result \eqn{F=J} we can also evaluate
the area operator when the surface elements~$S_I$ intersect a vertex.
In this case, the result depends on the particular intertwiner at that 
node and on the orientation of the surface element. Let us illustrate 
this on a four-valent vertex, with intertwiner 
$C^{j_1 j_2 k}_{\alpha_1 \alpha_2 \beta} 
C^{j_3 j_4 k}_{\alpha_3 \alpha_4 \beta}$ (see figure~\ref{f:area2}). 
Consider first the situation in which the
edges~$1$ and~$2$ are both located on the same side of the surface
element and the same is true for edges~$3$ and $4$
(figure~\ref{f:area2}a). In this case, one can use the invariance of
the Clebsch-Gordan coefficients~\eqn{e:CG_invariance} to move around
the inserted $\tau$-matrices, and re-write the action of the area operator 
in terms of a `virtual' edge. For our example, this yields the result
\begin{equation}
\hat{A}_S \,|\text{figure~\ref{f:area2}a}\rangle
 =  8 \pi l_p^2\hbar \gamma\, \sqrt{2\,k(k+1)}\, 
|\text{figure~\ref{f:area2}a}\rangle\,.
\end{equation}
Observe the relative factor of~$\sqrt{2}$ with respect
to~\eqn{Area-net}: the result for a virtual edge is not the same as
the one for a real edge.  The four-valent node is thus an eigenstate
of this area operator. Next, consider the situation in which the
surface element is oriented as in figure~\ref{f:area2}b. In this case,
one first has to rewrite the product of Clebsch-Gordan coefficients by
means of the recoupling relation~\eqn{e:recoupling}. Independently of
the normalisation of the $6j$-symbol, the important thing to note is
that the area operator will \emph{change} the relative coefficients of
the various terms in the sum~\eqn{e:recoupling}, picking up a
contribution $\propto \sqrt{m(m+1)}$ for each admissible value of the
angular momentum $|j_1 - j_3|\leq m \leq j_1 + j_3$ in the new channel
$1+3 \rightarrow 2+4$. Hence, the four-valent vertex is generally
\emph{not} an eigenstate of the area operator associated to the
surface of figure~\ref{f:area2}b. For higher valences, the number of
different contributions increases rapidly with the number of
accessible channels. Therefore, the area operator in general does
\emph{not} act diagonally on an arbitrary spin network state.

We conclude this subsection with three remarks. Although infinite spin
networks have been excluded \emph{by fiat}, it is nonetheless
instructive to see what would happen if we did include them. In this
case it may not be possible to achieve the refinement of one edge per
one cell in a \emph{finite} number of steps.  To see this, consider
the example of a strongly fluctuating network, of `shape' $z=0$,
$y=\sin(1/x)$ in the fiducial background coordinates. It is clear that
if one considers the area of the~$y=0$ surface, then no matter how
small the size of cells, the cell around zero will always be pierced
by an \emph{infinite} number of edges, and hence the area operator
would not be well defined.  Because classical space(-time?) presumably
emerges in the limit of infinite spin networks, the exclusion of such
`ill-behaved' networks really amounts to a prescription for the order
in which limits are to be taken, when they would otherwise produce an
ambiguous answer.

Secondly, the area operator is unbounded, and not every state in $\Hk$
possesses finite area (whereas every element of $\S$ has finite
area). This can happen even for compact spatial manifold $\Sigma$, and
the wave function can then be thought of as associated to an
`infinitely crumpled surface' in $\Sigma$.  To see this more
explicitly, consider a sequence of spin network states $\Psi_n$ with
edges labelled by $p=1,\dots ,L(\Gamma_n)$, and spins $j_p^{(n)}$;
then we have
\begin{equation} 
\label{Area-net1}
\hat{A}_S \left( \sum_{n=1}^\infty a_n  \Psi_n  \right)= 
8 \pi l_p^2\hbar \gamma \sum_{n=1}^\infty \left( \sum_{p=1}^{L(\Gamma_n)} 
\sqrt{j_p^{(n)}(j_p^{(n)} + 1)}\, 
 a_n  \Psi_n \right) \, .
\end{equation}
The expectation value of $\hat{A}_S$ can be made to diverge by letting 
the spins $\{ j_p^{(n)}\}$ grow sufficiently fast with $n$, even if 
$\sum_n |a_n|^2 ||\Psi_n||^2 < \infty$. 

Thirdly, the formula \eqn{Area-net} plays an important role in the
microscopic explanation of the Bekenstein-Hawking entropy of black
holes~\cite{Bekenstein:1973ur,hawk2}, which is now reduced to a
counting problem. As an important prerequisite for this calculation,
one must first \emph{define} the notion of (isolated) horizon in the
quantised theory, {\it i.e.\/} introduce a condition on the quantum
states that singles out the `horizon states'. This has been
accomplished in~\cite{Ashtekar:1997yu}.  Fitting the solution of this
problem to the known factor relating entropy and area of the horizon
then yields a prediction for the value of the parameter
$\gamma$. While the original value given was $\gamma =\frac{\ln
2}{\pi\sqrt{3}}\approx 0.12738402\ldots$, a more recent calculation gives
$\gamma = 0.23753295\ldots$~\cite{Domagala:2004jt,Meissner:2004ju} (see
also~\cite{Ghosh:2004wq,Khriplovich:2004kx}).

\subsubsection{The volume operator}
\label{s:volume}

\begin{figure}[t]
\begin{center}
\includegraphics*[width=.3\textwidth]{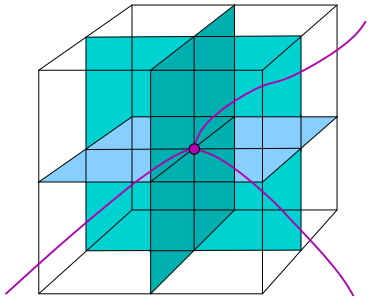}
\caption{An elementary cell in the Riemann sum~\eqn{e:vol-sum}, together
with the three surfaces on which the fluxes are evaluated.}
\label{f:cells}
\end{center}
\end{figure}

The construction of the volume operator is less `clean cut' than that
of the area operator, because it is fraught with ambiguities, which
can only be eliminated by making certain choices and by averaging,
see~\cite{Ashtekar:1997fb} for a comprehensive discussion. For a different
derivation using a more conventional point-splitting method, which yields 
the same final result, see~\cite{Thiemann:1996au}. In a first approximation 
one might try to use the functional differential operator \eqn{g}, but 
its direct application would lead to (square roots of) singular factors
$\delta^{(3)}(\x,\x)$. The `detour' in the construction to be
described, in particular the smearing over the surfaces $S^a_I$ in
\eqn{q} below, is necessary mainly in order to eliminate these
singular factors.  One starts with the classical expression for the
volume of a three-dimensional region $\Omega\subset\Sigma$,
\begin{equation} 
\label{e:vol}
V(\Omega) = \int_\Omega\! {\rm d}^3x \sqrt{\left|\frac{1}{3!} \epsilon_{abc} 
\epsilon^{mnp}\tilde{E}_m^a \tilde{E}_n^b \tilde{E}_p^c\right|}\,.
\end{equation}
Just as with the area operator, one partitions~$\Omega$ into small
cells $\Omega = \cup_I \Omega_I$, so that the integral can be replaced
with a Riemann sum. To express the volume operator in terms of
canonical quantities, we rewrite the volume element $\d V = \theta^1
\wedge \theta^2 \wedge \theta^3$ in terms of the area elements $\d F^1
= \theta^2\wedge \theta^3$,\ldots~(see~\eqn{dFa}) as
\begin{equation}\label{vol}
\d V= \sqrt{ \left| \d F^1\cdot \d F^2\cdot \d F^3 \right|}\,.
\end{equation}
In order to express this volume element in terms of the canonical
quantities introduced before, we next approximate the area elements 
${\rm d}F^a$ by the small but finite area operators $F^a_S[\tilde{E}]$, 
such that the volume is obtained as the limit 
of a Riemann sum
\begin{equation}
\label{e:vol-sum0}
V(\Omega) = \lim_{N \rightarrow \infty} \sum_{I=1}^N 
\sqrt{  \left| \frac1{3!} \epsilon_{abc}
F^a_{S^1_I}[\tilde{E}]\, F^b_{S^2_I}[\tilde{E}]\,
F^c_{S^3_I}[\tilde{E}]  \right|   } \,.
\end{equation}
Now, in order to make sense of this expression, we must for each cell
$\Omega_I$ choose three small non-coincident surfaces $S^a_I$, which 
subdivide $\Omega_I$, as shown in figure~\ref{f:cells}. This should
be done in such a way that the r.h.s. of \eqn{e:vol-sum0} reproduces the 
correct classical value. For instance, one can choose a point inside
each cube $\Omega_I$, then connect these points by lines and `fill in'
the faces. In each cell $\Omega_I$ we then have three lines labelled 
by $a=1,2,3$; the surface $S^a_I$ is then the one that is traversed 
by the $a$-th line. With this choice it can be shown that the result
is insensitive to small `wigglings' of the surfaces, hence independent of 
the shape of $S^a_I$, and the above expression converges to the desired
result~\footnote{Alternatively, one can partition $\Sigma$ with other 
types of cells, with analogous results.}.

Having constructed a sequence of refinements with the correct
classical limit, we can now attempt to promote \eqn{e:vol-sum0} to a
quantum operator by invoking the known action of each
factor~$\hat{F}^a_{S^a_I}$ on a spin network wave function, viz.
\begin{equation}
\label{e:vol-sum}
\hat{V}(\Omega) 
= \lim_{N\rightarrow\infty} \sum_{I=1}^N \hat{V}(\Omega_I) 
= \lim_{N \rightarrow \infty} \sum_{I=1}^N  
\sqrt{\big| \hat{q}_I^{(N)}(S^1_I,S^2_I,S^3_I)\big|}\, ,
\end{equation}
with the operator (= square of the local volume)
\begin{equation}\label{q}
\hat{q}_I^{{(N)}}(S^1_I,S^2_I,S^3_I) := \frac1{3!} \epsilon_{abc}
\hat{F}^a_{S^1_I}[\tilde{E}]\, \hat{F}^b_{S^2_I}[\tilde{E}]\, 
\hat{F}^c_{S^3_I}[\tilde{E}] \, .
\end{equation}
Ignoring the square root for the moment, \eqn{q} is indeed well defined 
as an operator at this point. However, unlike the classical volume,
the action of $\hat{q}_I^{{(N)}}$ on the spin network wave function 
depends on the way we partition $\Omega$, and more specifically on
the position of the faces $S^a_I$ in relation to the edges and vertices. 
As a simple example of the inherent ambiguities consider a cube which 
is traversed by one edge $e$, but not containing any vertex of the 
network. If the surfaces $S^a_I$ are positioned in such a way that 
the edge intersects each one of them with $\N(e,S^a_I) \neq 0$ at three 
different points in $\Omega_I$, the operator simply inserts the
matrices $\tau^1,\tau^2$ and $\tau^3$ at the appropriate places in 
$h_e[A]$. If the cell is sufficiently small, we can bring these 
matrices together, using $\tau^1 \tau^2 \tau^3 =i$, and thus pick up 
a contribution of order one. It is easy to construct sequences 
of refinements such that an arbitrarily large number of cells $\Omega_I$ 
gives such a contribution, no matter how small the cells are. In order 
to prevent catastrophic divergent sums from appearing on the r.h.s. 
of \eqn{e:vol-sum} one must therefore rule out such configurations 
`by hand' and stipulate that each cell $\Omega_I$ may contain at most 
two such intersections with any edge, in which case one of the three 
factors in \eqn{q} will give zero.

\begin{figure}[t]
\begin{center}
\includegraphics*[height=4cm]{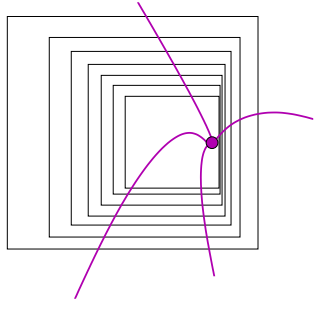}
\end{center}
\caption{A series of refining steps for one cell in a Riemann sum,
  such that the indicated node is never at the centre of any of the
  boxes except in the strict $N\rightarrow\infty$ limit. A volume
  operator based on cells of this kind would always yield a vanishing
  result for any finite cell size.}
\label{f:asymptotic}
\end{figure}

\begin{figure}[t]
\psfrag{OR }{or}
\begin{center}
\includegraphics*[height=4cm]{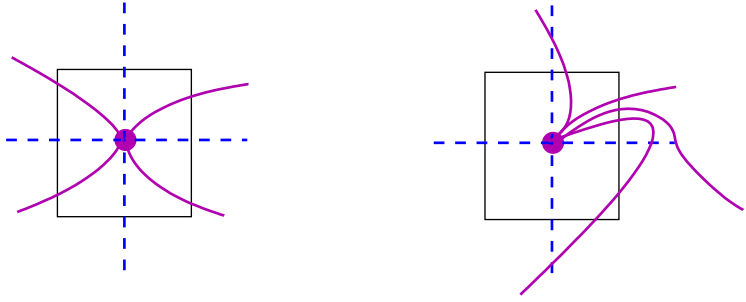}
\end{center}
\caption{Two different choices for the relative orientation of the
  edges emanating from a four-valent node and the cell used in the
  definition of the volume operator (prior to averaging over the
  positions of the surfaces~$S_I$). It is always possible to choose
  the division of the cell such that all four edges emanate from the
  same octant. We have here drawn the `active' version in which the
  edges rather than the surfaces are deformed. Note that these two
  configurations are not homotopically related.}
\label{f:ambiguity2}
\end{figure}

With this proviso, the sum receives contributions only from those 
cells $\Omega_I$ which contain a vertex $v$. If the vertex $v$ does 
not coincide with the intersection point of the three faces $S_I^a$,
further refinement will yield more cells, which do not contribute by
the above requirement. Hence, with sufficient refinement only those cells
will remain and contribute for which the vertex $v$ sits precisely
at the intersection: $v = S^1_I \cap S^2_I \cap S^3_I$. This means 
that we must also exclude refinements such as in figure \ref{f:asymptotic}
for which $S^1_I \cap S^2_I \cap S^3_I$ `misses' $v$ by an amount that 
gets smaller with $N$, but never vanishes for $N < \infty$. 
With these assumptions, the resulting action of the
operator~$\hat{q}_I$ on the vertex $v$ reduces to
\begin{equation}
\label{Js}
\hat{q}_I  = (8 \pi l_P^2 \gamma)^3 \frac{1}{48} \sum_{e_1, e_2, e_3}
 \epsilon(e_1,e_2,e_3) \epsilon_{abc} \hat{J}_a^{(e_1)} \hat{J}_b^{(e_2)} 
\hat{J}_c^{(e_3)} \, ,
\end{equation}
where the sum runs over all possible triplets of edges in the vertex
$v$, and the angular momentum operator $\hat{J}^{(e,v)}_a$ (defined
in~\eqn{F=J}) by definition inserts a matrix $\rho_{j_e} (\tau^a)$ at
the vertex $v$ into the spin network wave function. The coefficients
$\epsilon(e_1,e_2,e_3)$ are orientation factors which are equal to
$\pm 1$. They depend on how the faces $S^a_I$ cut across the vertex,
and on which edges are `above' and `below' $S^a_I$. More precisely, in
terms of the factors $\sigma(e,S)$ introduced in~\eqn{F=J} they are
given by
\begin{equation}\label{eps}
\epsilon(e_1,e_2,e_3) = 
   \sigma (e_1, S^1_I)\, \sigma (e_2, S^2_I)\, \sigma (e_3, S^3_I) 
\end{equation}
and thus completely fixed by the relative positions of the edges
and the surfaces. Let us also stress they are independent of any 
fiducial metric background structure, and purely combinatorial --
hence invariant under diffeomorphisms. \eqn{eps} is not yet the final 
answer, but let us first make the above formulas a little more concrete 
by working out the action of the above operator on a 4-valent 
vertex~\eqn{e:fourvalent}. Its net effect is the replacement of the 
intertwiner according to
\begin{multline}
C_{\alpha_1 \ldots \alpha_4}^{j_1 \ldots j_4} \rightarrow
\text{const}.\!\!\!\!\!\!\!\sum_{\text{perm.}\{1,2,3,4\}}\;
\epsilon(e_1,e_2,e_3) \; \epsilon_{abc}\\[-.5ex]
\times \Big(\rho_{j_{e_1}}(\tau^a)\Big)_{\alpha_1\beta_1} \,\, 
\Big(\rho_{j_{e_2}}(\tau^b)\Big)_{\alpha_2\beta_2} \, \,
\Big(\rho_{j_{e_3}}(\tau^c)\Big)_{\alpha_3 \beta_3} \,\, 
\delta_{\alpha_4 \beta_4}
C^{j_{e_1}\ldots j_{e_4}}_{\beta_1\ldots\beta_4} \, ,
\end{multline}
where the sum runs over all choices of three edges out of four. From this
formula it is obvious that in general the operator $\hat{q}_I$ acts 
\emph{non-diagonally} on a spin network wave function $|\Gamma,\{j\},\{C\}\rangle$, 
because it changes the intertwiners~$\{C\}$, although it does not affect 
the spin network itself ({\it i.e.\/}~the graph $\Gamma$). 
The operator vanishes
when all the edges enter $v$ through the same octant, in which case 
$\epsilon = +1$ for all choices, and
\begin{equation}
\epsilon_{abc} \big( \hat{J}_1^a \hat{J}_2^b \hat{J}_3^c +  \hat{J}_1^a \hat{J}_2^b \hat{J}_4^c + 
 \hat{J}_1^a \hat{J}_3^b \hat{J}_4^c  +  \hat{J}_2^a \hat{J}_3^b \hat{J}_4^c\big) = 0
\end{equation}
by angular momentum conservation, $\hat{J}_1^a + \hat{J}_2^a + \hat{J}_3^a + \hat{J}_4^a =0$.
Note that we can in principle arrange any 4-valent vertex to conform 
with this choice (see figure~\ref{f:ambiguity2}). Similar results hold
true when yet more edges enter the vertex from one octant.

\begin{figure}[t]
\psfrag{OR }{or}
\begin{center}
\includegraphics*[height=4cm]{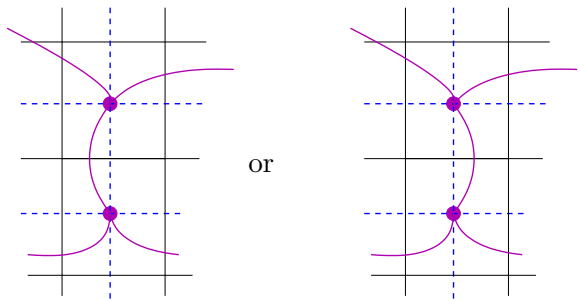}
\end{center}
\caption{Sign ambiguities in the volume operator, present because the
  relative location of the three surfaces~$S_I$ and the edges of the
  spin network can be chosen in different ways (again prior to
  averaging, as in figure~\ref{f:ambiguity2}).}
\label{f:ambiguity1}
\end{figure}

Let us now return to the coefficients $\epsilon(e_1,e_2,e_3)$ in
\eqn{eps}.  Though manifestly background independent they do depend on
how we choose to position the faces and edges: for instance, there is
no \emph{a priori} argument that would force a given edge linking two
vertices in adjacent cells to appear `to the left' or `to the right'
of the face connecting the two vertices, as illustrated in figure
\ref{f:ambiguity1}, and in fact almost any combination of factors $\pm
1$ is possible. To eliminate this remaining ambiguity in the
definition of the volume operator, one now averages over the various
possibilities~\footnote{We might note that this averaging is
`undemocratic' insofar as it is applied only to cells containing a
vertex. If it were extended to other cells, some of the unwanted
features which we just disposed of might re-appear. Vertex-less cells
which satisfy the condition that edges intersect at most two surfaces
do not necessarily satisfy this condition for all surfaces
orientations which appear in the averaging procedure.}.  More
specifically, this is done by integrating over the positions of the
surfaces with some measure $\d \mu (\theta_1,\theta_2,\theta_3)$,
where the $\theta_i$ are suitable angular coordinates, viz.
\begin{equation}\label{eps1}
\hat{\epsilon} (e_1,e_2,e_3) := \int \d \mu (\theta_1,\theta_2,\theta_3)\;
 \sigma (e_1, S^1_I(\theta_1))\, \sigma (e_2, S^2_I(\theta_2)) \,
 \sigma (e_3, S^3_I(\theta_3)) \,.
\end{equation}
This leads to a well defined combinatorial answer depending on the 
relative orientation of the three edges $(e_1,e_2,e_3)$ (= the sign
of the determinant of the dreibein spanned by the tangents to the
three edges at the vertex), but only modulo an overall factor which
encapsulates the freedom in the choice of the measure $\d\mu$. In 
principle such a factor can arise for each vertex, but in order to 
preserve diffeomorphism invariance of the volume these factors should be 
chosen to be \emph{the same} for all vertices. This factor is the one 
ambiguity which remains in the definition of the volume operator, and 
this ambiguity cannot be eliminated~\cite{Ashtekar:1997fb}.

Finally, given an assignment of $\hat\epsilon(e_1,e_2,e_3)$ for each 
vertex, how do we actually compute the spectrum of the volume operator 
in \eqn{e:vol-sum}? First of all, it is easy to see that this operator 
always vanishes on three-valent vertices which satisfy the Gauss 
constraint. This is a direct consequence of angular momentum conservation, 
or equivalently
\begin{equation}
\epsilon_{abc} \Big(\rho_{j_{e_1}}(\tau^a)\Big)_{\alpha_1\beta_1} 
               \Big(\rho_{j_{e_2}}(\tau^b)\Big)_{\alpha_2\beta_2} 
               \Big(\rho_{j_{e_3}}(\tau^c)\Big)_{\alpha_3\beta_3} 
\, C^{j_{e_1}\, j_{e_2}\, j_{e_3}}_{\beta_1\beta_2\beta_3} = 0\,.
\end{equation}
In other words, the volume operator vanishes on gauge invariant two-
and three-valent vertices (but \emph{not} on gauge variant ones), and
picks up contributions only from spin networks with vertices of
valence~$\geq 4$. Given any spin network wave function, the first step
is to work out the action of~$\hat{q}_I$ according to the above rules.
In a second step this action must be diagonalised for each $I$ on the
spin network wave functions. This can be done because the operators
$\hat{q}_I$ commute amongst each other, {\it i.e.\/}~$[\hat{q}_I ,
\hat{q}_{I'} ] = 0$ for $I\neq I'$. Only now can we define the
absolute value $|\hat{q}_I|$ and take the square root
$\sqrt{|\hat{q}_I|}$ using the spectral decomposition of this operator
(note that $\hat{q}_I$ is not necessarily a positive operator).
Unfortunately, these steps are all but straightforward to carry out in
practice~\cite{DePietri:1996pj}. The analysis can be simplified by
applying the operator identity~\cite{Thiemann:1996au,Brunnemann:2004xi}
\begin{equation}
\epsilon_{abc} J^1_a J^2_b J^3_c = 
\frac{i}4  \big[({\bf J}^1 + {\bf J}^2)^2 , ({\bf J}^2 + {\bf J}^3)^2 \big]
\end{equation}
to any three edges entering the vertex, enabling a numerical analysis
of simple configurations with low-valence nodes. This procedure has
recently been used to show that the operator corresponding to the
inverse volume is not bounded from above when acting on zero-volume
spin network states~\cite{Brunnemann:2005in} (in marked contrast to
the symmetry reduced models of~\cite{Bojowald:2001xe}, where the
avoidance of an initial singularity hinges on the fact that the
operator corresponding to the `inverse volume' is bounded from
above\footnote{The operator $\hat{X}$ corresponding to the inverse
volume is \emph{not} the same as the inverse of the volume
operator. The latter does not exist on all of $\Hk$ because of the
occurrence of zero eigenvalues in its (discrete) spectrum. Instead,
the operator $\hat{X}$ obeys $\hat{X}\circ \hat{V} = {\bf 1} +
\mathcal{O}(\hbar)$, and can therefore in principle stay bounded `near the
singularity'.}).  Nevertheless, the resulting formulas remain very
complicated. The lack of a more explicit representation for the
operators~$\sqrt{\hat{q}_I}$ is also a serious impediment towards a
better understanding of the Hamiltonian constraint.

\subsection{Some comments on regularisation dependence}
\label{s:regularisation}

At this point some comments regarding the regularisation procedures
for kinematical operators which we have outlined above, are in order.
Similar comments will apply to the regularisation of the Hamiltonian
constraint operator. First of all it should be evident from the 
foregoing discussion that the notions of `finiteness' and `regulator 
independence' as currently used in LQG are not the same as in conventional 
quantum field theory. There one insists that renormalised quantities 
should come out the same (modulo a finite number of normalisations) 
after removal of the regulator, \emph{no matter how the latter is chosen}. 
By contrast, LQG does not claim that different prescriptions for 
constructing an operator will always lead to the same physical result.
This is well illustrated by the volume operator, whose definition
and whose very existence hinge on various choices that have been 
made along the way.

Let us nevertheless briefly summarise what is meant here by
`regularisation'.  In order to regulate a given classical phase space
quantity defined by an integral (such as the volume), one first
approximates that integral by a Riemann sum (which is of course
assumed in the classical theory to converge to the desired quantity in
the limit of infinite refinement). One then replaces in this finite
Riemann sum all terms by operators, usually in such a way that the
various area or volume elements are absorbed into the definition of
these operators (as exemplified by the `small' operators
$\hat{F}^a_{S^a_I}[\tilde{E}]$ entering the definition of the area and
volume operators). Acting with a sequence of these discretised
operators on an arbitrary, but fixed spin network state, the resulting
sequence of refinements (labelled by $N$, say) eventually becomes
stationary ({\it i.e.\/}~constant for $N$ sufficiently large), because
the state is kept fixed as the limit $N\rightarrow\infty$ is
taken. This follows straightforwardly for the area operator, but
requires a choice of `preferred' regularisations for the volume
operator. With these choices, the `correct' result is always obtained
already after a finite number of refinements. A good analogy would be
the Riemann integral of a step function, for which a suitably refined
Riemann sum attains the exact value already after finitely many steps,
after which the result no longer depends on the `regulator' ({\it
i.e.\/}~the size of the intervals defining the Riemann sum). In this
sense, the results of LQG are indeed `regulator independent'.

From what has just been said, it is evident that infinities can 
never appear in the LQG regularisation procedure, and in this sense the 
resulting theory is `finite', at least as far as the kinematical 
operators are concerned. LQG nevertheless requires the regularisation 
of the area and volume operators in order even to be able to define the 
quantum counterparts of the classical constraints via Thiemann's trick. 
The outlined regularisation is, therefore, not introduced to remove 
divergences. Standard short distance, QFT divergences `disappear' in 
the LQG approach by the very construction of the theory: all states 
are discrete, and at any step of the calculation one deals only with a
finite number of objects. The price one pays are ambiguities of the type 
encountered above, some of which can only be eliminated by making 
\emph{ad hoc} choices, such as the rule prescribing the way in which
an edge has to traverse a cell. We shall encounter more such ambiguities
when we attempt to define the Hamiltonian constraint operator. 

Of course, these considerations do not necessarily imply any
inconsistency. Namely, one can always adopt the above prescriptions
for the kinematic operators as \emph{definitions}, which can be
plausibly related to the corresponding classical objects, and in this
sense the definitions of the above operators in LQG are perfectly
acceptable. What we would like to emphasise, however, is that with
this construction the work is hardly begun. The far more difficult
task is now to show that these definitions, when put together and
inserted in the constraints, yield a sensible quantum theory, and a
consistent quantum constraint algebra, in particular. The situation
here is reminiscent of the one in the early days of constructive
quantum field theory (see e.g.~\cite{b_glim1} and references
therein). The problem then was not so much finding non-perturbative
regulators and establishing the existence of selected limits when the
regulator is removed. Rather, the main problem was to establish that
the resulting theory has all the desired physical properties, to wit,
full Poincar\'e invariance, locality and causality. As is well known,
no interacting quantum field theory in four dimensions satisfying all
these requirements is known to this day~\footnote{Even the best
candidate model for realising the quantum field theorist's dream of a
UV finite theory, $N=4$ supersymmetric Yang-Mills theory, has not been
rigorously shown to exist beyond perturbation theory.}. And for those
low dimensional models which could actually be shown to exist and
possess all the requisite properties, the tightly knit construction
left no room for ambiguities other than the normalisation of a finite
number of physical parameters (masses and coupling constants).

\subsection{Coupling to matter fields}
\label{s:matter_couplings}

Any realistic theory of quantum gravity should at least allow for a
coupling to the matter fields of the Standard Model of elementary
particles. In the framework of geometrodynamics it is fairly easy 
to extend the relevant formulas to incorporate matter fields (but, of 
course, the difficulties with quantisation remain the same as before).
Likewise, it is possible to include gauge fields and fermions in the 
LQG formalism, essentially without any restriction on the gauge group 
or the number and structure of fermion families --- as we already pointed 
out LQG does not appear to put any restrictions on the matter couplings 
(such as renormalisability). Armed with the translation rules of 
section~\ref{s:ashtekar_variables}, the transition from the `old' metric 
or dreibein variables to the loop variables is in principle straightforward, 
though usually cumbersome (and, we may add, of considerably diminished 
esthetical appeal in comparison with Ashtekar's original reformulation 
of \emph{pure} gravity). 
Because there is an ample literature on this subject (see
e.g.~\cite{Baez:1997bw,Thiemann:1997rt,Varadarajan:1999it,Varadarajan:2001nm,Ashtekar:2001xp,Ashtekar:2002vh}), and because the introduction of
matter couplings does not alleviate the problems with the Hamiltonian 
constraint in LQG, we will be very brief and only mention two selected 
aspects here.

Consider, for example, the coupling of fermion fields in the LQG 
setting~\cite{Baez:1997bw,Thiemann:1997rq}. The starting point is the 
classical continuum action of a Dirac fermion coupled to gravity
\begin{equation}
\label{e:fermions_continuum}
S = \int\!{\rm d^4}x\, E\,\bar\chi \gamma^A E_A{}^\mu {\cal D}_{\mu} \chi   
\equiv \int\!{\rm d^4}x\, E\,\bar\chi \gamma^A E_A{}^\mu \big( \partial_\mu +
\omega_{\mu BC} \gamma^{BC} \big)\chi\, ,
\end{equation}
where we have omitted a possible mass term for simplicity, as its inclusion 
would pose no new conceptual problems. There are two immediate questions 
one faces when trying to construct quantum Hamiltonian for this system.
One is the choice of the conjugate pair of variables for fermions; finding
the proper fermionic variables may also require redefinitions of the other 
fields so as to `diagonalise' the canonical (Poisson or Dirac) brackets. 
For instance, for the above model it turns out~\cite{Thiemann:1997rt} that 
a good choice for the conjugate pair of fermionic variables is 
the pair of \emph{half-densitised} spinors
\begin{equation} 
\label{e:fermi-canon-var}
\tilde{\chi} \equiv \sqrt[4]{\det \tilde{E}}\, \psi \, , \quad \pi = i
{\tilde{\chi}}^\dagger \gamma^0 \, .
\end{equation}
If instead of densitised spinors we used bare spinors (without any factors 
of the determinant of the dreibein), one would have to modify the
gravitational connection~\cite{Thiemann:1997rt} in order to obtain 
diagonal canonical brackets, but at the expense of making the connection
complex again
\begin{equation} 
\label{ccon}
\tilde{A}_m^a = -\frac{1}{2}\epsilon^{abc} \omega_{m\,bc} +
\gamma K_m^a + \frac{i}{4\sqrt{\tilde{E}}} e_m^a \bar{\chi} \chi \, .
\end{equation}
This leads to a new reality problem, similar to the one already 
encountered in \cite{Matschull:1993hy}, and not different from the one
which eventually led LQG to abandon the `old' Ashtekar variables with 
$\gamma=\pm i$. The second question is how to express all variables in 
terms of the LQG variables $h_e[A]$ and $F^a_S[\tilde{E}]$. The answer
to this question can be obtained by employing the identities~\eqn{e:trick1} 
and~\eqn{e:trick2}. A similar procedure will be spelled out in the next 
section in order to construct the Hamiltonian constraint, so we will 
refrain here from giving more details.

When implementing the matter Hamiltonians in terms of spin networks,
the description of (abelian or non-abelian) gauge fields and matter
fields is somewhat similar to the description of standard lattice
gauge theories: matter fields (scalars or fermions) are attached to
the vertices of the spin network, while gauge fields are associated to
the edges (links), see
e.g.~\cite{Varadarajan:1999it,Varadarajan:2001nm,Ashtekar:2001xp} for
a treatment of electromagnetism and~\cite{Ashtekar:2002vh} for scalar
field theories in the LQG formalism. There are, however, two main
differences between lattice gauge theory and LQG. In the former, the
space or spacetime lattice is given, and hence there are only three
(or four) possible directions for the edges. The coupling between two
fermions $\chi_1$ and $\chi_2$ on adjacent vertices 1 and 2 connected
by an edge $e\equiv (12)$, is uniquely given by $\bar\chi_1
\gamma_{12} h_{12}[A] \chi_2$, where $\gamma_{12}$ is the
$\gamma$-matrix `along' the edge (12). In LQG (and in general
relativity), this simple formula fails because the direction of the
link is arbitrary, and a vielbein factor must be inserted to `align'
the $\gamma$-matrix with the holonomy. When rewriting the dreibein by
means of the formulas of section~\ref{s:ashtekar_variables}, this
usually leads to rather awkward expressions in terms of the spin
network variables $h_e[A]$ and $F^a_S[\tilde{E}]$. This problem is
further compounded by a second difficulty which goes back to the issue
of what the correct value is for the Barbero-Immirzi parameter. As we
mentioned already, for $\gamma\neq \pm i$, the connection $A$ is not
the pullback of the (chiral) spin connection. Therefore, the holonomy
$h_{12}[A]$ does not effect the correct parallel transport between
$\chi_1$ and $\chi_2$, and must accordingly be amended by further (and
again rather awkward) modifications.

We should stress that, \emph{even} when one has obtained an expression
for the matter action in terms of spin network operators, one does not
know whether the resulting quantum theory has any relation to the
usual Fock-space quantised theory. Showing that such a relation exist
requires the field-theory analogues of the `shadow states' for point
particles~\cite{Ashtekar:2002sn}. These states are complicated linear
combinations of elementary spin network states, approximating Fock
space coherent states. In flat space some results have been obtained
in the case of
Maxwell~\cite{Varadarajan:1999it,Varadarajan:2001nm,Ashtekar:2001xp}
and scalar fields~\cite{Ashtekar:2002vh}. Recent attempts to construct
coherent states in the presence of gravity
include~\cite{Thiemann:2000bw,Thiemann:2000ca,Thiemann:2000bx,Thiemann:2000by,Thiemann:2002vj,Sahlmann:2001nv,Sahlmann:2002qj,Sahlmann:2002qk}.

A completely different question concerns the role of the structure
group SU(2), and the possibility of generalising Ashtekar's variables
to higher dimensions, to achieve a Kaluza-Klein type unification of
matter and gravity. The closure of the classical constraint algebra in
the Ashtekar reformulation of gravity requires use of the identity
$\epsilon_{abe} \epsilon^{cde} = 2 \delta^{cd}_{ab}$, which is only
valid for the structure constants of the group SU(2) and its
non-compact form SO(1,2).  These properties no longer hold for SU($N$)
when $N>2$. For this reason, the LQG reformulation of gravity works
only in three and four spacetime dimensions. Nevertheless, one can try
to generalise the formalism to SU($N$)~\cite{Ashtekar:1995zh} and
thereby arrive at a novel type of coupling between gravity and
Yang-Mills fields~\cite{Peldan:1992iw,Chakraborty:1994vx}. A very
different attempt to generalise the formalism to include matter
couplings is based on the observation that an inverse densitised
vielbein also appears naturally in certain reformulations of $D=11$
supergravity~\cite{melo1}. It is not known whether any of these
\emph{ans\"atze} can be implemented in a spin network formulation
(although replacing SU(2) holonomies by SU($N$) holonomies seems
straightforward enough).

\section{Quantum constraints}

As we have seen in section~\ref{s:prelude}, the Hamiltonian framework
leads to three classical constraints: the Gauss, diffeomorphism and
Hamiltonian constraints. To implement these constraints at the quantum
level, one must first properly \emph{define} them, {\it
i.e.\/}~express them in terms of the elementary variables, the
holonomies and fluxes, and then investigate their properties. We will
first discuss the kinematical constraints, which are `easy'
(relatively speaking), whence we can be brief. We will then turn to
the truly difficult part of the story -- the Hamiltonian constraint.

\subsection{Kinematical constraints}
\label{s:kinematical_constraints}

The Gauss constraint is already expressed in canonical variables
$h_e[A]$ and $F_S[\tilde{E},f]$, and one can solve it exactly. This is
simply because fulfilment of the Gauss constraint is equivalent to the
gauge invariance of the spin network wave function: invariance is
simply achieved by contracting the SU(2) representation indices
entering a given vertex in an SU(2) invariant manner. Equivalently,
one says that \emph{angular momentum is conserved at each vertex},
viz.
\begin{equation}
\sum_e \hat{J}_{(e,v)}^a = 0
\end{equation}
for each $v\in\Gamma$. Recall (see the discussion around~\eqn{F=J})
that the operator $\hat{J}^{(e,v)}_a$ is \emph{defined} to act by inserting
a matrix $\rho_{j_e} (\tau^a)$ at the vertex $v$ in the representation
belonging to the (oriented) edge $e$, with appropriately contracted
SU(2) representation indices, and appropriate orientation.  For the
following discussion, we will tacitly assume that the wave function
satisfies the Gauss constraint, whenever appropriate.

The diffeomorphism constraint is more difficult to impose --- unlike
in geometrodynamics, one cannot immediately write down formal
expressions which are manifestly diffeomorphism invariant, because the
spin network functions are not supported on all of $\Sigma$, but only
on one-dimensional links. Nevertheless, it would seem most natural at
first sight to try to solve the diffeomorphism constraint on
the kinematical Hilbert space $\Hk$. However, this is impossible for
two reasons. First of all, the diffeomorphism generator does not even
exist as an operator: there is no way to differentiate finite
diffeomorphisms so as to obtain `infinitesimal' ones due to the lack
of weak continuity (and \emph{a fortiori},
differentiability). Secondly, with the exception of the trivial state
$\Psi={\bf 1}$ (the empty spin network), there is just \emph{no}
diffeomorphism invariant state in $\Hk$. This is intuitively clear
because every spin network state changes under a diffeomorphism which
moves the associated graph around.\footnote{This problem disappears
altogether in a \emph{topological theory}, such as
\mbox{(2+1)-dimensional} gravity, where the connection is flat and
diffeomorphism invariance is manifest from the outset.} For these
reasons, LQG invokes a variant of the so-called `group averaging'
method, but in a way that is different from the standard group
averaging procedure. Namely, application of the latter would require
knowing an integration measure ${\rm d}\mu(\phi)$ on the infinite
dimensional group Diff$(\Sigma)$. The formally diffeomorphism
invariant state would then be
\begin{equation}
\label{e:diffintegral}
\Psi^{\text{inv}}_{\Gamma,\psi}[A] \stackrel{?}{=} \int_{\text{Diff}(\Sigma)}
{\rm d}\mu(\phi)\, \Psi_{\Gamma,\psi}[A\circ\phi]\, ,
\end{equation}
and if the measure existed, this state would legitimately belong to
$\Hk$.  For the infinite dimensional group Diff$(\Sigma)$ no such
measure is known (the `counting measure' \eqn{diffaverage} introduced
below does not `live' on $\Hk$, but leads out of the kinematical
Hilbert space), hence the above expression is devoid of meaning ---
confirming the conclusion above that $\Hk$ does not admit non-trivial
solutions. LQG circumvents this difficulty by implementing `group
averaging' in a way that is more adapted to the non-separable Hilbert
space and the properties of the scalar product~\eqn{e:scalar_product}.
The key observation here is that the constraint can be formally solved
on a much larger space, namely \emph{the algebraic dual $\D^*$ of some
dense subspace} $\D$ of $\Hk$, such that
\begin{equation}
\D\subset \Hk \subset \D^*\,.
\end{equation}
Typically, the dual space $\D^*$ is thus a space of distributions. 
This is not as outlandish as it may seem at first, because for 
constrained systems, the solutions of the constraints are
generically distributions, see e.g.~\cite{Hajicek:1990eu}~\footnote{A
simple example is the solution of the Hamiltonian constraint for a
relativistic point particle -- the Klein Gordon equation -- which is
$\tilde\phi(p) = \delta(p^2 -m^2)$. As is well known, a scalar product and a
Hilbert space can be defined on this constraint hypersurface, but the
scalar product is not positive definite. These features can also be
made completely explicit in more complicated exactly solvable
theories, such as pure gravity~\cite{Witten:1988hc,Ashtekar:1989qd} 
and pure supergravity \cite{deWit:1993qn} in three dimensions.}. 
Given any (densely defined and closable)
operator ${\cal O}$ and its adjoint ${\cal O}^\dag$ with domain ${\cal
D}\subset\Hk$, the action of the \emph{dual operator} ${\cal O}^*$ on
the algebraic dual $\D^*$ of $\D$ is defined for any particular
element ${\cal X}\in\D^*$ by requiring
\begin{equation}\label{dual}
\langle {\cal O}^* {\cal X}\, | \, \Psi \rangle =
\langle {\cal X} \, |\, {\cal O}^\dag \Psi \rangle
\end{equation}
to hold for all $\Psi\in\D$. The space $\D^*$ is very large, and in
practice one restricts attention to a physically motivated vector
subspace thereof, the so-called `habitat' ${\cal V}^*\subset\D^*$, 
whose specification is closely related to the definition of the
Hamiltonian constraint. Observe also that the above definition 
does not require the operator ${\cal O}$ to be hermitean.
                                       
The obvious (`canonical') choice for $\D$ is
\begin{equation}
\D=\S
\end{equation}
that is, the space of finite linear combinations of spin network states.
However, there may exist other viable choices for~$\D$.
The dual space $\S^*$ is much larger than~$\Hk$, and actually too large 
for our purposes. What is important here is that~$\S^*$ contains a
certain subspace of diffeomorphism invariant states, and its completion 
w.r.t.~to the norm~\eqn{Norm2} below. This subspace is contained in
any of the proposed `habitats'~$\V^*$, and is obtained by averaging 
spin network states. More specifically, for any~$\Psi_\Gamma\in\S$ 
associated with a particular graph~$\Gamma$ we define the `group average' 
by the formal sum
\begin{equation}\label{diffaverage}
\eta(\Psi_\Gamma) \equiv \overline{\Psi}_\Gamma := 
\sum_{\phi\in {\rm Diff}(\Sigma |\Gamma)} \phi^*\circ \Psi_\Gamma
\end{equation}
where $(\phi^*\circ \Psi)[A] := \Psi[A\circ \phi]$, and where
Diff$(\Sigma|\Gamma)$ consists of the subgroup of Diff$(\Sigma)$ obtained
by dividing out the diffeomorphisms leaving invariant the graph $\Gamma$
on which $\Psi_\Gamma$ lives. While $\Psi_\Gamma\in\S$ by assumption, 
averaging the state in this way throws it out of $\S$ and also $\Hk$. 
At first sight, this seems to make matters worse: we have replaced the 
ill-defined integral over Diff$(\Sigma)$ by a highly formal continuous 
sum. However, the day is now saved by two facts: $(i)$ unlike the 
integral, the sum no longer requires unavailable detailed knowledge of 
the properties of Diff$(\Sigma)$, and $(ii)$ when applying an averaged 
spin network state $\eta(\Psi_{\Gamma'})$ (considered as a bra-vector) 
to any other spin network state $\Psi_{\Gamma}[A]$, the sum
\begin{equation}
\langle \eta(\Psi_{\Gamma'})\, |\, \Psi_{\Gamma}\rangle =
\sum_{\phi\in {\rm Diff}(\Sigma|\Gamma')} 
\langle \phi^* \circ \Psi_{\Gamma'}\,|\, \Psi_{\Gamma} \rangle
\end{equation}
consists at most of a \emph{finite number} of terms, and it is this
fact which ensures that $\eta(\Psi_\Gamma)$ is indeed well defined as
an element of the dual space $\S^*$. Namely, there is no contribution
at all if $\Gamma$ and $\Gamma'$ are not diffeomorphic. For $\Gamma$
and $\Gamma'$ diffeomorphic, only that term contributes for which
$\Gamma$ and $\phi^*\circ\Gamma'$ coincide. In the above equation, the
bracket on the r.h.s. is just the scalar product
\eqn{e:scalar_product}, while the bracket on the l.h.s.  denotes the
dual pairing between $\S$ and $\S^*$.

We can now promote the space of averaged diffeomorphism invariant 
spin network states to a pre-Hilbert space by `dividing out the volume 
of the gauge group' ( =  Diff($\Sigma$)). More precisely, we define 
the scalar product between two such states as
\begin{equation}\label{Norm2}
\langle\!\langle \eta(\Psi_{\Gamma',\psi'})\,\big| \, 
\eta(\Psi_{\Gamma,\psi})\rangle\!\rangle :=
\langle \eta(\Psi_{\Gamma',\psi})\,\big| \, \Psi_{\Gamma,\psi}\rangle\,.
\end{equation}
Here the bracket on the r.h.s.~is the dual pairing between $\S$ and
$\S^*$, and the bracket $\langle\!\langle .|. \rangle\!\rangle$ on the
l.h.s. denotes the new scalar product. The new Hilbert space $\Hd$ of
diffeomorphism averaged states is then obtained by completion
w.r.t. this norm, and can be thought of to consist of the averaged
versions of the states \eqn{Norm1} (but let us emphasise again that
it is \emph{not} a subspace of~$\Hk$). Observe that the Hilbert space
$\Hd$ is still a `small' subspace of $\S^*$ because the vast majority
of elements in $\S^*$ does not correspond to diffeomorphism averaged
spin network states, and cannot be obtained as limits of such states
w.r.t. to the above norm.  Moreover, $\Hd$ is `smaller' than the
original kinematical Hilbert space $\Hk$, because distinct elements of
the latter which are related by a diffeomorphism get mapped to the
\emph{same} element of $\S^*$; equivalently, the averaging map has a
very large kernel. There is no way to extend the above norm from $\Hd$ to 
all of $\S^*$, in accordance with the fact that distribution spaces 
cannot be made into Hilbert spaces, but at best into topological vector
spaces; however, there appears to be no consensus as to which of the many 
possible topologies is the right one for $\S^*$. While~$\Hk$ is non-separable,
it is expected that~$\Hd$ \emph{is} separable~\cite{Zapata:1997db,Fairbairn:2004qe}
provided one averages over \emph{homeomorphisms} rather than diffeomorphisms
(if one divides out only by analytic or~$C^\infty$ diffeomorphisms, 
$\Hd$ remains non-separable, see also~\cite{Ashtekar:2004eh} for a recent 
discussion of this point).

Which, then, is the arena, or `habitat', in which quantum gravity
takes place according to LQG, and where one must ultimately consider
the action of the Hamiltonian constraint? The foregoing discussion
shows that there is no easy answer to this question, and apparently
also no complete consensus in the community as to which space is
`right'. The Hilbert space $\Hd$ of diffeomorphism averaged states
certainly appears to be a good starting point for implementing and
solving the Hamiltonian constraint, but it may not be good enough. The
reason is that the action of the Hamiltonian constraint in general
does not preserve $\Hd$; in other words, it maps diffeomorphism
invariant states into ones that are not invariant.  Consequently, when
checking the commutator of two Hamiltonian constraints, one must again
pass to a larger space, which is preserved by the action of the
Hamiltonian, but which is no longer a Hilbert space. An explicit
example are the `vertex-smooth' states introduced
in~\cite{Lewandowski:1997ba}, obeying~\mbox{$\langle{\cal
X}|\Psi_1\rangle = \langle{\cal X}|\Psi_2\rangle$} for
all~\mbox{$\Psi_1, \Psi_2$} which are related by a diffeomorphism that
leaves the vertices of the spin network fixed. Alternatively, it has
been proposed to implement the Hamiltonian constraint on the
kinematical Hilbert space~$\Hk$ via a \wsl~\cite{Thiemann:1996ay}, see
also the following section. At any rate, the challenge here --- as
with any other constrained quantum mechanical system --- is to find a
subspace that is annihilated by \emph{all} the constraints, and to
define a \emph{physical inner product} which yields the `final'
\emph{physical Hilbert space}.

\subsection{Hamiltonian constraint}
\label{s:hamiltonian_constraint}

As we explained, the diffeomorphism constraint is imposed not in terms
of an operator constraint, but through the formal group averaging
method. By contrast, the Hamiltonian constraint is imposed as an
\emph{operator} constraint. Hence, as a zeroth step, one has to
construct this Hamiltonian operator. As we explain in more detail
below, in this derivation one is forced, at a certain stage, to make 
choices concerning the action of the Hamiltonian operator on a generic 
spin network state. Thus the resulting \emph{definition} is indeed
motivated by the classical expression for the Hamiltonian (expressed
in loop variables), but it is plagued by a very large number of
ambiguities~\footnote{We should caution readers that it is a matter of
debate as to what constitutes an `ambiguity', or whether a particular
choice narrowing down the number of possibilities should be viewed as
`natural' or even mandatory in order to obtain sensible expressions (one 
example being the `translation invariant' choice of measure factor in the 
definition of the volume operator). At any rate, LQG practitioners may
well disagree with our terminology here.}. To test the proposed prescription
and compare it with other possibilities, one would need to find states
which are annihilated by the Hamiltonian, and check that in the
semi-classical limit these states produce sensible results. This is,
however, currently out of reach: due to the great complexity of the
proposed Hamiltonian, not a single physically interpretable eigenstate
is known.  While this should not come as a surprise (after all, we
have no reason to expect to be able to solve the theory exactly), it
is remarkable that even the far more modest goal of working out in
complete detail the action of the (regulated) Hamiltonian on a general 
spin network wave function appears difficult to attain.

Even if we can work out the action of the Hamiltonian on any given
element of~$\S$, we are not done. As explained in the previous
subsection, in order to ensure satisfaction of the diffeomorphism
constraint, we must transfer the action of the Hamiltonian to the dual
space $\S^*$, or at least to some physically motivated subspace~$\V^*$
thereof.  Below, we will be mostly dealing with a \emph{regulated
Hamiltonian} $\hat{H}[N,\epsilon]$ (where $N$ is the lapse
function). Here, the regularisation parameter $\epsilon >0$ enters via
a plaquette $P(\epsilon)$, which is attached to a vertex and which
must be shrunk to zero at the end of the calculation. However, for a
given~$\Psi$, the limit~\mbox{$\epsilon\rightarrow 0$}
of~\mbox{$\hat{H}[N,\epsilon]\Psi$} does \emph{not} exist on~$\S$,
because wave functionals supported on the same network, but with an
extra loop~$\partial P(\epsilon)$ attached to one of the vertices, are
orthogonal to one another for different values of~$\epsilon$ by
\eqn{e:scalar_product}. For this reason, one must resort to a weaker
notion of limit by transferring the action of the Hamiltonian to the
dual space. More specifically, we define the
limit~\mbox{$\epsilon\rightarrow 0$} by demanding
\begin{equation}\label{dual1}
\langle \hat{H}^*[N]{\cal X}\, | \, \Psi\rangle = 
\lim_{\epsilon\rightarrow 0} \; \langle {\cal X} \,| \,
\hat{H}[N,\epsilon]\Psi\rangle
\end{equation} 
to hold for all~$\Psi\in\S$ and~${\cal X}\in\V^*$. Here~$\V^*\subset
\S^*$ is the `habitat' already mentioned in the previous subsection, a
vector subspace of~$\S^*$.  The choice of~${\V}^*$ should be
physically motivated.  Natural choices are~$\V^* = \Hd$ or the space
of vertex smooth states~\cite{Lewandowski:1997ba}, but there may be
others. An evident requirement here is, of course, that the choice
of~\mbox{$\V^*\subset \S^*$} should be such that the limit~\eqn{dual1}
exists. Although it is not entirely clear (to us) what precisely the
conditions on~${\cal V}^*$ should be for this limit to exist, and to
what extent they would determine~\mbox{${\hat H}^*[N]\chi \in {\cal V}^*$}
uniquely, it seems obvious that the limit~\eqref{dual1} will
\emph{not} exist if~${\cal V}^*$ is `much larger' than~${\cal
H}_{\text{diff}}$. There is an alternative point of view, which does
not require `habitats': according
to~\cite{Thiemann:1996ay,Thiemann:1996aw}, the choice~$\V^*=\Hd$ in
fact \emph{is} singled out, and formula~\eqn{dual1} should not be
viewed as defining a dual Hamiltonian on $\V^*$, but rather as
defining the limiting Hamiltonian as a \wsl on the \emph{kinematical}
Hilbert space~$\Hk$ w.r.t.~the \wst defined by~$\V^*=\Hd$
\footnote{There are actually infinitely many such operator topologies
w.r.t.~the dual pair~$\langle \S^*, \S\rangle$, depending on the
choice of $\V^*$ (see e.g.~\cite{Edwards}, p.88). The bigger the
subspace $\V^*$, the stronger the topology.}. In that interpretation,
only the r.h.s.~of \eqn{dual1} is meaningful --- but let us repeat
that there is \emph{no} state in~$\Hk$ that could be interpreted as
``$\lim_{\epsilon\rightarrow 0} \hat{H}[N,\epsilon]\Psi$''.\footnote{A
more recent proposal to circumvent these difficulties is the so-called
Master Constraint Programme of~\cite{Thiemann:2003zv}, where the
Hamiltonian constraint is replaced by another constraint (`master
constraint') $\propto \int\!{\rm d}^3x\, ({\tilde E})^{-1/2} \big( H_0
({\bf x}) \big)^2$ obtained by squaring the original Hamiltonian. This
operator can be implemented as a positive self-adjoint operator
directly on~$\Hd$ (after suitable regularisations), so questions about
`habitats' are sidestepped.}

Independently of whether or not one adopts the `habitat' point of
view, an essential ingredient in existing constructions is the
`diffeomorphism covariance' of the Hamiltonian. This means that by
transferring the action of~$\hat{H}$ to the space of diffeomorphism
invariant states, the existence of the above limit can be ensured by
the fact that the habitat~$\V^*$ is sensitive only to the
diffeomorphism invariant characteristics of the regulator. This is
also borne out by the construction of~\cite{Lewandowski:1997ba} which
takes~$\V^*$ to be the (larger) space of `vertex smooth states'. (But
note that from the point of view of the Hamiltonian action, going to
the slightly larger habitat of `vertex-smooth' states does not make
much of a difference; namely, when acting on a spin network state, the
Hamiltonian does not shift nodes around. Hence its action on~$\Hd$ or
on the larger habitat of `vertex-smooth' states is essentially the
same.) In addition to restricting the choice of~$\V^*$, requiring the
limit~\eqn{dual1} to exist eliminates at least some of the ambiguities
inherent in the definition of the Hamiltonian.
  
To sum up, the Hamiltonian $\hat{H}[N]$ can only be defined, via \eqn{dual1},
as a \emph{weak limit}. This is reminiscent of the LSZ formalism of 
ordinary quantum field theory, where one also defines the asymptotic
field operators as weak rather than strong limits. It is remarkable 
that the known constructions make essential use of the diffeomorphism 
covariance, and thus implicit use of the constraints, already in the very
definition of the Hamiltonian. However, for all we can tell, the question 
as to which ${\cal V}^*$ is `best' is not settled.
On a more practical note, and independently of the choice of $\V^*$,
we note that the concrete evaluation of the limit requires simultaneous 
knowledge of the action of the original regulated Hamiltonian 
$\hat{H}[N,\epsilon]$ on \emph{all} elements of $\S$. Furthermore, 
\emph{explicit} expressions for the dual Hamiltonian $\hat{H}^*[N]$ 
that could be used for any practical application of the formalism are 
even more difficult to obtain than for the Hamiltonian acting on spin 
network states, and certainly beyond the scope of our modest efforts 
in this section. Nevertheless, we should stress once more that the 
dual Hamiltonian $\hat{H}^*$ is our true object of interest, and that 
all further considerations (constraint algebra, semi-classical states, 
and so on) should be based on it, rather than on the regulated Hamiltonian 
acting on $\S$. With these caveats in mind, let us now proceed.

To motivate the form of the \emph{quantum Hamiltonian} one starts with
the classical expression, written in loop variables.  Despite the
simplifications brought about by equation~\eqn{e:EEF}, the
Hamiltonian constraint still looks formidable,
\begin{equation}
\label{e:clH}
H[N] = \int_\Sigma\!{\rm d}^3x\, 
N \frac{\tilde{E}_a^m \tilde{E}_b^n}{\sqrt{\det \tilde{E}}} 
\Big( \epsilon^{abc} F_{mnc} - \frac{1}{2} (1+\gamma^2) K_{[m}{}^a
  K_{n]}{}^b \Big)\,.
\end{equation}
In order to write the constraint in terms of only holonomies and
fluxes, one has to eliminate the inverse square root as well as the
extrinsic curvature factors. This can be done~\cite{Thiemann:1996aw}
using the relations~\eqn{e:trick1}--\eqn{e:trick3}. Inserting these
into the Hamiltonian constraint one obtains the expression
\begin{equation}
\label{e:H_start}
H[N] = \int_\Sigma\!{\rm d}^3x\, N \epsilon^{mnp} \Tr\Big( F_{mn}
\{ A_p, V\} - \frac{1}{2}(1+\gamma^2) 
 \{ A_m, \bar{K} \} \{ A_n, \bar{K}\} \{ A_p, V\} \Big)\,.
\end{equation}
This expression is the starting point for the construction of the
quantum constraint operator.

The next step is to express the Ashtekar connection~$A_m$ as well as
field strength~$F_{mn}$ and the extrinsic curvature~$\bar{K}$ in terms
of holonomies. As in lattice gauge theory, this requires the
introduction of a cell structure.  In the \emph{classical} expression
one splits the target space into cells of size~$\sim \epsilon^3$ in
\emph{coordinate} space.  In order to express connections and
curvatures in terms of holonomies, one now has to choose edges~$e_m$
within each cell, along which the holonomies are going to be
evaluated. Once the edges~$e_m$ are chosen, one uses that the
connection along a line of \emph{coordinate} distance~$\epsilon$ can
be approximated by
\begin{equation}
\begin{aligned}
{}\left\{ \int_e A, V \right\} &= - h_e[A]^{-1} \big\{ h_e[A], V \big\} 
+  o (\epsilon)\,, \\
{}\left\{ \int_e A, \bar{K} \right\} &= 
- h_e[A]^{-1} \big\{ h_e[A], \bar{K} \big\} + o (\epsilon) \, .
\end{aligned}
\end{equation}
For the field strength one has the standard lattice expression
\begin{equation}
\label{e:field-strength}
\int_{P(\epsilon)} F[A] = 
  \frac{1}{2} \Big( h_{\partial P(\epsilon)}^{-1}[A] -
   h_{\partial P(\epsilon)}[A]\Big) + o(\epsilon^3)  \,,
\end{equation}
where~$P(\epsilon)$ is a two-dimensional infinitesimal plaquette 
with boundary~$\partial P(\epsilon)$, that shrinks to zero as 
$\epsilon\rightarrow 0$. The extrinsic curvature is expressed using the
relation~\eqn{e:trick3}, in which in one
substitutes~\eqn{e:field-strength}. Finally, as for the volume
operator, one replaces integrals with Riemann sums over the cells,
which we will again label with~$\alpha$.  

In order to quantise the classical Hamiltonian \eqn{e:H_start} with the
above regularisation, we next elevate all classical objects to quantum 
operators as described in the foregoing sections, and replace the Poisson 
brackets in \eqn{e:H_start} by quantum commutators. The resulting 
\emph{regulated Hamiltonian} then reduces to a sum over the vertices  
$v_\alpha$ of the spin network with lapses $N(v_\alpha)$ 
\begin{multline}
\label{e:mess1}
\hat{H}[N,\epsilon] = \sum_{\alpha} \,  N(v_{\alpha})\, \epsilon^{mnp} 
  \Tr\Big( \big(h_{\partial{P_{mn}(\epsilon)}} 
  - h_{\partial{P_{mn}(\epsilon)}}^{-1} \big) 
\, h^{-1}_{p}\, \big[ h_{p}, \hat{V}\big]\Big)  \\[1ex]
- \frac{1}{2} (1 + \gamma^2) \sum_{\alpha}  N(v_\alpha)\,
  \epsilon^{mnp}  \Tr \Big( h^{-1}_{m}\,\big[ h_{m}, \bar{K} \big] \,
  h_{n}^{-1}\, \big[ h_{n}, \bar{K} \big]\, h_{p}^{-1} \big[h_{p}, \hat{V} \big]   \Big) \,,
\end{multline}
where we have already assumed a specific ordering of the
operators. The fact that the regularisation parameter~$\epsilon$ drops
out `miraculously' from this expression, and the
integral~\eqn{e:H_start} can be replaced by the sum~\eqn{e:mess1},
requires the Hamiltonian density to have the correct weight, and would
not work with the weight two density~\eqref{e:EEF},
cf.~footnote~\ref{densityfootnote} on
page~\pageref{densityfootnote}. If one does not want to make reference
to any particular spin network, the Hamiltonian can also be expressed
more abstractly as a continuous sum
\begin{equation}
\hat{H} [N,\epsilon] = \sum_{\x\in\Sigma} N(\x) \hat{H} (\x;\epsilon)
\end{equation}
which, on any given spin network, reduces to a finite sum as in~\eqn{e:mess1}.
In the remainder, we will keep the regularisation parameter $\epsilon >0$ 
fixed, and not always indicate it explicitly in the formulas below.

A first stumbling block here is the presence of the volume operator
$\hat{V}$ in~\eqn{e:mess1}. Although its square has been expressed in
terms of fluxes in section~\ref{s:volume}, the spectrum of this
operator is only partially known, see, however,
\cite{Thiemann:1996au,Brunnemann:2004xi,Brunnemann:2005in} for recent
progress (recall that $\hat{q}$ is not necessarily positive).  The
further evaluation of \eqn{e:mess1} on a given spin network wave
function would now in particular require a diagonal basis of spin
network states, on which we could determine the square root of
$|\hat{q}|$ for each node, cf.~\eqn{q}. To treat all the relevant
terms, this diagonalisation would have to be repeated several times,
along the lines of~\cite{Brunnemann:2004xi}, with intermittent
application of other operators. Although all the steps required in
this calculation are thus clearly laid out, presently available
technology does not, as far as we are aware, allow a complete
evaluation of $\hat{H}[N,\epsilon]$ on a given, but arbitrary spin
network state. For this reason, we will only be able to exploit
certain qualitative properties of the volume operator in our analysis
of the Hamiltonian constraint below. We will thus adopt a pragmatical
approach in the remainder, performing all computations with the local
trilinear operator~$\hat{q}$ instead of the volume operator itself,
and assume that we can divide by the relevant power of the volume at
the end (thereby postponing the most difficult part of the
calculation)~\footnote{We note that replacements such as $[\dots ,
\hat{V}] \longrightarrow \sum_v \sgn(\hat{q})\, \hat{|q|}^{-1/2}\, [
\dots , \hat{q} ]$ introduce further operator ordering ambiguities
into the definition of the Hamiltonian operator.}.  Accordingly, the
arguments in this section should be viewed as no more than a
\emph{gedanken calculation}, and there is no claim that this is the
way the calculation should ultimately be done.

Apart from obvious ordering ambiguities, and the difficulties with
volume operator just described, the action of the quantum
expression~\eqn{e:mess1} on a generic spin network state also depends
strongly on the choice of the coordinate system on~$\Sigma$.  To be
more precise, the action depends on whether or not the basis vectors
({\it i.e.\/}~the directions $m$,$n$ and $p$) are aligned with the
spin network edges at a given node. There exists a large number of
possibilities for these alignments, which are \emph{not}
diffeomorphically equivalent, and one might thus simply define the
action of the Hamiltonian by employing a particular alignment, as LQG
does. Some of these possibilities are explicitly listed below. We
regard these choices as genuine ambiguities in the quantisation
procedure, in the sense that observables quantities may depend on
them, if they can be shown to exist at all.  This potential
prescription dependence, and the abundance of possible outcomes for
\emph{physical quantities} are most difficult to digest from the
perspective of conventional quantum field theory, especially because,
the result for the corresponding classical quantity does not depend on
such choices when the limit $\epsilon\rightarrow 0$ is taken.

To get an idea of the complications, let us consider the action of the
first term in the regulated Hamiltonian constraint, which is occasionally
called the `Euclidean Hamiltonian' \footnote{And thus often designated as 
${\cal C}^{\text{Eucl}}$.} (because with a Euclidean signature and 
$\gamma = \pm 1$, the bothersome second term in \eqn{e:mess1} would be 
absent, as it is for~\mbox{$\gamma = \pm i$} with Lorentzian signature).
We will act with this operator on a
three-valent vertex~$v$ in a spin network, for given~$\epsilon
>0$. The spin network wave function on which we will act is given by
\begin{equation}
\psi = 
\big( h^{j_1}_{e_1}\big)_{\alpha_1\beta_1}
\big( h^{j_2}_{e_2}\big)_{\alpha_2\beta_2}
\big( h^{j_3}_{e_3}\big)_{\alpha_3\beta_3}\,
C^{j_1 j_2 j_3}_{\beta_1\beta_2\beta_3} \ldots\,,
\end{equation}
where the dots denote other parts of the spin network.  Since the
Hamiltonian acts node by node, we can restrict our attention to this
local structure at~$v$. Let us also write out all indices on the
Hamiltonian operator, restricted to that part which acts at~$v$,
\begin{equation}
\label{e:Hpart}
\hat{H} = \sgn(\hat{q})\,|\hat{q}|^{-1/2}\,\epsilon^{mnp}\,\big(F^{j}_{mn}\big)_{\lambda_1\lambda_2}
\big(h^{j}_{p}\big)^{-1}_{\lambda_2\lambda_3} 
\left[ \hat{q},\, \big(h^j_p\big)_{\lambda_3\lambda_1}\right] + \ldots\,,
\end{equation}
where the dots now denote terms which act at other vertices. The
choice of the spin~$j$ is a new ambiguity. In the first step, the
operator~\mbox{$h_{p}^{-1} [h_{p},\,{\hat q}] = {\hat q} - h^{-1}_p\,
{\hat q}\, h_p$} acts at the vertex of the network. When~${\hat q}$ in
the second term acts on three holonomies already present in the spin
network, the result cancels against the first term. (Note that the
result only vanishes because we are using~${\hat q}$ instead
of~$\hat{V}$, so that the local volume operator acts tri-linearly in
holonomy derivatives. This is no longer true when one uses the
square-root expression for~$\hat{V}$ in the Hamiltonian constraint).

When~$\hat{q}$ acts on two existing holonomies and on 
the~$h_p$ factor, however, the result is non-trivial. 
In order to work out this non-trivial action of~$\hat{q}$, it is
necessary to fix the orientation of the coordinate system
on~$\Sigma$. We will here choose the~$m,n$ and~$p$ directions to be
aligned with the three edges~$e_1, e_2$ and~$e_3$ of the spin network
respectively, as is usually done in the LQG literature. Note, however,
that this is not the only possibility; we will return to this
arbitrariness shortly. In effect, this means that the quantum
Hamiltonian is \emph{postulated} to be
\begin{multline}
\label{e:mess2}
\hat{H} = \sum_{\alpha} \sum_{I,J,K}\,  N(v_{\alpha})\, L^{IJK} 
  \Tr\Big( \big(h_{\partial{P_{IJ}}} - h_{\partial{P_{IJ}}}^{-1} \big) 
\, h^{-1}_{s_K}\, \big[ h_{s_K}, \hat{V}\big]\Big)  \\[1ex]
- \frac{1}{2} (1 + \gamma^2) \sum_{\alpha} \sum_{I,J,K} N(v_\alpha)\,
  L^{IJK}  \Tr \Big( h^{-1}_{s_I}\,\big[ h_{s_I}, \bar{K} \big] \,
  h_{s_{J}}^{-1}\, \big[ h_{s_J}, \bar{K} \big]\, h_{s_K}^{-1} \big[h_{s_K}, \hat{V} \big]   \Big) \,.
\end{multline}
The second sum is a sum over all triplets of edges emanating from the
vertex~$v_{\alpha}$ in the centre of the cell~$\alpha$. The boundary
of a plaquette in the plane spanned by the edges~$s_I$ and~$s_J$ is
denoted by~$\partial P_{IJ}$. Finally, the value of the constant,
fully anti-symmetric tensor~$L^{IJK}$ depends on the particular
vertex~$\alpha$. 

\begin{figure}[t]
\psfrag{a1}{\smaller\smaller$\alpha_1$}
\psfrag{a2}{\smaller\smaller$\alpha_2$}
\psfrag{a3}{\smaller\smaller$\alpha_3$}
\psfrag{b1}{\smaller\smaller$\beta_1$}
\psfrag{b2}{\smaller\smaller$\beta_2$}
\psfrag{b3}{\smaller\smaller$\beta_3$}
\psfrag{l1}{\smaller\smaller$\lambda_1$}
\psfrag{l2}{\smaller\smaller$\lambda_2$}
\psfrag{l3}{\smaller\smaller$\lambda_3$}
\psfrag{j}{$j$}
\psfrag{j1}{$j_1$}
\psfrag{j2}{$j_2$}
\psfrag{j3}{$j_3$}
\psfrag{k1}{$k_1$}
\psfrag{k2}{$k_2$}
\psfrag{k3}{$k_3$}
\psfrag{C}{$C$}
\psfrag{V}{$V$}
\begin{center}
\includegraphics*[width=.95\textwidth]{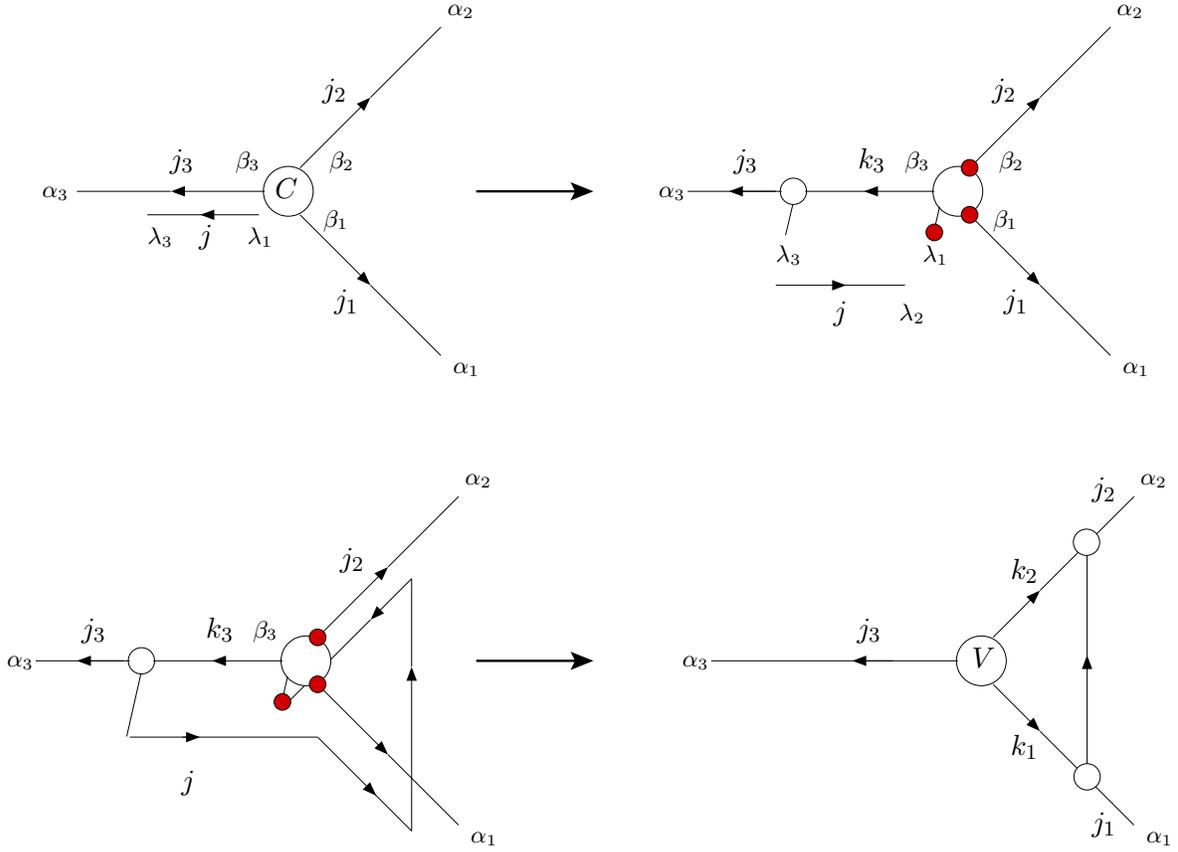}
\end{center}
\caption{The action of the first term in the Hamiltonian constraint on
a spin network, in the special case that the~$m$,~$n$ and~$p$ directions
are aligned with three existing edges of the spin network. The three
edges are drawn in the plane for convenience, but should span a
tetrahedron in order for the volume operator to act non-trivially.
Filled dots denote insertions of~$\tau$ matrices. In the last two
steps, we have only drawn one term in the plaquette~\eqn{e:plaquette}.
\label{f:Haction1}}
\end{figure}

Having made this coordinate choice, we can now return to our
calculation. The situation is as in the first graph of
figure~\ref{f:Haction1}. Because the small edge~$p$ is extending in
the $e_3$ direction, we can use the tensor product decomposition rule
to first `merge' the~$h_p$ holonomy with (a segment of) the~$h_{e_3}$
holonomy of the spin network. For two arbitrary holonomies along
identical edges, this decomposition rule reads
\begin{equation}
\label{e:tensorprod}
\big(h^{j_1}_e\big)_{\alpha_1\beta_1}
\big(h^{j_2}_e\big)_{\alpha_2\beta_2} 
= \sum_k C^{j_1 j_2 k}_{\alpha_1\alpha_2\gamma_1} 
         C^{j_1 j_2 k}_{\beta_1\beta_2\gamma_2} \, \big(h^{k}_e\big)_{\gamma_1\gamma_2}\,.
\end{equation}
By splitting the edge~$e_3$ into a `small' piece, overlapping with the~$p$
edge, and a `larger' \footnote{We put quotation marks here, because there
is no \emph{a priori} reference metric w.r.t.~which the lengths of these 
subedges could be measured.} remainder, we can write
\begin{equation}
\label{e:splitedge}
\big(h^{j_3}_{e_3})_{\alpha_3\beta_3} =
\big(h^{j_3}_{e_3'}\big)_{\alpha_3\gamma_3}
\big(h^{j_3}_{e_3''}\big)_{\gamma_3\beta_3}\,,
\end{equation}
where the edge~$e_3''$ is the small piece of the edge, equal to~$p$.
Applying the tensor product rule to the holonomies~$h_p$ and~$h_{e_3''}$
one now obtains (recalling that also~$m$ and~$n$ are aligned
with~$e_1$ and~$e_2$ respectively)
\begin{multline}
\hat{H}\psi = \sgn(\hat{q})\,|\hat{q}|^{-1/2}\,\big(F^{j}_{e_1 e_2}\big)_{\lambda_1\lambda_2}
 \big(h_{e_3''}^j\big)^{-1}_{\lambda_2\lambda_3}
 \big(h_{e_3'}^{j_3}\big)_{\alpha_3\gamma_3}
 \hat{q} \;\\[1ex]
 \times \left(\sum_{k_3} C^{j\; j_3\; k_3}_{\lambda_3 \gamma_3 \rho_3}
          C^{j\; j_3\; k_3}_{\lambda_1 \beta_3 \sigma_3}\,
          \big( h^{k_3}_{e_3''} \big)_{\rho_3\sigma_3}\right)\;
\big( h^{j_1}_{e_1}\big)_{\alpha_1\beta_1}
\big( h^{j_2}_{e_2}\big)_{\alpha_2\beta_2}
\, C^{j_1 j_2 j_3}_{\beta_1\beta_2\beta_3}\, 
\ldots\,.
\end{multline}
The factor in brackets corresponds to the two merged holonomies.
We can now work out the action of the first line of this expression.

As explained in section~\ref{s:volume}, the square of the local volume
operator inserts a~$\tau$ matrix at the end of the three holonomies on
which it acts. In our case, this means that the matrices are connected
to the indices~$\lambda_1$, $\beta_1$ and $\beta_2$, as these indices
correspond to the end-points of the three holonomies which are located
at the vertex~$v$ (remember that in expression~\eqn{e:Hpart}, the only
non-zero contribution arises when $\hat{q}$ acts on two holonomies
of the spin network and on the holonomy~$(h_p)_{\lambda_3\lambda_1}$
inside the Hamiltonian constraint; the $\lambda_1$ index is the one
located at the vertex). Up to a normalisation factor, this leads to
\begin{multline}
\label{e:beforeplaquette}
\hat{H}\psi = \sgn(\hat{q})\,|\hat{q}|^{-1/2}\,\epsilon^{abc}\,\big(F^{j}_{e_1 e_2}\big)_{\lambda_1\lambda_2}
 \big(h_{e_3''}^j\big)^{-1}_{\lambda_2\lambda_3}
 \big(h_{e_3'}^{j_3}\big)_{\alpha_3\gamma_3}
 \;\\[1ex]
 \times \big(\tau^a\big)_{\lambda_1\tau_3} 
     \left(\sum_{k_3} C^{j\; j_3\; k_3}_{\lambda_3 \gamma_3 \rho_3}
          C^{j\; j_3\; k_3}_{\tau_3 \beta_3 \sigma_3}\,
          \big( h^{k_3}_{e_3''} \big)_{\rho_3\sigma_3}\right)\;
\big( h^{j_1}_{e_1}\, \tau^b\big)_{\alpha_1\beta_1}
\big( h^{j_2}_{e_2}\, \tau^c\big)_{\alpha_2\beta_2}
\, C^{j_1 j_2 j_3}_{\beta_1\beta_2\beta_3}\, 
\ldots\,.
\end{multline}
What is left to do is to insert the plaquette. Because of our choice
of coordinate system on~$\Sigma$, the plaquette extends in the plane
spanned by the~$e_1$ and~$e_2$ edges. In the
limit~$\epsilon\rightarrow 0$, the field strength can be expressed in
terms of elementary holonomies as
\begin{equation}
\label{e:plaquette}
\big( F_{e_1e_2}^j \big)_{\lambda_1\lambda_2}
  \simeq \big( h^j_{e_1} \big)_{\lambda_1\kappa_1}
    \big( h^j_{e_1e_2} \big)_{\kappa_1\kappa_2}
    \big( h^j_{e_2} \big)_{\kappa_2\lambda_2} 
 \; - \;\; \text{(inverted loop)}\,,
\end{equation}
where $h^j_{e_1e_2}$ denotes a holonomy in the direction which
connects the edge~$e_1$ with the edge~$e_2$. When we insert this
expression in~\eqn{e:beforeplaquette}, we can again split the edges in
a small and large piece as in~\eqn{e:splitedge} and then use the tensor
product decomposition~\eqn{e:tensorprod} to merge the two holonomies
in the~$e_1$ direction and the two holonomies in the~$e_2$ direction.
This introduces four new Clebsch-Gordan coefficients, with the result
\begin{multline}
\hat{H}\psi = \sgn(\hat{q})\,|\hat{q}|^{-1/2}\,\epsilon^{abc}\,
 \big(h_{e_1'}^{j_1}\big)_{\alpha_1\gamma_1}
 \big(h_{e_2'}^{j_2}\big)_{\alpha_2\gamma_2}
 \big(h_{e_3'}^{j_3}\big)_{\alpha_3\gamma_3}\;
\\[1ex]
\times \big(h_{e_3''}^j\big)^{-1}_{\lambda_2\lambda_3}
\, \big(h_{e_1 e_2}^j\big)_{\kappa_1\kappa_2}\,
\, C^{j_1 j_2 j_3}_{\tau_1\tau_2\beta_3}\, 
  \big(\tau^c\big)_{\beta_1\tau_1}
  \big(\tau^b\big)_{\beta_2\tau_2}
  \big(\tau^a\big)_{\lambda_1\tau_3}
 \left(\sum_{k_3} 
          C^{j\; j_3\; k_3}_{\lambda_3 \gamma_3 \rho_3}
          C^{j\; j_3\; k_3}_{\tau_3 \beta_3 \sigma_3}\,
          \big( h^{k_3}_{e_3''} \big)_{\rho_3\sigma_3}\right)\\[1ex]
\times \left(\sum_{k_1} 
          C^{j\; j_1 k_1}_{\lambda_1  \beta_1  \sigma_1}
          C^{j\; j_1 k_1}_{\kappa_1 \gamma_1 \rho_1}
          \big(h^{k_1}_{e_1''}\big)_{\rho_1\sigma_1}
 \right)
 \left(\sum_{k_2} 
          C^{j\; j_2 k_2}_{\lambda_2 \beta_2  \sigma_2}
          C^{j\; j_2 k_2}_{\kappa_2  \gamma_2 \rho_2}
          \big(h^{k_2}_{e_2''}\big)_{\rho_2\sigma_2}        
 \right)\,\ldots\,.
\end{multline}
Altogether, these manipulations have introduced three new vertices, at
locations~$\epsilon$ away along each edge, and they have also
modified the original vertex. The three new vertices are given by
\begin{equation}
\begin{aligned}
(V^{(1)})^{k_1 j_1 j}_{\rho_1 \gamma_1 \kappa_1}  &= 
  C^{j j_1 k_1}_{\kappa_1 \gamma_1 \rho_1}\,, \\[1ex]
(V^{(2)})^{k_2 j_2 j}_{\rho_2 \gamma_2 \kappa_2}  &=
  C^{j\;\; j_2 k_2}_{\kappa_2\gamma_2\rho_2}\,,\\[1ex]
(V^{(3)})^{k_3 j_3 j}_{\rho_3 \gamma_3 \lambda_3} &=
  C^{j\;\; j_3 k_3}_{\lambda_3\gamma_3\rho_3}\,,
\end{aligned}
\end{equation}
while the old vertex has been changed to 
\begin{equation}
V^{k_1k_2k_3 \;j}_{\sigma_1\sigma_2\sigma_3\lambda_2} = 
  C^{j\; j_1 k_1}_{\lambda_1  \beta_1  \sigma_1}
  \,C^{j\; j_2 k_2}_{\lambda_2 \beta_2  \sigma_2}
  \,C^{j\; j_3\; k_3}_{\tau_3 \beta_3 \sigma_3}\times
   \epsilon^{abc}\,\big(\tau^b\big)_{\beta_1\tau_1}
   \big(\tau^c\big)_{\beta_2\tau_2}
   \big(\tau^a\big)_{\lambda_1\tau_3}\times
   C^{j_1 j_2 j_3}_{\tau_1\tau_2\beta_3}\,.
\end{equation}
One can still merge~$\big(h_{e_3''}^j\big)^{-1}_{\lambda_2\lambda_3}$
and~$\big( h^{k_3}_{e_3''} \big)_{\rho_3\sigma_3}$ using the tensor
product decomposition rule, which will reduce the vertex~$V^{(3)}$ to
an overall factor times a Kronecker delta. We will refrain from
spelling out these details.

\begin{figure}[t]
\medskip
\psfrag{a1}{\smaller\smaller$\alpha_1$}
\psfrag{a2}{\smaller\smaller$\alpha_2$}
\psfrag{a3}{\smaller\smaller$\alpha_3$}
\psfrag{b1}{\smaller\smaller$\beta_1$}
\psfrag{b2}{\smaller\smaller$\beta_2$}
\psfrag{b3}{\smaller\smaller$\beta_3$}
\psfrag{l1}{\smaller\smaller$\lambda_1$}
\psfrag{l2}{\smaller\smaller$\lambda_2$}
\psfrag{l3}{\smaller\smaller$\lambda_3$}
\psfrag{j}{$j$}
\psfrag{j1}{$j_1$}
\psfrag{j2}{$j_2$}
\psfrag{j3}{$j_3$}
\psfrag{k1}{$k_1$}
\psfrag{k2}{$k_2$}
\psfrag{k3}{$k_3$}
\psfrag{C}{$C$}
\psfrag{V}{$V$}
\begin{center}
\includegraphics*[height=4.5cm]{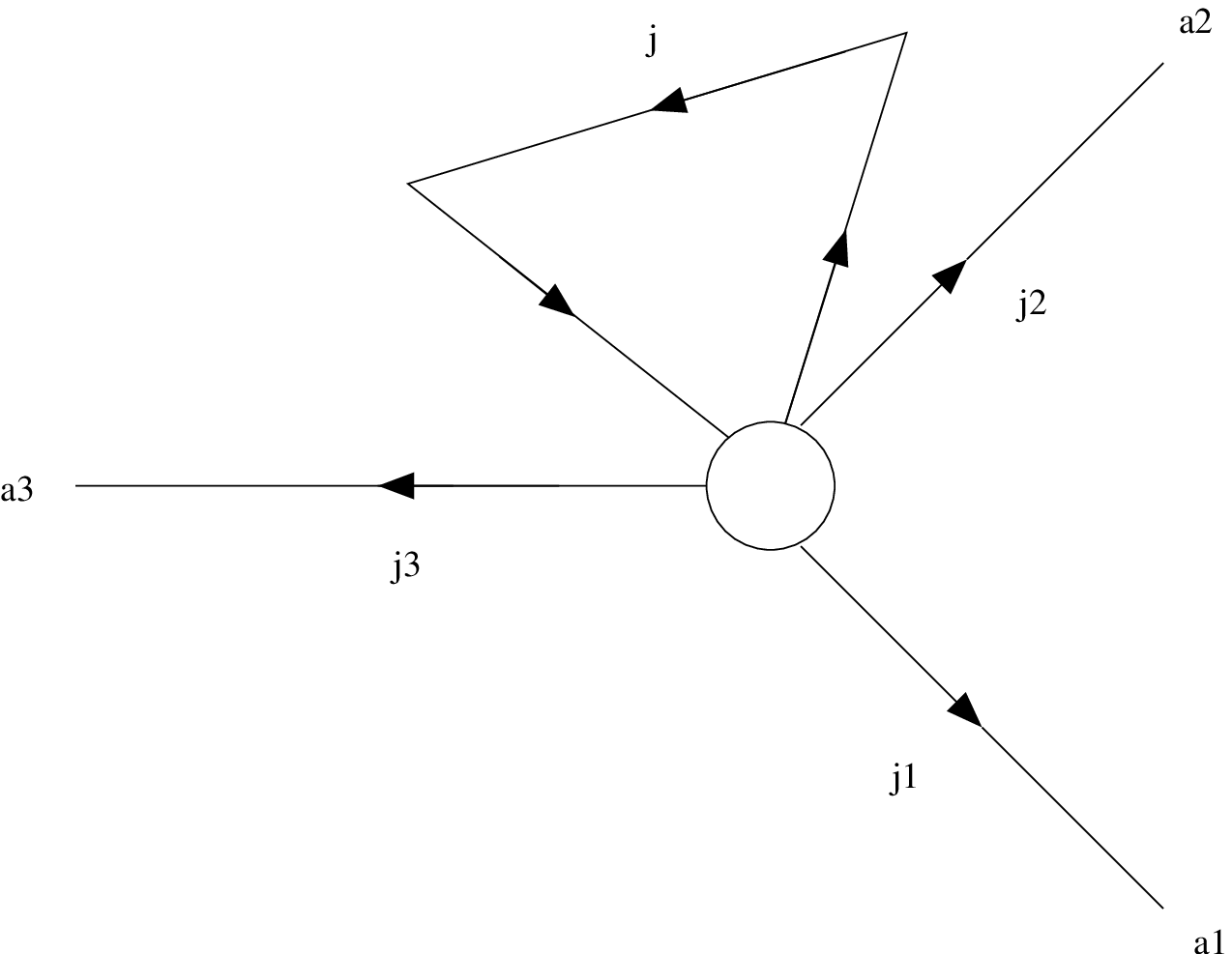}\vspace{-2ex}
\end{center}%
\caption{The action of the first term in the Hamiltonian constraint,
  for the case in which the $m$ and $n$ directions  on~$\Sigma$ are
  not aligned with any of the edges emanating from the vertex.}%
\label{f:Haction_misaligned}
\end{figure}

Let us now comment on other possible coordinate systems and
alignments.  The use of the tensor product
decomposition~\eqn{e:tensorprod} depends crucially on the edges of the
spin network being aligned with the edges of the plaquette (and,
through~\eqn{e:mess1}, with the coordinate system on~$\Sigma$). If
this is not the case for the $m$,$n$ and $e_1$,$e_2$ edges, the
plaquette will not get glued into the spin network, but instead
remains freely floating, only connected at the original vertex (see
figure~\ref{f:Haction_misaligned}). Similarly, if the $p$~edge is not
aligned with~$e_3$, the plaquette will only be connected to the spin
network through the~$\tau^a$ matrix in the construction above. In the
LQG literature, such cases are excluded by hand, by always choosing a
local coordinate system around each vertex which is completely aligned
with (three of the) edges emanating from the vertex. This choice is
usually justified by invoking background independence, which would be
violated otherwise~\cite{Thiemann:2001yy}.

We have so far not discussed the second term in the
constraint~\eqn{e:mess1}, which arose from the product of extrinsic
curvatures in~\eqn{e:clH}.  This term is hardly ever discussed in 
detail, and with good reason: it is far more complicated
than the first one. In fact, the technical complications which we have
just described pile up in such a way as to make any concrete
calculation almost unfeasible.  The complexity of this term stems from
the complexity of the intermediate expression
\begin{equation}
{}[ A_I, \bar{K} ] = h^{-1}_{e_I} \left[ h_{e_I}, 
  \epsilon^{JKL} \left[ (h_{\alpha_{JK}} - h^{-1}_{\alpha_{JK}})
  h^{-1}_{e_L}\, \big[ h_{e_L}, \hat{V} \big],\; \hat{V} \right] \right]\, ,
\end{equation}
which is obtained using~\eqn{e:trick1}--\eqn{e:trick3}. One recognises
the combination of operators which opens up the vertex and inserts a
plaquette, but there there is an additional volume operator which
applies further changes to the node. If the coordinate system
on~$\Sigma$ is aligned in such a way that all three directions
coincide with an existing edge, the action of the second term in the
constraint is roughly to glue in three new edges, introducing six new
vertices. But as the discussion above shows, there is a plethora of
other possibilities.

Let us, for clarity, summarise the problems which we see with the
Hamiltonian constraint.
\begin{itemize}
\item {\it Ambiguities.} The construction of the Hamiltonian outlined
  here is plagued with a variety of ambiguities. Firstly, a choice has
  to be made for the coordinate system on~$\Sigma$. The action of the
  Hamiltonian depends in a qualitative way on this choice. Secondly,
  there is an ambiguity in choosing the representation in which the
  trace is evaluated ({\it i.e.\/}~the representation in which the
  field strength~$F_{mn}$ and the holonomy~$h_p$ are taken). Thirdly,
  there are operator ordering ambiguities. Finally, and less
  conspicuously, there may exist $\hbar$-dependent `higher order'
  corrections to the Hamiltonian~\eqn{e:mess1}, which are still
  compatible with diffeomorphism covariance, but would not affect the
  classical limit.  Not only will physical predictions depend on the
  choices made to fix these ambiguities, but these choices are
  important in order to determine whether or not the
  limit~$\epsilon\rightarrow 0$ of~\eqn{dual1} exists at all.

\item {\it Ad-hoc choices.} The quantum constraint should be well-defined
  and unambiguous on the full space of spin-network states. If this is
  not the case, it should be shown that the choices do not matter for
  the physical predictions, or alternatively that a particular choice
  is singled out by physical consistency. By focussing only on the
  operator~\eqn{e:mess2}, it becomes impossible to tell whether the
  constraint operator is indeed independent of the coordinate choice
  on~$\Sigma$.

\item {\it Ultralocality.} From the explicit discussion of the
  Hamiltonian constraint it should be clear that its action is always
  `ultralocal', in the sense that all changes to the spin network are
  made in an $\epsilon \rightarrow 0$ neighbourhood of a given vertex,
  while the spin network graph is kept
  fixed~\cite{Smolin:1996fz,Neville:1998eb,Loll:1997iw}.  More
  specifically, it has been argued \cite{Lewandowski:1997ba} that the
  Hamiltonian acts at a particular vertex only by changing the
  intertwiners at that vertex, viz.
\begin{equation}\label{ultralocal}
\hat{H}(v)\, | \Gamma, \{j\}, \{C\} \rangle = 
\sum_{I,J} U^a |\Gamma, \{j\}, C_a(\Gamma,\{j\},I,J)\rangle  
\end{equation}
  where $U^a$ is the traceless part of the holonomy associated with the
  plaquette $P_{IJ}$ and $C_a$ a triplet of new intertwiners at $v$, and
  the sum is over pairs of edges $I,J$ entering the vertex.
  This is in contrast to what happens in lattice field theories. There 
  the action of the Hamiltonian always non-trivially links two existing
  nodes, the plaquettes are by construction always spanned between
  existing nodes, and the continuum limit involves the lattice as 
  a whole, not only certain sub-plaquettes. 
\end{itemize}

\section{Constraint algebra}
\label{s:constraint_algebra}

As we already emphasised in the introduction, the proper closure of
the quantum constraint algebra is perhaps \emph{the} crucial
consistency requirement that must be met by any candidate theory of
quantum gravity.  Of course, the devil hides in the meaning of the
word `proper'. To find out where it might possibly hide, we now return
to the points raised at the end of
section~\ref{s:kinematical_constraints}. We will not bother here with
the commutators involving the Gauss constraints, which are
straightforward to take care of. The canonical (Poisson) brackets
involving the diffeomorphism generators read
\begin{align}
\label{e:DiffDiff}
\Big\{ D_m[\vec{M}]\,,\, D_n[\vec{N}] \Big\}_{\text{PB}} &=
\int_{\Sigma}\!{\rm d}^3 x\,
  (M^m \partial_m N^n - N^m \partial_m M^n) D_n\,,\\[1ex]
\label{e:DiffHam}
\Big\{ D_m[\vec{M}]\,,\, H_0[N] \Big\}_{\text{PB}} &= \int_{\Sigma}\!{\rm d}^3 x
\, (M^m \partial_m N - N \partial_m M^m) H_0\,.
\end{align}
Here $D_m \equiv {e_m}^a D_a$ is the diffeomorphism generator, while the
smeared Hamiltonian and smeared diffeomorphism generators are defined as
\begin{equation}
H_0 [M] := \int_\Sigma\!\d^3 x \, H_0(\x) M(\x)\,,\quad
D_m [\vec{M}] := \int_\Sigma\!{\rm d}^3x\, D_m(\x) M^m(\x)\,.
\end{equation}
Because infinitesimal diffeomorphisms do not actually exist on $\Hk$,
only the exponentiated version of~\eqn{e:DiffDiff} makes sense in LQG.
For the same reason, relation~\eqn{e:DiffHam} is implemented by 
considering only finite (exponentiated) diffeomorphisms~$\varphi$. 
Given a realisation of $\varphi$ on the Hilbert space through a unitary 
operator $\hat{U}[\varphi]$, the bracket~\eqn{e:DiffHam} can 
be exponentiated formally to\footnote{When this relation 
is understood to act on a state in $\Hd$, one can
use invariance of the state under diffeomorphisms, 
$\hat{U}[\varphi] |\psi\rangle = |\psi\rangle$, to rewrite 
equation~\eqn{e:finiteDiff} as~\cite{Thiemann:1997rv}
\begin{equation}
\Big[ \hat{U}[\varphi]\,,\, \hat{H}_0[N] \Big] =
  \hat{H}_0[ \varphi^*N - N ]\,.
\end{equation}}
\begin{equation}
\label{e:finiteDiff}
\hat{U}[\varphi]\, \hat{H}_0[N]\, \hat{U}[\varphi]^{-1} = 
\hat{H}_0[\varphi^* N]\,.
\end{equation}
These relations constitute what we would call the `kinematical 
part' of the constraint algebra.

The crucial, remaining relation is the commutator of two Hamiltonian
constraints, whose classical analog reads
\begin{equation}\label{HH}
\Big\{ H_0 [M] \, , \, H_0 [N] \Big\}_{\text{PB}} =
\int_\Sigma\!\d^3 x \, (M \partial_m N - N \partial_m M) g^{mn} D_n \,.
\end{equation}
Together with the `kinematical relations' above, this relation
encapsulates the closure of the classical algebra of spacetime
diffeomorphisms in the Hamiltonian formulation (we should note,
however, that the relation between timelike diffeomorphisms on the one
hand, and the `motion' generated by the Hamiltonian constraint on the
other, is quite subtle~\cite{Mukhanov:1994zn}).  To be sure, the
proper closure of this part of the quantum constraint algebra, {\it
i.e.\/}~the issue of promoting \eqn{HH} to a relation between
operators acting on a suitable Hilbert space of states, is also one of
the major unsolved problems of the conventional geometrodynamics
approach. There it has not been possible to resolve the ordering and
other ambiguities, and to get rid of the infinities resulting from the
clash of functional differential operators (the situation is partly
better, though, in certain supergravity models whose constraint
superalgebra always contains the bosonic constraint algebra as a
subalgebra \cite{Matschull:1993hy}).  

\subsection{The bosonic string as an example}

Before continuing our discussion of the quantum constraint algebra in
canonical gravity, we would like to recall that there does exist a
model which can be solved exactly, and all the way. Hence it can serve
as an example of how quantisation should work. This model is the
bosonic string (in a Polyakov type formulation), when viewed as a
\emph{bona fide} model of matter coupled quantum gravity in two
dimensions (see also~\cite{Kouletsis:2001zb} for an earlier discussion
of the bosonic string in this context).  For this purpose, we only
need to forget about the target space interpretation and its
ramifications, such as string scattering, or higher order corrections
in $g_s$, involving splitting and joining of strings (in the present
context, the latter processes would amount to changes of the spatial
topology, `baby universes' and the like, hence, `third quantisation'). 
The resulting theory is manifestly independent of the two-dimensional 
space-time background, free of divergences and other pathologies, and 
also possesses nice observables (vertex operators).  More specifically, 
the diffeomorphism and scalar constraints $D$ and $H$ are given by 
(see e.g.~\cite{b_gree12} or~\cite{b_polc12})
\begin{equation}
\begin{aligned}
D[f] &= \frac{1}{2}\int_0^{2\pi}\!{\rm d}\sigma\,  f(\sigma) P_\mu(\sigma)
\partial_\sigma X^\mu(\sigma) \,,\\
H[N] &= \frac{1}{4}\int_0^{2\pi}\!{\rm d}\sigma\, N(\sigma) \Big( 
   P_\mu(\sigma) P^\mu(\sigma) + \partial_\sigma X_\mu(\sigma)
	\partial_\sigma X^\mu(\sigma)\Big)\,.
\end{aligned}
\end{equation}
with shift $f(\sigma)$ and lapse $N(\sigma)$. The classical constraint
algebra, given by~\eqn{e:DiffDiff}, \eqn{e:DiffHam} and~\eqn{HH} in
four dimensions, reduces to
\begin{equation}
\label{e:alg2d}
\begin{aligned}
{}\Big\{ D[f_1],\; D[f_2] \Big\}_{\text{PB}} &= \tfrac{1}{2} D[f_1\partial f_2 
- f_2 \partial f_1]\,,\\[1ex]
{}\Big\{ D[f],\; H[N] \Big\}_{\text{PB}} &= \tfrac{1}{2}
H[f\partial N - N \partial f] \,,\\[1ex]
{}\Big\{ H[N_1],\; H[N_2] \Big\}_{\text{PB}} &= \tfrac{1}{2}
D[N_1\partial N_2 - N_2 \partial N_1]\,.
\end{aligned}
\end{equation}
Canonical quantisation is straightforward in the conformal gauge. The
state space of the theory ${\cal F}$ is simply the tensor product of
the space of one-particle wave functions and an infinite number of
harmonic oscillator Hilbert spaces associated with the excited string
modes.  Composite operators become well defined, {\it i.e.\/}~have
finite matrix elements between arbitrary states in ${\cal F}$, through
normal ordering.  Both the kinematical and the Hamiltonian constraint
are hermitean operators.  With the normal ordered energy momentum
tensors $T_{\pm\pm}(\sigma) \equiv \; :({H}(\sigma) \pm {D}
(\sigma)):$, the classical constraint algebra is modified by the
anomaly (the central term), which is responsible for almost everything
that is non-trivial about string theory,
\begin{equation}\label{Vir}
\big[ T_{\pm\pm} (\sigma) \, , \, T_{\pm\pm} (\sigma') \big] = 
\delta'(\sigma,\sigma') 
\big(T_{\pm\pm}(\sigma)+ T_{\pm\pm}(\sigma')\big)
+ \hbar c\,  \delta'''(\sigma,\sigma')\,.
\end{equation}
Our main point in repeating this well known story here is that this
algebra holds on the full space of states, {\it i.e.\/}~prior to the
imposition of any constraints, hence closes \emph{off-shell}. Only at
this point can we impose and solve the quantum constraints to obtain
the subspace ${\cal F}_0\subset {\cal F}$ of \emph{physical states}
--- with the well known result (`No Ghost Theorem') that negative norm
states are absent if the number of scalar fields is $\leq 26$. Because
the Hilbert spaces ${\cal F}_0$ and ${\cal F}$ of physical and
unphysical states, respectively, are well defined, there is
no need to discuss `habitats' here.  Of course, one key simplifying
feature of the string model must be pointed out here, which is not
shared by (matter coupled) gravity in higher dimensions:~\eqn{e:alg2d}
is an honest (albeit infinite dimensional) Lie algebra (and~\eqn{Vir}
closes in Dirac's sense),
whereas~\eqn{e:DiffDiff},~\eqn{e:DiffHam}, and~\eqn{HH} are not. This Lie
algebra furthermore possesses a triangular decomposition, allowing for
a systematic study of its representations, and this fact is a crucial
ingredient in establishing the above results.

Very recently, LQG methods have been applied to string theory, but have 
been shown to lead to results which are at variance with those obtained 
using Fock-space quantisation~\cite{Starodubtsev:2002xe,Thiemann:2004qu}. 
It therefore does not appear that LQG can recover conventional results;
instead it gives rise to an inequivalent quantisation of the string model, 
in which the quantum constraint algebra is realised without a central term,
and without any restriction on the target space dimension. The meaning 
of these results is currently subject of an intense debate,
see~\cite{Schreiber:2004ys,Helling:2004tb}.

Another simple model which can be quantised all the way is dilaton
gravity in 1+1~dimensions. For this model the constraint algebra has
been worked out for various inequivalent representations of the
operators, and the physical consequences have been analysed in
detail. See for
instance~\cite{Louis-Martinez:1993eh,Benedict:1996qy,Cangemi:1995yz}.
See also~\cite{Grumiller:2002nm} for a review of dilaton gravity with
many further references.

\subsection{On-shell vs.~off-shell closure}

We have seen that LQG treats the diffeomorphism and Hamiltonian 
constraints in a very different manner. Diffeomorphism invariance is
implemented by an averaging procedure that makes the states invariant
under \emph{finite} ({\it i.e.\/}~`exponentiated') diffeomorphisms.
In LQG a canonical generator of diffeomorphisms simply does not exist,
since the lack of weak continuity that goes with the scalar product
\eqn{e:scalar_product} prevents us from differentiating finite 
diffeomorphisms so as to obtain `infinitesimal' ones. By contrast, at
least so far, group averaging is of no use, even at the formal level
and allowing for whatever approximation scheme, when it comes to 
imposing the Hamiltonian constraint. Therefore the latter, whose
cumbersome form we have exhibited in the foregoing section, must be 
analysed `the traditional way'~\footnote{To overcome the difficulties, 
it has even been suggested that in analogy with the diffeomorphism 
generator, the Hamiltonian itself may actually not exist as an 
infinitesimal generator, but that only `finite' translations in time  
might be well defined~\cite{Reisenberger:1997pu,Markopoulou:1997hu}.}.
This restriction applies also to the commutator of two Hamiltonian 
constraint operators. A further (and well known) difficulty is that, 
even for the classical theory, this commutator does not generate 
an ordinary Lie algebra, in the sense that there appear \emph{field 
dependent} structure `constants'; as is well known, these lead to  
ordering problems and ambiguities in the quantum constraint algebra. 

While there is general agreement as to what one means when one speaks
of `closure of the constraint algebra' in classical gravity (or any
other classical constrained system), this notion is more subtle in the
quantised theory. Let us therefore clarify first the various notions
of closure that can arise: we see at least three different
possibilities. As we explained in the foregoing section, the `true'
Hamiltonian $\hH^*$ is defined as a weak limit on some habitat ${\cal
V}^*\subset\S^*$ by means of~\eqn{dual1}.  Assuming that this subspace
is preserved by the action of $\hH^*$, the strongest notion is
`off-shell closure' (or `strong closure'), where one seeks to
calculate
\begin{equation}\label{HH1}
\big[ \hH^*[N_1] \, , \, \hH^* [N_2] \big] = \hO^*(N_1;N_2)
\end{equation}
without further restrictions on the states, on which this relation is
supposed to hold  --- in other words on a sufficiently large `habitat' 
${\cal V}^*$ that does not only contain diffeomorphism invariant states.
The operator $\hO^*(N_1;N_2)$ here is the quantum
analogue of the r.h.s. of~\eqn{HH}, and would be proportional to the
diffeomorphism generator (but this commutator might also differ
from the classical one by certain quantum modifications). Together with 
the `kinematical part' of the quantum constraint algebra, the above 
relation would express the \emph{quantum space-time covariance} of the 
theory. Strong closure is realised for the string model, with the well 
known result \eqn{Vir}. If one could resolve the ordering ambiguities 
and eliminate all singularities, this is also the type of closure 
expected to work in conventional geometrodynamics. On the other hand, 
although the calculation in principle does make sense in LQG, too, 
the r.h.s.~of \eqn{HH1} may not exist as an operator in LQG, as we 
explained above. Of course, even if the diffeomorphism generator does 
not make sense as an operator, it is still conceivable that a combination 
like $g^{mn} H_n$ does exist as a `composite operator'. In this case,
however, this operator would not generate a closed algebra: repeated 
commutation would generate an infinite tower of new operators, and 
one would therefore no longer be discussing the algebra of constraints. 

At any rate, it appears that the goal of determining $\hO^*(N_1;N_2)$ 
as a \emph{bona fide} `off-shell' operator on $\V^*$, and prior to
the imposition of any constraints, is unattainable within the current
framework of LQG. For this reason, LQG must resort to weaker
notions of closure, by making partial use of the constraints. More
specifically, equation~\eqn{HH1} can be relaxed substantially by 
demanding only 
\begin{equation}\label{HH2}
\big[ \hH^*[N_1] \, , \, \hH^* [N_2] \big] {\cal{X}} = 0\,.
\end{equation}
This `weak closure' should hold for all states $\cal{X}$ in a restricted 
habitat ${\cal V}^*$ of states that are `naturally' expected to be 
annihilated by the r.h.s. of \eqn{HH1}, and that are subject to the 
further requirement that the Hamiltonian can be applied twice without 
leaving the `habitat'. The latter condition is, for instance, met by
the `vertex smooth' states of~\cite{Lewandowski:1997ba}. As shown in
~\cite{Gambini:1997bc,Lewandowski:1997ba}, the commutator of two 
Hamiltonians indeed vanishes on this `habitat', and one is therefore 
led to conclude that the full constraint algebra closes `without anomalies'. 

The same conclusion was already arrived at in an earlier computation
of the constraint algebra in~\cite{Thiemann:1996aw,Thiemann:1996ay},
which was done from a different perspective (no `habitats'), and is
based on the choice $\V^* = \Hd$, the `natural' kernel of the r.h.s.
of \eqn{HH1}. Here the first step consists in calculating the
commutator of two regulated Hamiltonians (see
section~\ref{s:hamiltonian_constraint})
\begin{equation}\label{hO}
\hO(N_1;N_2;\epsilon) :=
\big[ \hH[N_1,\epsilon] \, , \, \hH [N_2,\epsilon] \big] \,,
\end{equation}
for fixed but arbitrary $\epsilon >0$ on the space $\S$ of spin 
network states. As for the Hamiltonian itself, letting $\epsilon\rightarrow 0$
in this commutator produces an uncountable sequence 
$\hO(N_1,N_2;\epsilon)\Psi$ of mutually orthogonal states w.r.t.~the 
scalar product \eqn{e:scalar_product}; 
consequently, $\lim_{\epsilon\rightarrow 0}\hO(N_1,N_2;\epsilon)$ 
again does not exist in the usual sense, but only as a \wsl. 
More specifically, one shows that~\cite{Thiemann:1996aw,Thiemann:1996ay}
\begin{equation}\label{HH3}
\lim_{\epsilon\rightarrow 0}\;  \langle {\cal X}\, | \,
\hO(N_1;N_2;\epsilon) \Psi\rangle = 0 
\end{equation}
for all ${\cal X}\in\Hd$, and for all $\Psi\in\S$ (the `diffeomorphism
covariance' of the Hamiltonian is again essential for this result); in
this sense, the constraint algebra is free of anomalies. Let us stress
that \eqn{HH2} and \eqn{HH3} are by no means the same: in \eqn{HH2}
one uses the dual Hamiltonian defined in \eqn{dual1} (where the limit
$\epsilon\rightarrow 0$ has already been taken) and performs the
calculation on a subspace ${\cal V}^*\subset\S^*$, whereas the
calculation of the commutator in \eqn{HH3} takes place on the space
$\S$, that is, inside~$\Hk$, and the limit $\epsilon\rightarrow 0$ is
taken only \emph{after} computing the commutator of two regulated
Hamiltonians on $\S$. In other words, these two operations (taking the
limit $\epsilon\rightarrow 0$, and calculating the commutator) need
not commute. Because with both \eqn{HH2} and \eqn{HH3}, one forgoes
the aim of finding an operatorial expression for the commutator
$\big[\hat{H}^*[N_1], \hat{H}^*[N_2] \big]$ as an operator on $\V^*$,
or the commutator \eqn{hO} on the kinematical Hilbert space, making
partial use of the constraints, we say (in a partly supergravity
inspired terminology) that the algebra closes `on-shell'.

At this point we should like to remark that computations done so 
far make only little use of the detailed structure of the terms in the 
Hamiltonian. In particular, (on-shell) closure is achieved independently of 
whether or not one takes into account the second term in \eqn{e:mess1}, and 
there may exist (diffeomorphism covariant) modifications of the type 
discussed at the end of this section, or `higher order' modifications of 
the Hamiltonian~\eqn{e:mess1}, which do not alter this result. The 
analysis of \cite{Lewandowski:1997ba} uses only schematic formulas like 
\eqn{ultralocal} rather than \emph{explicit} operatorial expressions 
for the Hamiltonian. All this is contrary to naive expectations, because
computations of more complicated gravity-matter systems (such as the
supergravity constraint algebra studied in~\cite{Nicolai:1990vb}) 
show that the detailed structure and the interplay of all terms in the 
constraints \emph{is} indispensable already for the proper closure of the 
classical constraint algebra. This may indicate that crucial information is 
lost, and that, in a sense, the on-shell closure is already `built into' 
the formalism via the diffeomorphism covariance of the Hamiltonian, and by
restricting the calculation to special states (or matrix elements)
only. Still, one would hope to be able in principle to extract from this 
computation some kind of explicit operatorial expression for the commutator.
However, (at least our) attempts to actually \emph{evaluate} \eqn{hO} on 
a given spin network wave function $\Psi$ along the lines of 
section~\ref{s:hamiltonian_constraint}, and before transferring the result 
to $\V^*$, get stuck quickly: the resulting state would be substantially 
more complicated than the (incomplete) expressions displayed there. 

The procedure of making partial use of the constraints \emph{prior} to
the evaluation of the constraint algebra, as well as in the very
construction of the Hamiltonian constraint, is very different from the
conventional treatment of quantised canonical
systems~\cite{Henneaux:1992ig}, as exemplified by the bosonic string
in the previous subsection.  There, one first studies the constraint
operators and tries to make them well defined as quantum operators
without constraining the allowed states, and \emph{before} checking
the closure of the constraint algebra. When one then in a second step
computes the algebra, one usually encounters subtle quantum ({\it
i.e.\/}~$\hbar$-dependent) modifications of this algebra, whose
determination requires great care. Because these modifications are
subject to certain constraints (Wess-Zumino type consistency
conditions and their offspring in BRST cohomology), they can often be
determined in a representation and gauge independent way, as is the
case for the group Diff$(S^1)$. If one succeeds in working out this
algebra and demonstrating its consistency, the quantum constraint
algebra closes `off-shell'.  Only \emph{after} ensuring the closure of
the quantum constraint algebra does one proceed to impose and solve
the quantum constraints. Let us also remind readers that, at least
according to conventional (quantum field theory) wisdom, `on-shell'
calculations of the algebra might be `empty': in general it is not
correct to use a symmetry before it has been shown that it can be
implemented without anomalies~\cite{DeJonghe:1992um}.

Why do we emphasise off-shell closure, despite the difficulties
(ordering ambiguities, field dependent structure constants, and the
like), which likewise plague other approaches to canonical quantum
gravity? Our main reason for emphasising this requirement is two-fold:
first and most importantly, the off-shell closure of the quantum
constraint algebra (which, as we pointed out, need not coincide with
the classical algebra) encapsulates the key property of quantum
gravity, namely \emph{quantum space-time covariance}. Secondly, this
property furnishes an excellent means to distinguish between the
`correct' theory and a mere regularisation: consider for instance
replacing the kinetic part of the WDW operator by a smeared
point-split operator.  With this modification, all constraint
operators become formally well defined, so one might contemplate
`defining' the quantised theory in terms of them, and simply proceed
to study the solutions of these modified constraints. This procedure
would, of course, suffer from the same kind of ambiguities encountered
in section~\ref{s:hamiltonian_constraint} and before, but it might
still take us a long way if we did not pay due attention to the
constraint algebra. If, however, we do subject the modified WDW
Hamiltonian to this acid test, it is immediately clear that the
resulting operator (or any other obtained by such a regularisation) is
`wrong': it does not satisfy the correct algebra, whence the quantised
theory would not satisfy the basic requirement of quantum space-time
covariance. 

As another example, consider modifying the Hamiltonian constraint 
of string theory by multiplying it with an operator which commutes with 
all Virasoro generators,
\begin{equation}
(\Hat{T}_{++} + \Hat{T}_{--})\; \longrightarrow\; (\Hat{T}_{++} + 
  \Hat{T}_{--})\, \hat{C}\,,
\end{equation}
where $[\Hat{C},\Hat{T}_{++}]=0$ and $[\Hat{C},\Hat{T}_{--}]=0$, while
keeping the diffeomorphism constraint \mbox{$(\Hat{T}_{++} -
\Hat{T}_{--})$} unchanged. There are many such operators $\Hat{C}$
in string theory. The simplest is to take $\Hat{C}$ to be equal to
the mass operator $P^\mu P_\mu$ minus some arbitrary positive integer,
where $P^\mu$ is the conserved global momentum of the string (which indeed 
commutes with all Virasoro operators). In this way, we arrive at a
realisation of the constraint operators which is very similar to the
one used in LQG: the algebra of spatial diffeomorphisms is realised
via a (projective) unitary representation, and the Hamiltonian 
constraint transforms covariantly as in~\eqn{e:finiteDiff} 
(the extra factor~$\Hat{C}$ does not matter, because it commutes 
with all constraints). In a first step, one can restrict 
attention to the subspace of states annihilated by the diffeomorphism
constraint, the analog of the space~$\Hd$. Imposing 
\mbox{$(\Hat{T}_{++} + \Hat{T}_{--})\, \hat{C}|\text{phys}\rangle=0$} 
on this subspace would now produce a `non-standard' spectrum by allowing 
extra diffeomorphism invariant states of a certain prescribed mass, but 
the algebra would still close on-shell, i.e.~on the `habitat' of states 
annihilated by the diffeomorphism constraint. The point here is not so 
much whether this new spectrum is `right' or `wrong', but rather that in 
allowing such modifications which are 
compatible with on-shell closure of the constraint algebra, we introduce 
an infinite ambiguity and arbitrariness into the definition of the physical 
states. In other words, if we only demand on-shell closure as in LQG, 
there is no way of telling whether or not the vanishing of a commutator 
is merely accidental, that is, not really due to the diffeomorphism 
invariance of the state, but caused by some other circumstance. 

This, then, is our main point: by weakening the requirements on the 
constraint algebra and by no longer insisting on off-shell closure, 
crucial information gets lost. This loss of information is reflected in 
the ambiguities inherent in the construction of the LQG Hamiltonian. 
It is quite possible that the LQG Hamiltonian admits many further 
modifications on top of the ones we have already discussed, for which the 
commutator continues to vanish on a suitably restricted habitat of 
states --- in which case neither \eqn{HH2} nor \eqn{HH3} would amount 
to much of a consistency test.

\section{Conclusions}

String theory and LQG pursue the same goal, a consistent theory of
quantum gravity, though with very different means. Both approaches
address core issues of quantum gravity, but concentrate on
complementary aspects of the problem, and have led to valuable
insights.  For this reason, the opinion has been voiced that string
theory and LQG may ultimately merge together in a grand synthesis, or
that LQG might become part of string
theory~\cite{Musser:2003,Overbye:2004a}. \footnote{Finding
a connection between \emph{classical} two-form gravity in four
dimensions and topological string theory~\cite{Dijkgraaf:2004te} can
hardly be considered evidence for the connection between LQG and
string theory. String theory provides a large variety of topological 
theories, of which the ones that appear in LQG are only special examples.
More importantly, the quantisation methods employed there
are entirely different from those of LQG.}
On the basis of the available evidence, this does not appear a 
likely outcome to us: the basic hypotheses underlying string theory and 
LQG seem impossible to reconcile without major modifications in either 
approach. There \emph{is} a basic clash here, as reflected in such questions 
as to whether or not supersymmetry is necessary for a consistent quantisation 
of gravity. One might thus say that the differences between the 
two approaches are such that, in the end, at most one `can be right'.

We have reported here on the status of loop quantum gravity from a
somewhat uncommon perspective. Our general conclusion is that, despite
the optimism prevalent in many other reviews, more attention should be
paid to basic aspects and unresolved problems of the theory. As many
LQG practitioners may find this assessment too harsh, let us therefore
summarise once more the basic point we have been trying to make in
this review. The main issue here is not so much specific details
of the LQG approach, where there has been a lot of serious work and 
considerable progress, but rather the question: \emph{what does 
one mean when one speaks of a consistent theory of quantum gravity?} And 
what are the basic properties that such a theory should satisfy? It is 
here that opinions differ. In the introduction, we have spelled out  
some of the criteria that \emph{we} believe such a theory should meet. 
Apart from the question of the semi-classical limit (which is generally 
acknowledged to be a main outstanding problem of the LQG approach), 
and our insistence on the key role of the perturbative two-loop
divergence~\eqn{e:2loopdiv}, we have emphasised the importance of
finding a good criterion for ensuring the \emph{space-time} covariance
of the theory at the quantum level --- after all, this, rather than merely 
covariance under spatial diffeomorphisms, is the essence of Einstein's theory.
We have recalled the well known fact that, in the canonical approach, it 
is the constraint algebra that encodes the information about space-time
covariance, and that everything hinges on how this algebra is to be
implemented in the quantum theory. Whereas LQG proponents generally
seem to be content with~\eqn{HH2} or~\eqn{HH3} as an expression of
covariance, as well as the procedure of solving constraints `in steps', 
using the diffeomorphism constraint already in the very definition of the
Hamiltonian, we have tried to explain why we consider both~\eqn{HH2}
and~\eqn{HH3} as too weak, hence insufficient, in this regard. Instead, 
we have proposed the \emph{off-shell closure} of the quantum constraint 
algebra as a criterion of space-time covariance. We have also argued that 
imposing off-shell closure may further reduce the large number of ambiguities 
present in the formalism, and in the definition of the Hamiltonian constraint,
in particular. As we have stressed repeatedly, it is this constraint 
which lies at the core of canonical quantum gravity.

In summary, we feel that there are still important problems at a basic
level that need to be addressed and resolved before one can tell
whether or not the loop quantum gravity programme can succeed. We hope
that the present paper will provide an incentive to re-focus attention
to these basic issues, and contribute to the debate between the
different approaches.

\bigskip
\section*{Acknowledgements}

This work grew out of the Workshop ``Strings meet Loops'' at the AEI
in October 2003, and the numerous discussions that followed it. We
would especially like to thank Abhay Ashtekar, \mbox{Martin} Bojowald,
Laurent Freidel, Robert Helling, Marc Henneaux, Chris Isham, Claus
Kiefer, Karel Kucha\v{r}, Jurek Lewandowski, \mbox{Renate} Loll,
\mbox{Dieter} Maison, Don Marolf, Slava Mukhanov, Heide Narnhofer,
Hendryk Pfeiffer, Jan Plefka, \mbox{Nicolai} Reshetikhin, Ingo Runkel,
Urs Schreiber, \mbox{Aureliano} Skirzewski, Lee Smolin, Matthias
\mbox{Staudacher} and \mbox{Walter} Thirring for useful discussions
and correspondence. Special thanks go to Kirill \mbox{Krasnov} and
Thomas \mbox{Thiemann} for insightful comments and critical remarks on the
manuscript, and especially to Thomas \mbox{Thiemann} for suggesting numerous
improvements and for patiently explaining his point of view to us.
\mbox{Finally}, we thank the referees for many comments which we believe have 
substantially improved this paper.

\newpage
\begin{small}
\setlength{\bibsep}{3.1pt}

\begingroup\raggedright\endgroup

\end{small}
\end{document}